\documentclass[reprint,aps,pra,amsmath,amssymb,superscriptaddress]{revtex4-1}
\usepackage[english]{babel}
\usepackage[utf8x]{inputenc}
\usepackage[T1]{fontenc}

\usepackage{graphicx}
\usepackage[caption=false]{subfig}
\usepackage[colorlinks=true,allcolors=blue]{hyperref}

\usepackage{physics}
\usepackage{mathtools}
\usepackage{siunitx}
\usepackage{dsfont}
\usepackage{xspace}
\usepackage{xcolor}

\usepackage[normalem]{ulem}

\newcommand{\vect}[1]{\vb*{#1}}
\newcommand{\transpose}{\mathsf{T}}
\newcommand{\dr}{\dd{\vect{r}}}

\newcommand{\imag}{\mathrm{i}}
\newcommand{\e}{\mathrm{e}}
\newcommand{\hc}{\text{h.c.}}

\DeclareMathOperator{\diag}{diag}
\DeclareMathOperator{\spn}{span}

\DeclareMathOperator{\triangleSquare}{triangleSquare}

\newcommand{\R}{\mathbb{R}}

\newcommand{\Z}{\mathbb{Z}}
\newcommand{\LBK}{L_\text{BK}}
\newcommand{\SLS}{SL_S}

\newcommand{\vF}{v_\text{F}}
\newcommand{\gruneisen}{\beta_\text{G}}
\newcommand{\kB}{k_\text{B}}
\newcommand{\const}{\text{const}}

\newcommand{\Tc}{T_\text{c}}
\newcommand{\Thalf}{T_{\SI{50}{\percent}}}
\newcommand{\muc}{\mu_\text{c}}
\newcommand{\muhalf}{\mu_{\SI{50}{\percent}}}
\newcommand{\muopt}{\mu_\text{opt}}
\newcommand{\Ds}{D^\text{s}}
\newcommand{\TBKT}{T_\text{BKT}}

\newcommand{\omegac}{\omega_\text{c}}

\newcommand{\halfcolwidth}{0.47\columnwidth}

\newcommand{\ie}{\emph{i.e.}\xspace}
\newcommand{\eg}{\emph{e.g.}\xspace}
\newcommand{\etal}{\emph{et al.}\xspace}

\newcommand{\BvK}{Born--von K\'arm\'an\xspace}
\newcommand{\moire}{moir\'e\xspace}

\definecolor{TTH-color}{named}{green}
\definecolor{TP-color}{named}{orange}

\definecolor{TTH-color2}{named}{green}
\definecolor{TP-color2}{named}{orange}

\begin{document}

\title{\titlemain}
\newcommand{\titlemain}{Flat-band superconductivity in periodically strained graphene: mean-field and Berezinskii--Kosterlitz--Thouless transition}
\author{Teemu J. Peltonen}
\author{Tero T. Heikkilä}
\affiliation{University of Jyväskylä, Department of Physics and Nanoscience Center,
P.O. Box 35 (YFL), FI-40014 University of Jyväskylä, Finland}

\begin{abstract}
In the search of high-temperature superconductivity one option is to focus on increasing the density of electronic states. Here we study both the normal and $s$-wave superconducting state properties of periodically strained graphene, which exhibits approximate flat bands with a high density of states, with the flatness tunable by the strain profile. We generalize earlier results regarding a one-dimensional harmonic strain to arbitrary periodic strain fields, and further extend the results by calculating the superfluid weight and the Berezinskii--Kosterlitz--Thouless (BKT) transition temperature $\TBKT$ to determine the true transition point. By numerically solving the self-consistency equation, we find a strongly inhomogeneous superconducting order parameter, similarly to twisted bilayer graphene. In the flat-band regime the order parameter magnitude, critical chemical potential, critical temperature, superfluid weight, and BKT transition temperature are all approximately linear in the interaction strength, which suggests that high-temperature superconductivity might be feasible in this system. We especially show that by using realistic strain strengths $\TBKT$ can be made much larger than in twisted bilayer graphene, if using similar interaction strengths. We also calculate properties such as the local density of states that could serve as experimental fingerprints for the presented model.
\end{abstract}

\maketitle


\section{Introduction}

Graphene was long waiting for superconductivity to be added to its long list of miraculous properties. It took over ten years after its discovery before superconductivity was demonstrated in chemically doped graphene \cite{Ludbrook2015,Chapman2016,Ichinokura2016,Tiwari2017} with a critical temperature $\Tc$ of a few kelvin. Recently the experiments on magic-angle twisted bilayer graphene (TBG) \cite{Cao2018b,Yankowitz2016,Lu2019} have drawn much more attention, demonstrating superconductivity in a carbon-only material (although the role of the hexagonal boron nitride substrates is being disputed \cite{Moriyama2019}) similarly with a $\Tc$ of a few kelvin.

Lack of superconductivity in pristine graphene can be understood from the small-$\nu$ limit of the standard Bardeen--Cooper--Schrieffer (BCS) result for the critical temperature, $\Tc\sim\omegac\e^{-1/(\abs{\lambda}\nu)}$ \cite{Bardeen1957,Heikkila2011}, with $\abs{\lambda}$ describing the strength of the attractive electron--electron interaction, $\nu$ being the density of states (DOS) at the Fermi level, and $\omegac$ being the cutoff (Debye) frequency. Since for intrinsic, undoped, graphene the density of states at the Fermi level is $\nu=0$, according to this result we have also $\Tc=0$. The doping experiments can be understood from the same result. Since close to the Dirac point $\nu$ increases linearly with chemical potential, doping can be utilized to render $\Tc$ finite. But due to the exponential suppression of the critical temperature, to produce $\Tc$ of a few kelvin, the chemical potential shift has to be of the order of $\si{eV}$ \cite{Ludbrook2015,Ichinokura2016}, corresponding to a very heavy doping level.

TBG provides an alternative mean to render $\Tc$ finite: increase the density of states by flattening the electronic bands through \moire-modulated interlayer coupling. In the limit of a large $\nu$ (the flat-band limit), BCS theory gives a linear relationship $\Tc\sim\abs{\lambda}\Omega$ \cite{Heikkila2011}, where $\Omega$ is the area of the flat band, instead of the exponential one. The linear relation allows in principle to increase $\Tc$ much higher even with a small interaction $\abs{\lambda}$. Here the limiting factor seems to be the area $\Omega$ of the flat band, which in the case of TBG is roughly the superlattice (\moire) Brillouin zone (SBZ), fixed by the rotation angle $\theta$. Since $\theta$ fixes also the interlayer coupling modulation, the whole dispersion is fixed by the rotation alone. From experiments \cite{Cao2018b,Yankowitz2019} and theories \cite{Bistritzer2011,LopesdosSantos2012} we know that in order to yield flat bands $\theta$ has to be close to the magic angle $\theta^*\approx\SI{1}{\degree}$, for which $\Omega$ is only about $\SI{0.04}{\%}$ \cite{Peltonen2018} of the original graphene Brillouin zone (BZ). An increase of a few kelvin in $\Tc$ has been successfully demonstrated \cite{Yankowitz2016} by applying high pressure to slightly increase $\theta^*$ and thus also $\Omega$. In TBG the flat bands are in fact not exactly at zero energy, but of the order of $\si{meV}$ higher and lower. But compared to chemically doped graphene where $\sim\si{eV}$ doping levels are needed, a thousand-fold reduction in the needed chemical potentials allows using much simpler and more easily tunable electrical doping.

In this paper we study yet another mechanism to produce flat bands in graphene, which is possibly free of the limitations in TBG: periodic strain \cite{Guinea2008,Tang2014,Venderbos2016,Kauppila2016,Tahir2018,Jiang2019a}. Instead of periodically modulating interlayer hopping in TBG, we modulate the intralayer hopping in monolayer graphene by periodic strain. In this system we can, in principle, separately choose the strain period $d$ (and thus the SBZ area $\sim\Omega$) and its strength $\beta$ (and thus the flatness of the bands), potentially allowing us to increase $\Tc$ higher than in TBG by engineering strains with high amplitude and small period.

At low energies, near the $\vect{K}$ and $\vect{K}'=-\vect{K}$ points where graphene can be described as a Dirac material, strain is modelled by a pseudo vector potential $\vect{A}$ \cite{supplement,Suzuura2002,Vozmediano2010,Venderbos2016,Ilan2019}, similarly to an external magnetic field. But while the external magnetic field breaks the time-reversal symmetry and usually suppresses superconductivity, the strain-induced $\vect{A}$ has opposite signs on different valleys, preserving time-reversal symmetry and thus preserving and even promoting spin-singlet superconductivity. Moreover, strain-induced pseudo vector potentials can easily reach an effective magnetic field strength of tens \cite{Guinea2010} or even hundreds \cite{Levy2010,Jiang2019a} of tesla, opening the possibility for extreme tuning of electronic properties.

Possibilities for experimentally producing periodic strain in graphene are numerous. In fact, flat bands have already been observed in an experiment by Jiang \etal \cite{Jiang2019a}, where both 1D and 2D periodic strains were created by boundary conditions. In this experiment the displacement amplitude was of the order of $\SI{1}{\angstrom}$ and the period $d$ was tunable between $8$ and $\SI{25}{nm}$. Even better control of the strain pattern could perhaps be achieved by optical forging \cite{Johansson2017}, which allows drawing arbitrary out-of-plane strain patterns in graphene, even below the diffraction limit \cite{Koskinen2018}. On the other hand the small secondary ripples observed in the simulations \cite{Johansson2017} could be exploited, similarly to the Jiang \etal experiment \cite{Jiang2019a}, but with better control.

Another experimentally demonstrated method is to use an AFM tip to evaporate hydrogen from a Ge-H substrate to produce a pressurized $\text{H}_2$ gas under specific locations of graphene \cite{Jia2019}. One option could be graphene on a corrugated surface \cite{Aitken2010,Jiang2017}. Applying in-plane compression has been predicted to produce periodic wrinkles both in simply-supported \cite{Aitken2010,Yang2016} and encapsulated \cite{Androulidakis2014,Koukaras2016} graphene, with amplitude and period of the order of $\SI{0.2}{\angstrom}$ and $\SI{2}{nm}$, respectively. In the same spirit the proposed graphene cardboard material could be manufactured \cite{Koskinen2014}. Also an ultracold atom gas in a tunable optical honeycomb lattice \cite{Tarruell2012} could be used.

It has been predicted \cite{vanWijk2015,Dai2016b,Nam2017,Lin2018a,Fang2019} and observed \cite{Yoo2018} that TBG exhibits \moire-periodic strain due to lattice mismatch and the following structural relaxation. The relative magnitude of the \moire and strain effects can be, however, difficult to disentangle, as superconductivity by both effects has been predicted by BCS theory \cite{Kauppila2016,Peltonen2018}. But if the \moire effect is enhancing for superconductivity, as it seems to be, we get a lower bound for $\Tc$ by studying the strain effects. Similarly periodic strain can be expected with other mismatch lattices, such as graphene on hBN \cite{Yankowitz2016}.

In this work we generalize the model and results of Kauppila \etal \cite{Kauppila2016}, where both the normal and superconducting $s$-wave state in periodically strained graphene (PSG) have been studied in the case of a cosine-like 1D potential $\vect{A}(x,y)=\frac{\beta}{d}(0,\cos(2\pi x/d))$, to arbitrary periodic pseudo vector potentials $\vect{A}$. This generalization is motivated by the experiment of Jiang \etal \cite{Jiang2019a}, where a variety of periodic strain patterns, both 1D and 2D, were manufactured. On the other hand generalizing the theory to 2D strains bridges the gap between PSG \cite{Kauppila2016} and TBG \cite{Peltonen2018} by showing how similar these two systems are in many aspects.

The main conclusions of Kauppila \etal are that (i) approximate flat bands are formed in the normal state, (ii) the superconducting order parameter $\Delta(x)$ becomes inhomogeneous and is peaked near the minima/maxima of $\nabla\cross\vect{A}$, similarly to the local density of states (LDOS), (iii) magnitude of $\Delta$ can be tuned by the amplitude $\beta$, (iv) $\Tc$ is linear in $\abs{\lambda}$ in the flat-band regime (large $\lambda$ or $\beta$), and (v) even though $\Delta$ is strongly inhomogeneous and anisotropic, supercurrent is only slightly anisotropic. We show that these results continue to hold even when we change the shape of $\vect{A}$ and move to 2D potentials. In addition we show how the shape of $\vect{A}$ and its dimensionality affect the superconducting order parameter $\Delta$, the critical chemical potential $\muc$, and the critical temperature $\Tc$. We furthermore extend the calculations by calculating the superfluid weight \cite{Peotta2015,Liang2017} $\Ds$ and the Berezinskii--Kosterlitz--Thouless (BKT) transition temperature $\TBKT$ to determine the proper transition temperature in a 2D system.

This article is organized as follows. In section~\ref{sec:model} we derive the Bogoliubov--de Gennes (BdG) theory to describe the superconducting state of PSG at low energies, details of which are shown in the Supplementary Material \cite{supplement}. In section~\ref{sec:results} we present the results of applying some selected periodic pseudo vector potentials $\vect{A}$ by numerically solving the self-consistency equation. In section~\ref{sec:conclusions} we summarize the main results and discuss open questions and future prospects.


\section{Model}
\label{sec:model}
In the low-energy limit, after adding an in-plane displacement field $\vect{u}$ and an out-of-plane displacement field $h$, the graphene continuum Hamiltonian for valley $\rho\in\{+,-\}$ is
\begin{equation}
	\mathcal{H}^\rho(\vect{r}) = \hbar\vF\vect{\sigma}^\rho\vdot(-\imag\nabla+\rho \vect{A}(\vect{r})) - \mu,
    \label{eq:H0r}
\end{equation}
where the pseudo vector potential is given by \cite{Suzuura2002,Vozmediano2010,supplement}
\begin{equation}
	\vect{A} = -\frac{\gruneisen}{2a_0} \left(u_{xx}-u_{yy}, -2u_{xy}\right)
    \label{eq:A}
\end{equation}
and the strain tensor is
\begin{equation}
	u_{ij} = \frac{1}{2}(\partial_iu_j+\partial_ju_i) + \frac{1}{2}\partial_ih\partial_jh.
    \label{eq:uij}
\end{equation}
Here $\vF$ is the graphene Fermi velocity, $\mu$ is the chemical potential, $\gruneisen=-\dv*{\ln{t}}{\ln{a_0}}\approx 2$ is the graphene Gr\"uneisen parameter \cite{Vozmediano2010}, $a_0$ is the carbon-carbon bond length, $\vect{\sigma}^\rho = (\rho\sigma_x,\sigma_y)$ is a vector of sublattice-space Pauli matrices, and the graphene zigzag direction is assumed to be in the $x$ direction. Note that $\vect{A}$ works exactly like a vector potential related to an external magnetic field, but with the important difference that it changes sign on valley exchange $\rho\mapsto\bar{\rho}$, preserving time-reversal symmetry $\mathcal{H}^{\bar{\rho}*}=\mathcal{H}^\rho$. Because of the relation \eqref{eq:A} we use the words ``strain'' and ``pseudo vector potential'' interchangeably. Note that for the linear elasticity theory to be valid we should have \cite{supplement}
\begin{equation}
    \norm{\vect{u}(\vect{r}+\vect{\delta}_j)-\vect{u}(\vect{r})}, \norm{h(\vect{r}+\vect{\delta}_j)-h(\vect{r})} \ll a_0,
    \label{eq:validity}
\end{equation}
where $\vect{\delta}_1$, $\vect{\delta}_2$, and $\vect{\delta}_3$ are the graphene nearest neighbor vectors.

We model the possible superconducting state by a (slightly generalized) BCS theory using BdG formalism. We assume an intervalley, local (also in sublattice) interaction of strength $\lambda$ (negative for attractive interaction considered here), which has been widely used in the past graphene literature \cite{Zhao2006,Uchoa2007,Kopnin2008,Kopnin2010,Hosseini2015,Wu2018} to model $s$-wave superconductivity. In this case the effective interacting mean-field continuum Hamiltonian can be shown to be \cite{supplement}
\begin{align}
	H_\text{BdG} &= \sum_{\sigma\rho} \int\dr \psi_{\sigma\rho}^\dagger(\vect{r}) \mathcal{H}^\rho(\vect{r}) \psi_{\sigma\rho}(\vect{r}) \notag\\
    &+ \frac{1}{2} \sum_{\sigma\rho} \int\dr \psi_{\sigma\rho}^\dagger(\vect{r}) \Delta_\sigma(\vect{r}) \psi_{\bar{\sigma}\bar{\rho}}^{\dagger\transpose}(\vect{r}) +\hc +\const,
    \label{eq:H_psi}
\end{align}
where $\sigma\in\{\uparrow,\downarrow\}$ denotes spin, the real space integrals are over the \BvK cell $\R^2/\LBK$, and $\psi_{\sigma\rho}(\vect{r}) = (\psi_{\sigma\rho,A}(\vect{r}), \psi_{\sigma\rho,B}(\vect{r}))^\transpose$ is a sublattice-space vector of the electron annihilation operators. Furthermore the superconducting order parameter in the sublattice space is $\Delta_\sigma(\vect{r}) = \diag(\Delta_{\sigma,A}(\vect{r}), \Delta_{\sigma,B}(\vect{r}))$, where
\begin{equation}
	\Delta_{\sigma,\alpha} = \lambda \sum_{\rho} \expval{ \psi_{\bar{\sigma}\bar{\rho},\alpha} \psi_{\sigma\rho,\alpha} }
    \label{eq:Delta_psi}
\end{equation}
with angle brackets denoting the thermal average and $\alpha\in\{A,B\}$ denoting the sublattice. Note that this kind of a local interaction corresponds to spin-singlet type of superconductivity, since from the fermionic anticommutation relations it directly follows that $\Delta_{\bar{\sigma},\alpha} = -\Delta_{\sigma,\alpha}$. Furthermore due to locality $\vect{r}$ denotes the center-of-mass coordinate of the Cooper pair, while the relative coordinate is always zero, meaning that this interaction corresponds to $s$-wave superconductivity.

Utilizing the fermionic anticommutation relations and by doubling the basis set we can bring $H_\text{BdG}$ in \eqref{eq:H_psi} into the Nambu form
\begin{equation}
	H_\text{BdG} = \frac{1}{2} \sum_{\sigma\rho} \int\dr \Psi_{\sigma\rho}^\dagger(\vect{r}) \mathcal{H}_\text{BdG}^\rho(\vect{r}) \Psi_{\sigma\rho}(\vect{r}) + \const,
\end{equation}
where the BdG Hamiltonian in Nambu space and the Nambu-vector are
\begin{equation}
	\mathcal{H}_\text{BdG}^\rho =
    \begin{pmatrix}
    	\mathcal{H}^\rho & \Delta \\
        \Delta^* & -\mathcal{H}^\rho
    \end{pmatrix}, \quad
	\Psi_{\sigma\rho} =
    \begin{pmatrix}
    	\psi_{\sigma\rho} \\
        s(\sigma)\psi_{\bar{\sigma}\bar{\rho}}^{\dagger\transpose}
    \end{pmatrix},
\end{equation}
respectively. Here the spin-independent order parameter is $\Delta = \Delta_\uparrow = s(\sigma)\Delta_\sigma$, $s(\uparrow)=1$, and $s(\downarrow)=-1$.

Using the spectral theorem, a symmetry between the positive and negative energy states, and defining the fermionic Bogoliubon operators as
\begin{equation}
	\gamma_{\sigma\rho b\vect{k}} = \frac{1}{\sqrt{V}} \int\dr w_{\rho b\vect{k}}^\dagger(\vect{r}) \Psi_{\sigma\rho}(\vect{r}),
    \label{eq:gamma}
\end{equation}
we may bring $H_\text{BdG}$ into the diagonal form \cite{supplement}
\begin{equation}
	H = \frac{1}{2} \sum_{\sigma\rho b\vect{k}} E_{\rho b\vect{k}} \gamma_{\sigma\rho b\vect{k}}^\dagger \gamma_{\sigma\rho b\vect{k}} + \const.
    \label{eq:H_gamma}
\end{equation}
Here $\vect{k}$ together with the band index $b$ enumerate the \emph{positive-energy} solutions of the BdG equation
\begin{equation}
	\mathcal{H}_\text{BdG}^\rho(\vect{r}) w_{\rho b\vect{k}}(\vect{r}) = E_{\rho b\vect{k}} w_{\rho b\vect{k}}(\vect{r})
    \label{eq:BdG_equation_realspace}
\end{equation}
and $V=\abs{\R^2/\LBK}$ is the area of the \BvK cell. According to the calculation above, diagonalizing $H_\text{BdG}$, \ie bringing it to the form \eqref{eq:H_gamma}, is equivalent to solving the BdG equation \eqref{eq:BdG_equation_realspace}.

By inverting the Bogoliubov transformation \eqref{eq:gamma} we may write the definition of the order parameter \eqref{eq:Delta_psi} as the self-consistency equation \cite{supplement}
\begin{equation}
	\Delta_\alpha(\vect{r}) = -\frac{\lambda}{V} \sum_{\rho b\vect{k}} u_{\rho b\vect{k},\alpha}(\vect{r}) v_{\rho b\vect{k},\alpha}^*(\vect{r}) \tanh(\frac{E_{\rho b\vect{k}}}{2\kB T}),
    \label{eq:self-consistency_equation_realspace}
\end{equation}
at temperature $T$, where we denoted the Nambu components of $w$ as $w = (u,v)^\transpose$. Note that $\Delta_\alpha$ might depend on sublattice $\alpha$, while Kauppila \etal \cite{Kauppila2016} defined $\Delta$ by summing over $\alpha$. As we see below, the self-consistent $\Delta_\alpha$ is, in fact, sublattice dependent, leading to a different $\vect{r}$ dependence than in \cite{Kauppila2016}.

In real space the self-consistency equation \eqref{eq:self-consistency_equation_realspace} is local in space but the BdG equation \eqref{eq:BdG_equation_realspace} is a group of 2 difficult differential eigenvalue equations. The equations can be made easier to solve by utilizing periodicity of $\vect{A}$ and writing them in Fourier space. We assume both the pseudo vector potential $\vect{A}:\R^2/SL \rightarrow \R^2$ (and thus the strain) and the order parameter $\Delta$ to be periodic in translations of the arbitrary superlattice $SL=\spn_\Z\{\vect{t}_1,\vect{t}_2\}\subset\R^2$, allowing us to use the Fourier series \cite{supplement}
\begin{equation}
	\vect{A}(\vect{r}) = \sum_{\vect{G}} \e^{\imag\vect{G}\vdot\vect{r}} \tilde{\vect{A}}(\vect{G}), \quad
    \Delta(\vect{r}) = \sum_{\vect{G}} \e^{\imag\vect{G}\vdot\vect{r}} \tilde{\Delta}(\vect{G}).
    \label{eq:A+Delta_fourier_series}
\end{equation}
Here the sums are over $\SLS^*$, where $SL_\text{RZ}^*=SL^*=\spn_\Z\{\vect{G}_1,\vect{G}_2\}$ is the reciprocal lattice of $SL$, $SL_\text{MZ}^*=\spn_\Z\{\vect{G}_1\}$ is a one-dimensional sublattice of $SL^*$, and $S\in\{\text{RZ},\text{MZ}\}$ denotes either the \emph{reduced zone scheme} or the \emph{mixed zone scheme} (the terms are justified below), the latter of which being applicable only if $\vect{A}$ and $\Delta$ are constant in the $\vect{t}_2$ direction, which we call the \emph{1D potential} case. Otherwise we call $\vect{A}$ a $\emph{2D potential}$.

Together with the assumption of the eigenfunctions $w_{\rho b\vect{k}}$ being periodic in the \BvK cell, the Fourier series \eqref{eq:A+Delta_fourier_series} imply the existence of the Bloch-type Fourier series
\begin{equation}
	w_{\rho b\vect{k}}(\vect{r}) = \e^{\imag\vect{k}\vdot\vect{r}} \sum_{\vect{G}} \e^{\imag\vect{G}\vdot\vect{r}} \tilde{w}_{\rho b\vect{k}}(\vect{k}+\vect{G})
    \label{eq:w_bloch_fourier_series}
\end{equation}
and the Fourier space version of the BdG equation \cite{supplement}
\begin{equation}
	\sum_{\vect{G}'} \tilde{\mathcal{H}}_{\text{BdG},\vect{G}\vect{G}'}^\rho(\vect{k}) \tilde{w}_{\rho b\vect{k}}(\vect{k}+\vect{G}') = E_{\rho b\vect{k}} \tilde{w}_{\rho b\vect{k}}(\vect{k}+\vect{G}).
	\label{eq:BdG_equation_kspace_Gcomponents}
\end{equation}
In the matrix form \eqref{eq:BdG_equation_kspace_Gcomponents} can be written as
\begin{equation}
    \tilde{\underline{\mathcal{H}}}_\text{BdG}^\rho(\vect{k}) \tilde{\underline{w}}_{\rho b\vect{k}} = E_{\rho b\vect{k}} \tilde{\underline{w}}_{\rho b\vect{k}},
    \label{eq:BdG_equation_kspace}
\end{equation}
where the underlined variables are matrices or vectors in the $\vect{G}$ space. Here $\vect{k}\in\LBK^*/\SLS^*$ belongs to the superlattice Brillouin zone (SBZ) in the scheme $S$, $b$ enumerates the positive-energy bands for each $\vect{k}$, and the Nambu-space BdG Hamiltonian is
\begin{equation}
	\tilde{\mathcal{H}}_{\text{BdG},\vect{G}\vect{G}'}^\rho(\vect{k}) =
    \begin{pmatrix}
		\tilde{\mathcal{H}}_{\vect{G}\vect{G}'}^\rho(\vect{k}) & \tilde{\Delta}(\vect{G}-\vect{G}') \\
        \tilde{\Delta}^*(\vect{G}'-\vect{G}) & -\tilde{\mathcal{H}}_{\vect{G}\vect{G}'}^\rho(\vect{k})
	\end{pmatrix}
	\label{eq:HGGp}
\end{equation}
with the noninteracting (normal state) Hamiltonian
\begin{align}
	\tilde{\mathcal{H}}_{\vect{G}\vect{G}'}^\rho&(\vect{k}) =
	\label{eq:H0GGp} \\
	&\hbar\vF\vect{\sigma}^\rho\vdot\left[ (\vect{k}+\vect{G})\delta_{\vect{G}\vect{G}'} + \rho\tilde{\vect{A}}(\vect{G}-\vect{G}') \right] - \mu\delta_{\vect{G}\vect{G}'}. \notag
\end{align}
Note the similarity to the Dirac-point low-energy TBG model in \cite{Peltonen2018,Wu2018,Julku2019}: while here $\tilde{\vect{A}}$ couples the sublattices and $\vect{G}$ vectors within the layer, in TBG the Hamiltonian \eqref{eq:H0GGp} has a two-layer structure, $\tilde{\vect{A}}$ is absent, and the interlayer coupling $\tilde{t}_\perp$ couples sublattices and $\vect{G}$ vectors between the layers. As we show in this paper, the second layer is not necessary for yielding flat bands, but what seems to be enough is coupling in the $\vect{G}$ space. To generalize the theory to study the combined effect of periodic strain and \moire physics, which should yield even more pronounced flat bands, would thus be easy: add the second rotated layer to the noninteracting Hamiltonian \eqref{eq:H0GGp} and couple the layers by $\tilde{t}_\perp(\vect{G}-\vect{G}')$.

Let us discuss the notion of the reduced and the mixed zone schemes. In the reduced zone scheme $\vect{k}=k_1\vect{G}_1+k_2\vect{G}_2\in\LBK^*/SL_\text{RZ}^*$ is periodic both in the $\vect{G}_1$ and $\vect{G}_2$ directions, with both $k_1,k_2\in[-\frac{1}{2},\frac{1}{2}[$ being periodic Bloch momenta. This is also traditionally called the reduced zone (or the repeated zone) scheme. In the case of $\vect{A}$ and $\Delta$ being constant in the $\vect{t}_2$ direction (the 1D potential case) we are also allowed to use the mixed zone scheme, where $\vect{k}=k_1\vect{G}_1+k_2\vect{G}_2\in\LBK^*/SL_\text{MZ}^*$ is periodic only in the $\vect{G}_1$ direction but not in the $\vect{G}_2$ direction, with $k_1\in[-\frac{1}{2},\frac{1}{2}[$ being a periodic Bloch momentum and $k_2\in]-\infty,\infty[$ being a nonperiodic real momentum. Thus in the traditional notion the $\vect{G}_1$ direction is in the reduced (or repeated) zone and the $\vect{G}_2$ direction in the extended zone scheme, justifying the term mixed zone scheme.

The reduced zone scheme is convenient if one wants to compare the effects of the 1D and 2D potentials, since the dispersions look similar and the notion of a band is the same, but the calculations are heavy due to the $\vect{G}$ space being two-dimensional. On the other hand the mixed zone scheme produces cleaner-looking dispersions and is computationally much lighter due to the $\vect{G}$ space being only one-dimensional, but with the cost of more difficult comparison between the 1D and 2D potentials. Thus in all the 1D potential calculations we use the mixed zone scheme unless otherwise stated. Also Kauppila \etal \cite{Kauppila2016} used the mixed zone scheme in all the calculations and visualizations.

Using the Fourier series \eqref{eq:A+Delta_fourier_series} and \eqref{eq:w_bloch_fourier_series} in \eqref{eq:self-consistency_equation_realspace} and approximating the $\vect{k}$ sum as an integral (assuming the \BvK cell to be large), the Fourier-space self-consistency equation becomes \cite{supplement}
\begin{align}
	\tilde{\Delta}_\alpha(\vect{G}) = -&\frac{\lambda}{(2\pi)^2} \sum_{\rho b\vect{G}'} \int\dd{\vect{k}} \tanh(\frac{E_{\rho b\vect{k}}}{2\kB T}) \times \notag\\
    \times&\tilde{u}_{\rho b\vect{k},\alpha}(\vect{k}+\vect{G}') \tilde{v}_{\rho b\vect{k},\alpha}^*(\vect{k}+\vect{G}'-\vect{G}), \label{eq:self-consistency_equation_kspace}
\end{align}
where the integral is over the continuum superlattice Brillouin zone $\R^2/\SLS^*$ in the scheme $S$, which in the reduced zone scheme can be interpreted as the parallelogram defined by $\vect{G}_1$ and $\vect{G}_2$, and in the mixed zone scheme as the semi-infinite parallelogram with the finite side being $\vect{G}_1$ and the infinite side being in the direction of $\vect{G}_2$.

In summary, in Fourier space we are solving the BdG equation \eqref{eq:BdG_equation_kspace_Gcomponents} together with the self-consistency equation \eqref{eq:self-consistency_equation_kspace}. Now the BdG equation is a normal matrix eigenvalue equation, but the price to pay is that the corresponding matrix has countably infinite dimension ($2\times 2\times\abs{\SLS^*}$), and the self-consistency equation becomes nonlocal in the Fourier components. Numerically, however, they are easy to solve, provided we truncate the Fourier-component set $\SLS^*$ and the band sum, and in the case of 1D potential add a momentum cutoff in the $\vect{k}$ integral in the $\vect{G}_2$ direction. These cutoffs we choose so large that the results (dispersion, $\Delta$) start to become saturated, and together they correspond to the energy cutoff $\hbar\omegac$ introduced earlier.

In a 2D system, however, we know that the superconducting transition is not properly described by the mean-field critical temperature $\Tc$ determined from the order parameter $\Delta$, but by the BKT transition temperature determined from the superfluid weight $\Ds$, which describes the linearized supercurrent density response $\expval{\vect{j}} = (\frac{\si{e}}{\hbar})^2 \Ds\expval{\vect{\mathcal{A}}}$ to an external (real) vector potential $\vect{\mathcal{A}}$, where the angle brackets denote average over position. For the present model we may calculate the $\mu,\nu\in\{x,y\}$ component of the superfluid weight from \cite{Liang2017,supplement}
\begin{align}
    \Ds_{\mu\nu} &= \frac{(\hbar\vF)^2}{(2\pi)^2} \sum_{\rho bb'} \int\dd{\vect{k}} \frac{f(E_{\rho b})-f(E_{\rho b'})}{E_{\rho b}-E_{\rho b'}} \times
    \label{eq:Ds} \\
    \times&\left( \tilde{\underline{w}}_{\rho b}^\dagger \sigma_\mu^\rho \tilde{\underline{w}}_{\rho b'} \tilde{\underline{w}}_{\rho b'}^\dagger \sigma_\nu^\rho \tilde{\underline{w}}_{\rho b} - \tilde{\underline{w}}_{\rho b}^\dagger \tau_z\sigma_\mu^\rho \tilde{\underline{w}}_{\rho b'} \tilde{\underline{w}}_{\rho b'}^\dagger \tau_z\sigma_\nu^\rho \tilde{\underline{w}}_{\rho b} \right), \notag
\end{align}
where the $b,b'$ band sums are calculated over both the positive and negative energy bands, $\tau_z$ is the Pauli-$z$ matrix in Nambu space, $f$ is the Fermi--Dirac distribution, the difference quotient is interpreted as the derivative $f'(E_{\rho b})$ if $E_{\rho b}=E_{\rho b'}$, and where we suppressed the $\vect{k}$ dependence.

From the temperature dependence of $\Ds$ we can then calculate the BKT transition temperature $\TBKT$ from the generalized KT--Nelson criterion \cite{Nelson1977,Cao2014,Xu2015}
\begin{equation}
    \kB\TBKT = \frac{\pi}{8} \sqrt{\det\Ds(\TBKT)}
    \label{eq:TBKT}
\end{equation}
for an anisotropic superfluid weight, which also needs to be calculated self-consistently, unless $\Ds(\TBKT)\approx\Ds(0)$.


\section{Results}
\label{sec:results}
We solve \footnote{The \textsc{Mathematica} code we have used to compute the normal-state properties, the mean-field order parameter, superfluid weight, and the BKT transition temperature will be made available upon publication of this manuscript.} the order parameter $\Delta$, the superfluid weight $\Ds$, and the Berezinskii--Kosterlitz--Thouless transition temperature $\TBKT$ for a selection of periodic pseudo vector potentials $\vect{A}$ with the period $d$. $\Delta$ is solved from the self-consistency equation \eqref{eq:self-consistency_equation_kspace} by the fixed-point iteration method with the initial guess of a constant order parameter $\Delta_A=\Delta_B$ \cite{supplement}, $\Ds$ is calculated from \eqref{eq:Ds}, and $\TBKT$ is calculated by interpolating \eqref{eq:TBKT} in a predetermined temperature mesh.

\begin{figure}
	\includegraphics[width=\halfcolwidth]{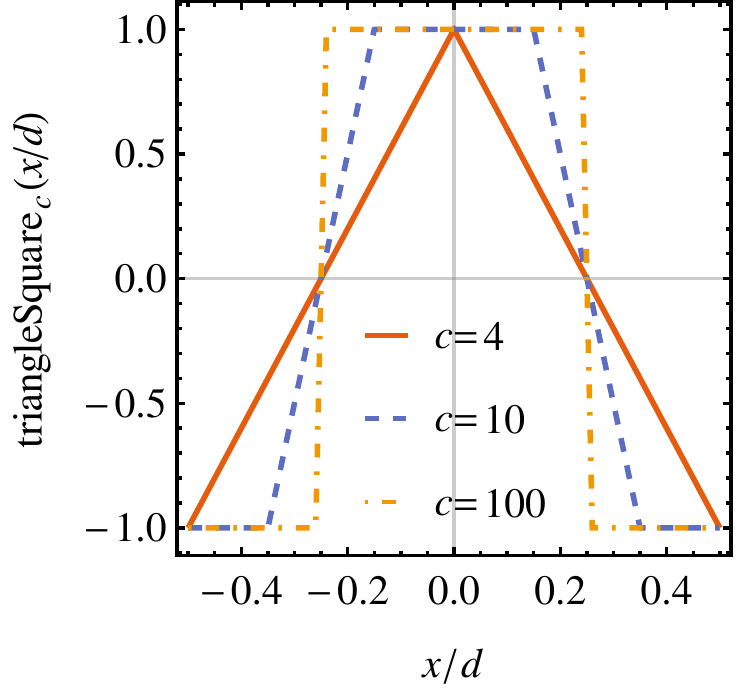}
    \caption{A plot of the $d$-periodic function $\triangleSquare_c$ used for defining the potential $\vect{A}_c^\text{1D}$, shown here for three values of $c$. The slope near the points $x=\mp d/4$ is given by $\pm c/d$.}
    \label{fig:triangleSquare_slope}
\end{figure}

In the case of a 1D potential we concentrate on the potentials
\begin{align}
	\vect{A}_{\cos}^\text{1D}(x,y) &= \frac{\beta}{d} (0, \cos(2\pi x/d)), \label{eq:A_cos1D} \\
    \vect{A}_c^\text{1D}(x,y) &= \frac{\beta}{d} (0, \triangleSquare_c(x/d)), \label{eq:A_c}
\end{align}
both periodic in translations of the square superlattice $SL=\spn_\Z\{\vect{t}_1,\vect{t}_2\}$ with the primitive vectors $\vect{t}_1=(d,0)$ and $\vect{t}_2=(0,d)$ (or any multiple of $\vect{t}_2$). The latter utilizes the function $\triangleSquare_c$, shown in figure~\ref{fig:triangleSquare_slope}, which is a $d$-periodic waveform where the slope parameter $c\in[4,\infty[$ can be used to interpolate between the triangle and square waveforms. This allows controlling the slope $\pm\beta c/d^2$ of $\vect{A}_c^\text{1D}$ at the lines $x=\mp d/4$. Note that the potential $\vect{A}_{2\pi}^\text{1D}$ has exactly the same slope as $\vect{A}_{\cos}^\text{1D}$ at the points $x=\pm d/4$ and also otherwise approximates that potential rather well, so all the following results are more or less indistinguishable between these two potentials. Since both the potentials $\vect{A}_{\cos}^\text{1D}$ and $\vect{A}_c^\text{1D}$ are constant in the $\vect{t}_2$ direction, this allows us to use either the reduced zone or the mixed zone scheme in the theory.

To concretize the difference between the two schemes we write the Fourier components of the cosine potential. In the reduced zone scheme they are \cite{supplement} 
\begin{align}
    \tilde{\vect{A}}_{\cos}^\text{1D}(m_1\vect{G}_1&+m_2\vect{G}_2) = \notag\\
    &\frac{\beta}{2d}(0, \delta_{m_1,-1}+\delta_{m_1,1})\delta_{m_2,0},
\end{align}
for the cosine potential and for $\vect{A}_c^\text{1D}$ they are given in the Supplementary Material \cite{supplement}.
Here $m_1\vect{G}_1+m_2\vect{G}_2\in SL_\text{RZ}^*=SL^*$ belongs to SBZ in the reduced zone scheme, where the SBZ primitive vectors are $\vect{G}_1=(2\pi/d,0)$ and $\vect{G}_2=(0,2\pi/d)$. But since for the 1D potentials the components are multiplied by $\delta_{m_2,0}$, we may as well use a one-dimensional Fourier series \cite{supplement} and define in the mixed zone scheme
\begin{equation}
    \tilde{\vect{A}}_{\cos}^\text{1D}(m_1\vect{G}_1) = \frac{\beta}{2d}(0, \delta_{m_1,-1}+\delta_{m_1,1}),
\end{equation}
where $m_1\vect{G}_1\in SL_\text{MZ}^*$ belongs to SBZ in the mixed zone scheme.

On the other hand in the 2D case we concentrate on the simplest generalization of the 1D cosine-like potential $\vect{A}_{\cos}^\text{1D}$, the potential
\begin{equation}
	\vect{A}_{\cos}^\text{2D}(x,y) = \frac{\beta}{d} (\cos(2\pi y/d), \cos(2\pi x/d))
	\label{eq:A_cos2D}
\end{equation}
with the lattice of periodicity being the square superlattice $SL=\spn_\Z\{\vect{t}_1, \vect{t}_2\}$, with the primitive vectors $\vect{t}_1=(d,0)$, $\vect{t}_2=(0,d)$. Note that we are allowed to choose a potential periodic in any superlattice, whereas in TBG the (\moire) superlattice is fixed by the rotation angle. Thus in principle the periodic strain allows much more freedom in tuning the system. The Fourier components of the 2D cosine potential are
\begin{align}
    \tilde{\vect{A}}_{\cos}^\text{2D}(m_1\vect{G}_1&+m_2\vect{G}_2) = \notag\\
    &\frac{\beta}{2d}(\delta_{m_2,-1}+\delta_{m_2,1}, \delta_{m_1,-1}+\delta_{m_1,1})
\end{align}
in the reduced zone scheme, where $m_1\vect{G}_1+m_2\vect{G}_2\in SL_\text{RZ}^*$ with $\vect{G}_1=(2\pi/d,0)$ and $\vect{G}_2=(0,2\pi/d)$.

\begin{figure}
    \includegraphics[width=\columnwidth]{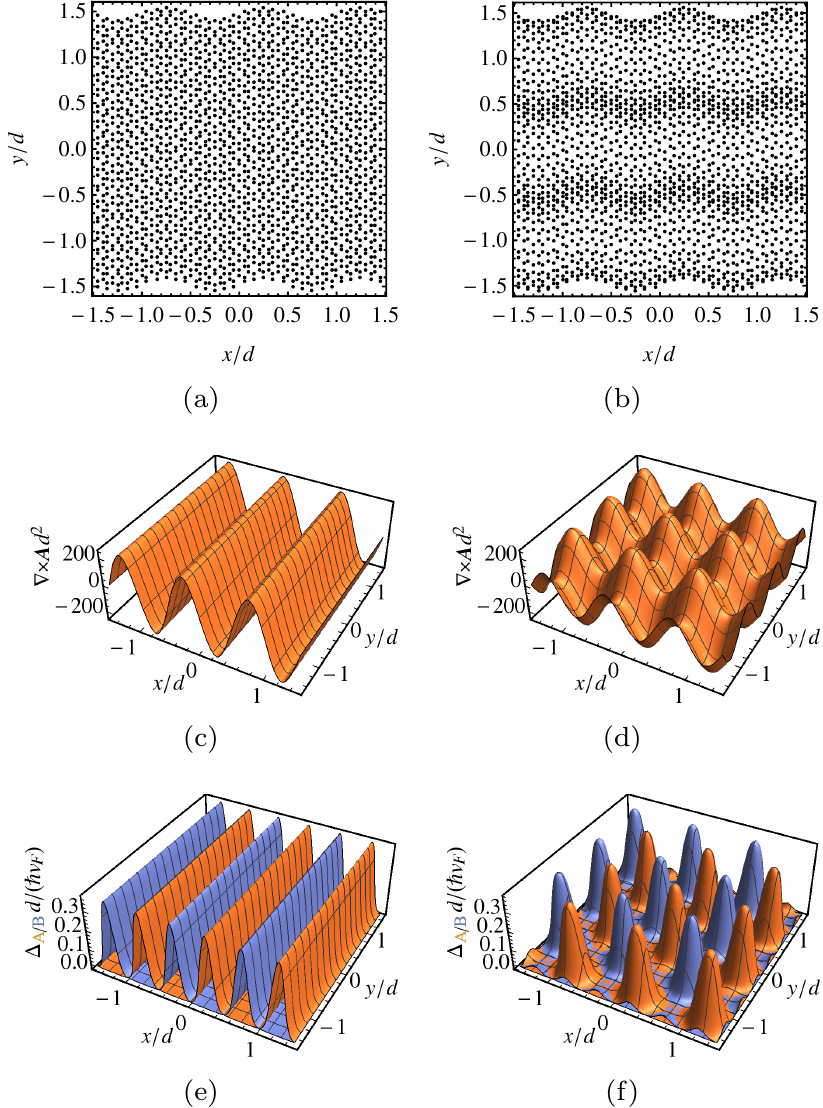}
    \caption{(a,b) Example in-plane displacement fields, defined in \eqref{eq:u_cos1D} and \eqref{eq:u_cos2D}, producing the studied pseudo vector potentials $\vect{A}_{\cos}^\text{1D}$ and $\vect{A}_{\cos}^\text{2D}$ through \eqref{eq:A} with exaggeratedly large amplitude and small period. (c,d) The corresponding pseudomagnetic fields $\vect{B}=\nabla\cross\vect{A}$ with $\beta=40$ and $\beta=20$, respectively. (e,f) Corresponding typical profiles of the self-consistent superconducting order parameter $\Delta_{A/B}$ ($A$ orange, $B$ blue), which is always peaked at the minima/maxima of $\nabla\cross\vect{A}$. The parameters for calculating $\Delta$ are $T=0$, $\lambda/(\hbar\vF d)=-0.01$, and optimal doping $\mu=\muopt$ yielding a maximal $\Delta$ ($\mu=0$ produces the same $\Delta$ for such large $\lambda$).}
    \label{fig:uvsr+curlAvsr+Deltavsr}
\end{figure}

According to \eqref{eq:A} the potentials $\vect{A}_{\cos}^\text{1D}$ and $\vect{A}_{\cos}^\text{2D}$ can be produced for example by the in-plane displacement fields
\begin{align}
    \vect{u}_{\cos}^\text{1D}(x,y) &= \frac{\beta a_0}{\gruneisen\pi} (0,\sin(2\pi x/d)), \label{eq:u_cos1D} \\
    \vect{u}_{\cos}^\text{2D}(x,y) &= \frac{\beta a_0}{\gruneisen\pi} (0,\sin(2\pi x/d)+\sin(2\pi y/d)), \label{eq:u_cos2D}
\end{align}
respectively. The pseudomagnetic fields $\vect{B}=\nabla\cross\vect{A} = \partial_xA_y-\partial_yA_x$ produced by the 1D and 2D cosine potentials, together with these example displacement fields, are shown in figures~\ref{fig:uvsr+curlAvsr+Deltavsr}(a--d). The amplitude $B$ of $\vect{B}$, which is an important factor determining the flatness of the bands and the magnitude of the superconducting order parameter $\Delta_{A/B}$, is
\begin{equation}
    B_{\cos}^\text{1D} = \frac{2\pi\beta}{d^2}, \quad B_c^\text{1D} = \frac{c\beta}{d^2}, \quad B_{\cos}^\text{2D} = \frac{4\pi\beta}{d^2}
\end{equation}
for the potential $\vect{A}_{\cos}^\text{1D}$, $\vect{A}_c^\text{1D}$, or $\vect{A}_{\cos}^\text{2D}$, respectively. To give a realistic scale for $\beta$, in the experiment by Jiang \etal \cite{Jiang2019a} a pseudomagnetic field of $\frac{\hbar}{e}B\approx\SI{100}{T}$ was observed for a strain period of $d=\SI{14}{nm}$, which corresponds to $\beta\approx 5$ for the 1D cosine potential. To be better in the flat-band regime, we mostly use a factor of $4$ to $8$ times larger values of $\beta$ in this study.

Corresponding typical profiles of $\Delta_{A/B}$ for the cosine potentials are shown in figures~\ref{fig:uvsr+curlAvsr+Deltavsr}(e--f), from where it is clear that $\Delta_{A/B}$ is always peaked at the minima/maxima of the pseudomagnetic field $\vect{B}$. For comparison in TBG \cite{Peltonen2018} $\Delta$ is localized around the $AA$ stacking regions and is independent of the sublattice and layer. Note that the sublattice dependence was not present in the work by Kauppila \etal \cite{Kauppila2016} due to sublattice-summation in the self-consistency equation. As we see below, it is approximately the maximum (over position $\vect{r}$) of the order parameter that is important in describing the strength of the superconducting state. As for all the studied potentials the maximum of the order parameter is independent of the sublattice, we simply denote $\max\Delta \coloneqq \max\Delta_A = \max\Delta_B$.

\begin{figure}
    \includegraphics[width=\columnwidth]{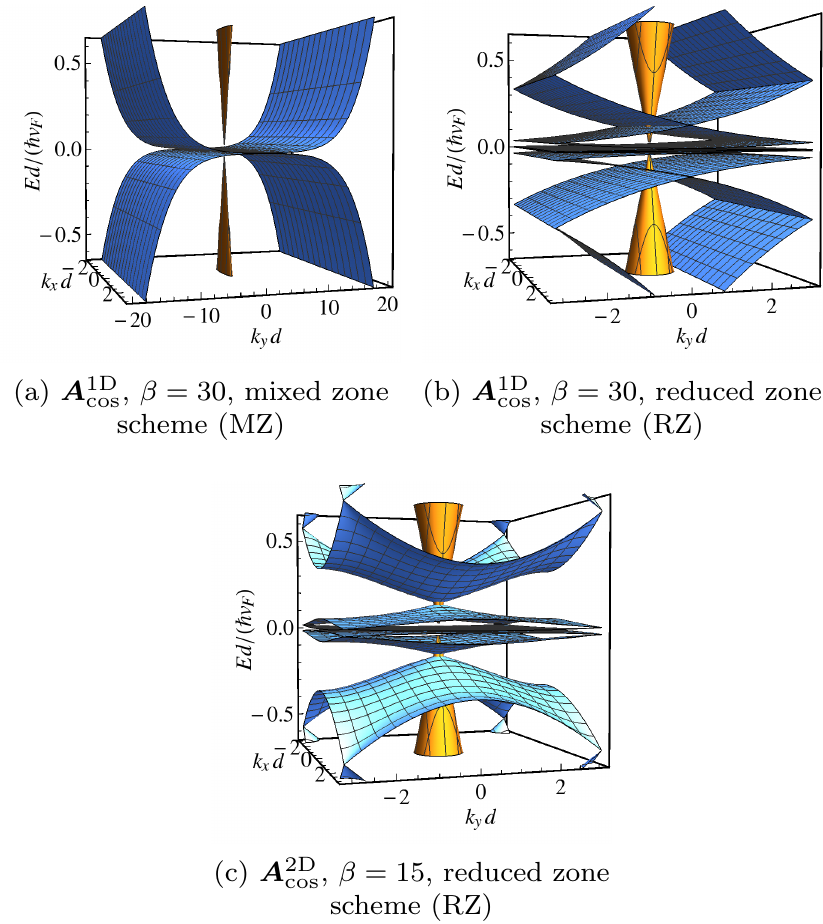}
    \caption{Typical dispersions in the normal state at the valley $\rho=\pm$ with $\mu=0$. (a,b) The 1D cosine potential $\vect{A}_{\cos}^\text{1D}$ (shown for $\beta=30$) (a) in the mixed zone scheme (MZ) and (b) in the reduced zone scheme (RZ). (c) The 2D cosine potential $\vect{A}_{\cos}^\text{2D}$ (shown for $\beta=15$) in the reduced zone scheme. The strained dispersions are shown in blue and for comparison the conical unstrained graphene dispersions in orange.}
    \label{fig:Evsk}
\end{figure}

The typical dispersion relations in the normal state are shown in figure~\ref{fig:Evsk} together with the conical unstrained graphene dispersions. For an easier comparison the 1D potential $\vect{A}_{\cos}^\text{1D}$ dispersion is shown both in the mixed zone and reduced zone schemes, while the 2D potential $\vect{A}_{\cos}^\text{2D}$ dispersion only in the reduced zone scheme (the only possibility in this case). We find similar-looking approximate flat bands as in TBG \cite{Peltonen2018,Julku2019}, with the difference that here the number and the flatness of the flat bands can be controlled by $\beta$ and $c$. Also all the successive bands are touching, while in TBG many models predict the flat bands to be isolated \cite{Nam2017,Tarnopolsky2019,Fang2019,Julku2019}.

\begin{figure}
    \includegraphics[width=\columnwidth]{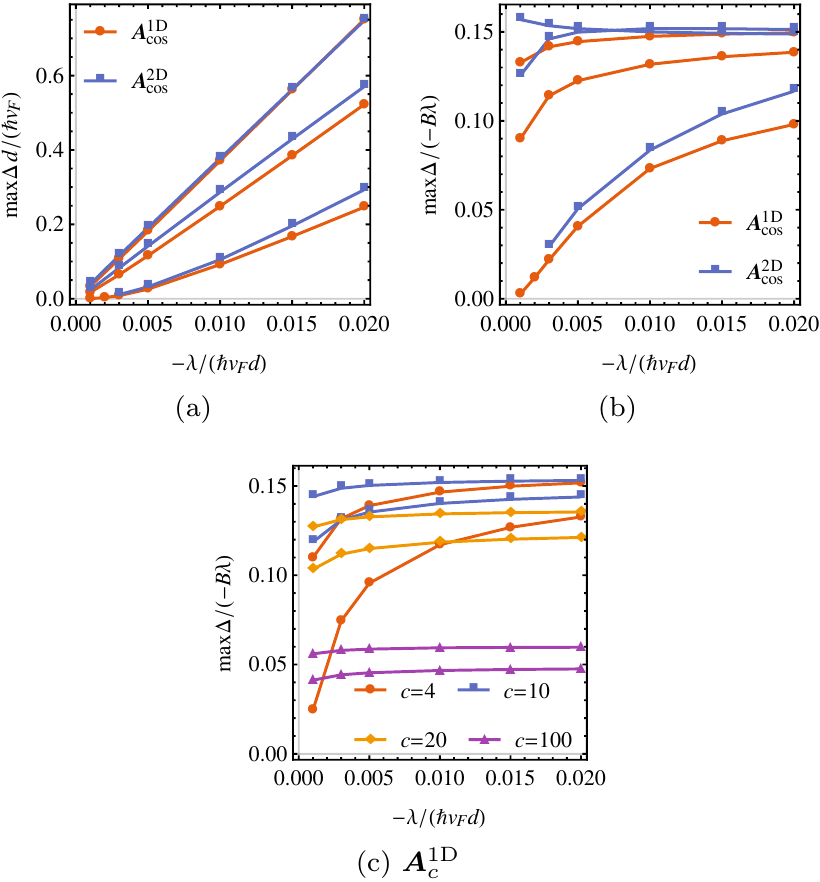}
    \caption{Behavior of the maximum of the superconducting order parameter $\Delta$ as a function of the interaction strength $\lambda$ at optimal doping $\mu=\muopt$ and $T=0$. (a) Linearity of $\max\Delta$ in $\lambda$ in the flat-band regime for the cosine potentials. Each potential has three curves corresponding to $\beta=20,30,40$ (1D potential) or $\beta=10,15,20$ (2D potential), from bottom to top. (b,c) The ratio $\max\Delta/(-\lambda B)$ as a function of $\lambda$ for (b) the cosine potentials and (c) $\vect{A}_c^\text{1D}$ with varying $c$, where $B$ is the amplitude of the pseudomagnetic field $\vect{B}$. In (b) the curves are the same as in (a) while in (c) each $c$ has two curves corresponding to $\beta=30,40$, from bottom to top. In the flat-band regime the ratio tends approximately to a constant as in \eqref{eq:Delta_result}.}
    \label{fig:Deltavslambda}
\end{figure}

We calculate most of the superconducting state results at \emph{optimal doping} $\mu=\muopt$, which is the energy of the density of states peak as discussed in Sec.~\ref{sec:order_parameter_profile_dispersion_and_density_of_states}, and is thus the doping level with the highest $\Delta$. We start discussing the superconducting state results by calculating $\max\Delta$ as a function of the interaction strength $\lambda$ for the different potentials $\vect{A}$, as shown in figures~\ref{fig:Deltavslambda}(a) for the cosine potentials. The most important conclusion is that for large enough $\lambda$, $\beta$, or $c$, which we call the \emph{flat-band regime} due to the energy scale of $\Delta$ exceeding the flat-band bandwidth, the dependence is linear in $\lambda$ as we would expect for any flat-band superconductor \cite{Heikkila2011}. On the other hand for small enough $\lambda$, $\beta$, and $c$ the dispersive behavior of the lowest energy bands starts playing a role, which we call the \emph{dispersive regime}. In the dispersive regime the order parameter is exponentially suppressed and we also start seeing quantum critical points \cite{Kopnin2008}. We further see how in the flat-band regime the behavior of $\vect{A}_{\cos}^\text{2D}$ with $\beta$ is similar to that of $\vect{A}_{\cos}^\text{1D}$ with $2\beta$. Since in this paper we are mostly interested in the flat-band regime, we choose to calculate many of the following results at the fixed interaction strength $\lambda/(\hbar\vF d)=-0.01$, which is clearly in the flat-band regime except for $\vect{A}_{\cos}^\text{2D}$ with $\beta=10$, which is at the interface of the dispersive and flat-band regimes.

To further confirm that in the flat-band regime $\max\Delta$ is linear both in the interaction strength $\lambda$ and the amplitude $B$ of the pseudomagnetic field $\vect{B}$,
\begin{equation}
    \max\Delta = -\zeta B\lambda,
    \label{eq:Delta_result}
\end{equation}
we show the ratio $\zeta$ for all the potentials in figures~\ref{fig:Deltavslambda}(b,c) at $\mu=\muopt$ and $T=0$. In the flat-band regime $\zeta$ tends approximately to a constant $\zeta\approx 0.15$, which holds as long as $c\lesssim 20$. For $c\gtrsim 20$ we start seeing deviations from this result, with $\zeta\approx 0.05$ for the extreme case of $c=100$. The small variation in $\zeta$ due to $c$ even in the flat-band regime is most likely due to the fact that the maximum of $\Delta$ is not exactly the correct quantity to calculate, but it gives a very good estimate. We may compare this to the exact-flat-band result \cite{Peltonen2018} with a constant $\Delta^\text{FB}$, for which $\Delta^\text{FB} = -\frac{1}{(2\pi)^2}n\Omega\lambda$ with $\Omega=1/d^2$ and $n$ being the number of flat bands. In PSG it is the amplitude $B$ of the pseudomagnetic field $\vect{B}$ that effectively determines $n\Omega$, the number of approximate flat bands in the system with the SBZ area of $1/d^2$.

\subsection{Order parameter profile, dispersion, and density of states}
\label{sec:order_parameter_profile_dispersion_and_density_of_states}

\begin{figure}
    \includegraphics[width=\columnwidth]{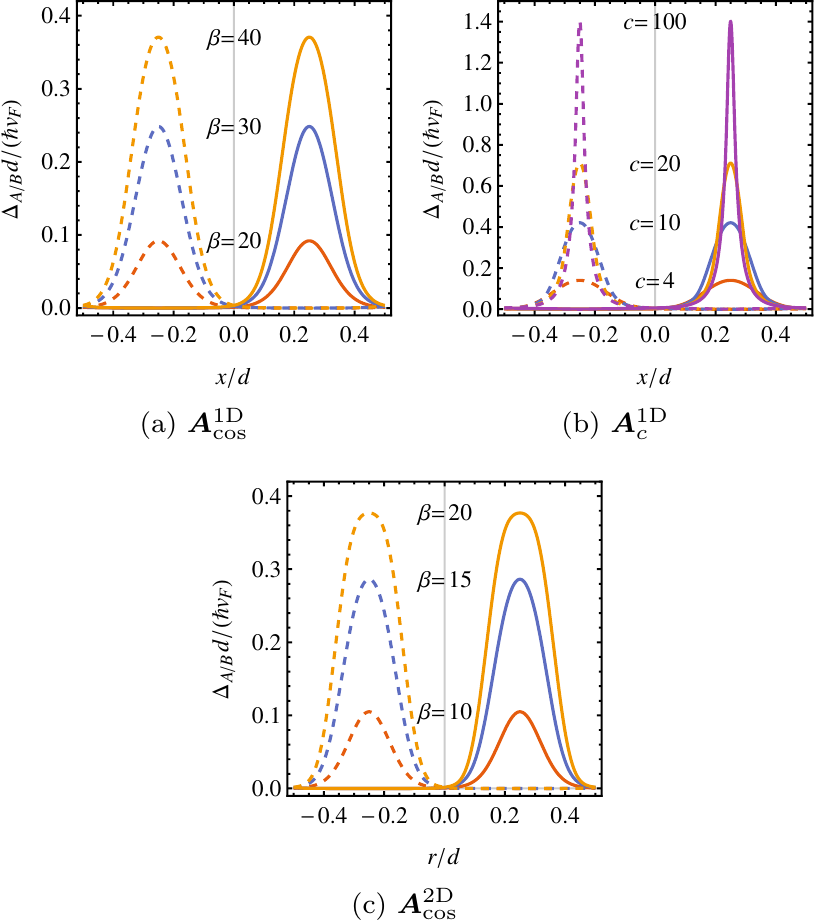}
    \caption{Effect of the amplitude $\beta$ and the slope parameter $c$ on $\Delta_{A/B}$ ($A$ solid, $B$ dashed lines) at $\lambda/(\hbar\vF d)=-0.01$, and optimal doping $\mu=\muopt$ ($\mu=0$ produces the same $\Delta$ for such large $\lambda$). (a) Varying $\beta$ of the 1D cosine potential $\vect{A}_{\cos}^\text{1D}$. (b) Varying the slope parameter $c$ of the 1D potential $\vect{A}_c^\text{1D}$ with $\beta=30$. (c) Varying $\beta$ of the 2D cosine potential $\vect{A}_{\cos}^\text{2D}$. $\Delta_{A/B}$ is drawn along the line $(x,0)$ [1D potentials] or $r(1,-1)$ [2D potential].}
    \label{fig:Deltavsx}
\end{figure}

In figure~\ref{fig:Deltavsx} we show a cross section of the self-consistent $\Delta_{A/B}$ [as in figures~\ref{fig:uvsr+curlAvsr+Deltavsr}(e,f)] along the line $(x,0)$ [1D potentials] or $r(1,-1)$ [2D potential] for different potentials $\vect{A}$, strain strengths $\beta$, and slope parameters $c$. The effect of $\beta$ is to simply linearly increase the amplitude of $\Delta_{A/B}$. On the other hand increasing $c$ not only increases the amplitude of $\Delta_{A/B}$, but makes it also more localized. We also see that for the 2D potential $\vect{A}_{\cos}^\text{2D}$, $\Delta_{A/B}$ with the strain strength $\beta$ along the diagonal behaves similarly as $\Delta_{A/B}$ in the $x$ direction for the 1D potential $\vect{A}_{\cos}^\text{1D}$ with $2\beta$.

\begin{figure}
    \includegraphics[width=\columnwidth]{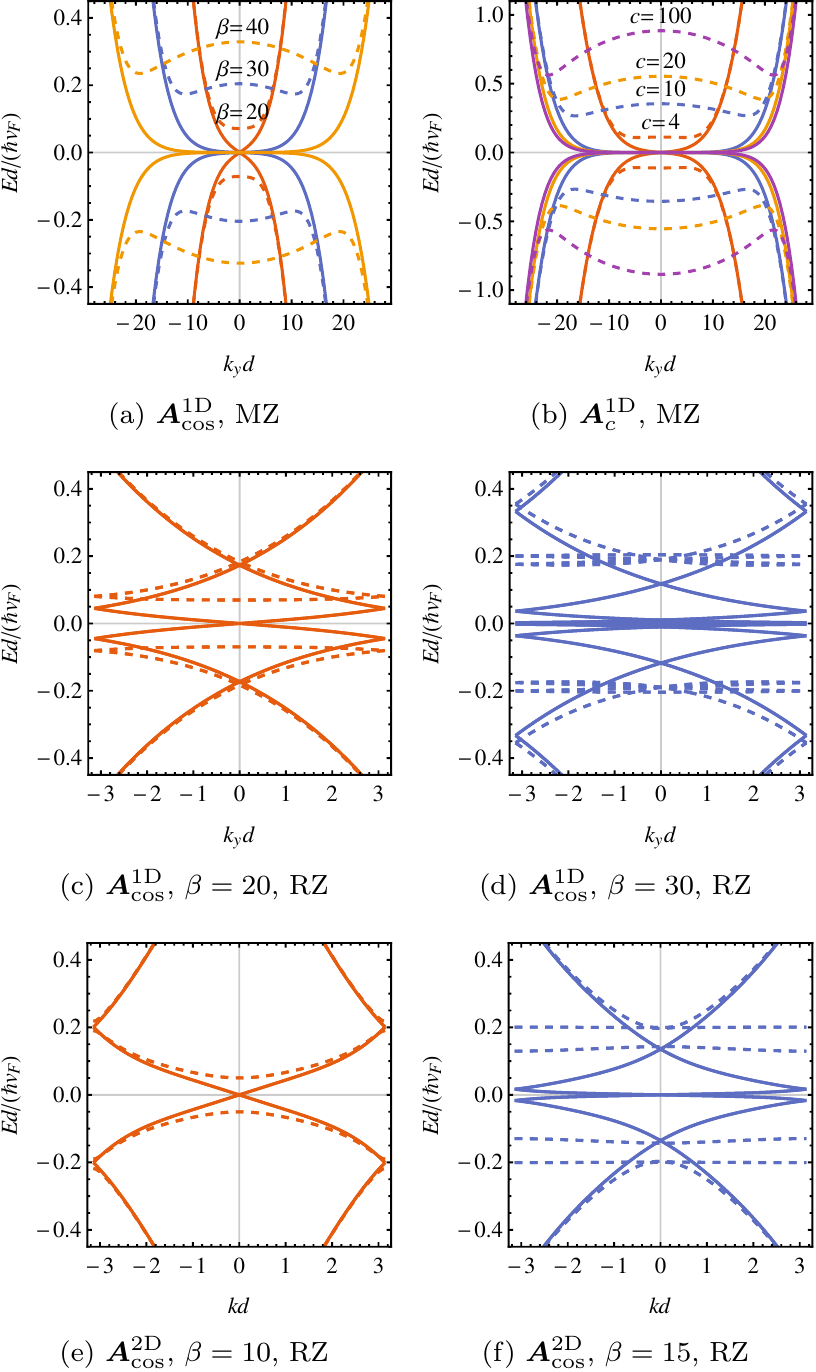}
    \caption{Effect of the strain strength $\beta$ and the slope parameter $c$ on the dispersion (normal state: solid, superconducting state: dashed lines) for the different potentials at $\mu=0$. In the superconducting state $T=0$ and $\lambda/(\hbar\vF d)=-0.01$. (a,b) Dispersions in the mixed zone scheme (MZ) along the line $(k_y,0)$ for (a) $\vect{A}_{\cos}^\text{1D}$ with various $\beta$ and for (b) $\vect{A}_c^\text{1D}$ with various $c$ and fixed $\beta=30$. (c,d) Corresponding dispersions in the reduced zone scheme (RZ) along the line $(0,k_y)$ for $\vect{A}_{\cos}^\text{1D}$ with $\beta=20$ and $30$, respectively. (e,f) Dispersions for $\vect{A}_{\cos}^\text{2D}$ along the diagonal line $k(1,1)$ in the reduced zone scheme for $\beta=10$ and $15$, respectively.}
    \label{fig:Evsky+Evskdiag}
\end{figure}

\begin{figure}
    \includegraphics[width=\columnwidth]{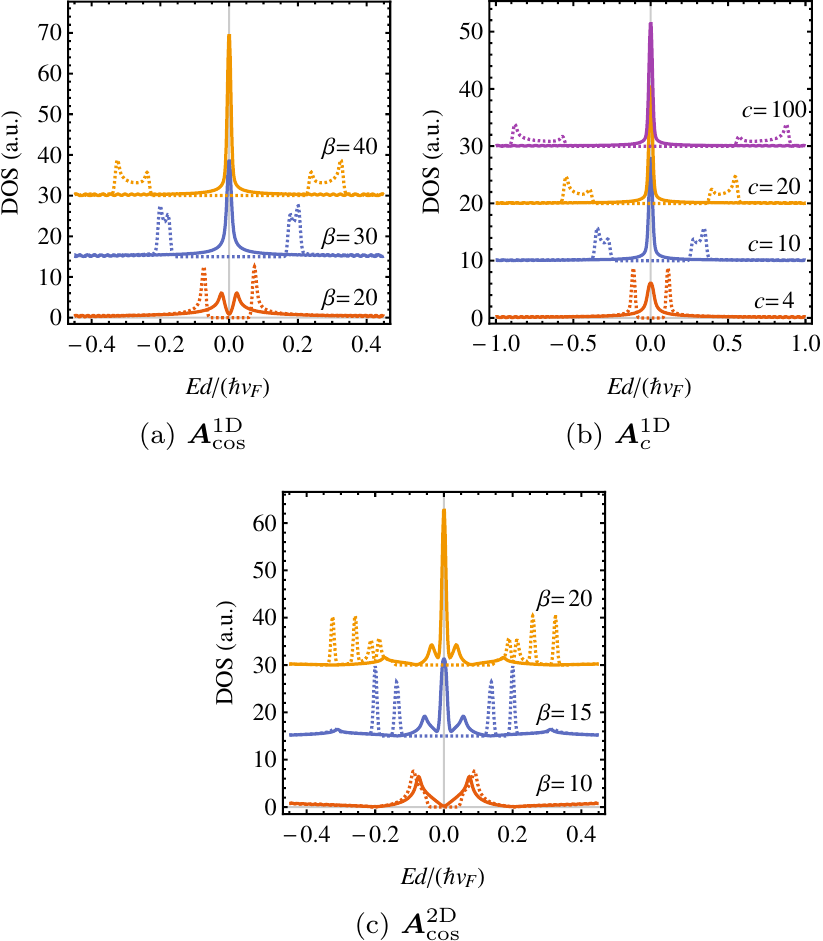}
    \caption{Effect of the strain strength $\beta$ and the slope parameter $c$ on the density of states (DOS) at $\mu=0$, $T=0$, and $\lambda/(\hbar\vF d)=-0.01$ (normal state: solid, superconducting state: dashed lines) for (a) $\vect{A}_{\cos}^\text{1D}$, (b) $\vect{A}_c^\text{1D}$, and (c) $\vect{A}_{\cos}^\text{2D}$. For clarity the successive curves in the DOS plots are shifted vertically by $15$ in (a,c) and by $7$ in (b). Each curve is normalized such that the shown area integrates to unity.}
    \label{fig:DOSvsE}
\end{figure}

These effects we can further see in the dispersions and densities of states in figures~\ref{fig:Evsky+Evskdiag} and \ref{fig:DOSvsE}, respectively, which are plotted at $\mu=0$ for clarity. In figure~\ref{fig:Evsky+Evskdiag} we show the cross section of the dispersions in figure.~\ref{fig:Evsk} along the line $(0,k_y)$ [1D potentials] or $k(1,1)$ [2D potential], both in the normal and superconducting states, and in the different schemes to allow for easier comparison between the 1D and 2D potentials. In figure~\ref{fig:DOSvsE} we show the corresponding densities of states (DOS). We clearly see in the normal state how increasing $\beta$ and $c$ both suppress the group velocity, thus increasing flatness of the bands. The density of states becomes correspondingly more and more peaked at zero energy. The superconducting energy gap also increases both with increasing $\beta$ and $c$, and the peculiar double-peak structure in the superconducting DOS is also better revealed for higher $\beta$ or $c$. In the 2D case it is notable how increasing $\beta$ generates multiple peaks in the normal state DOS, and thus also in the superconducting state DOS, in a way that separates it from the 1D potentials.

\begin{figure}
    \includegraphics{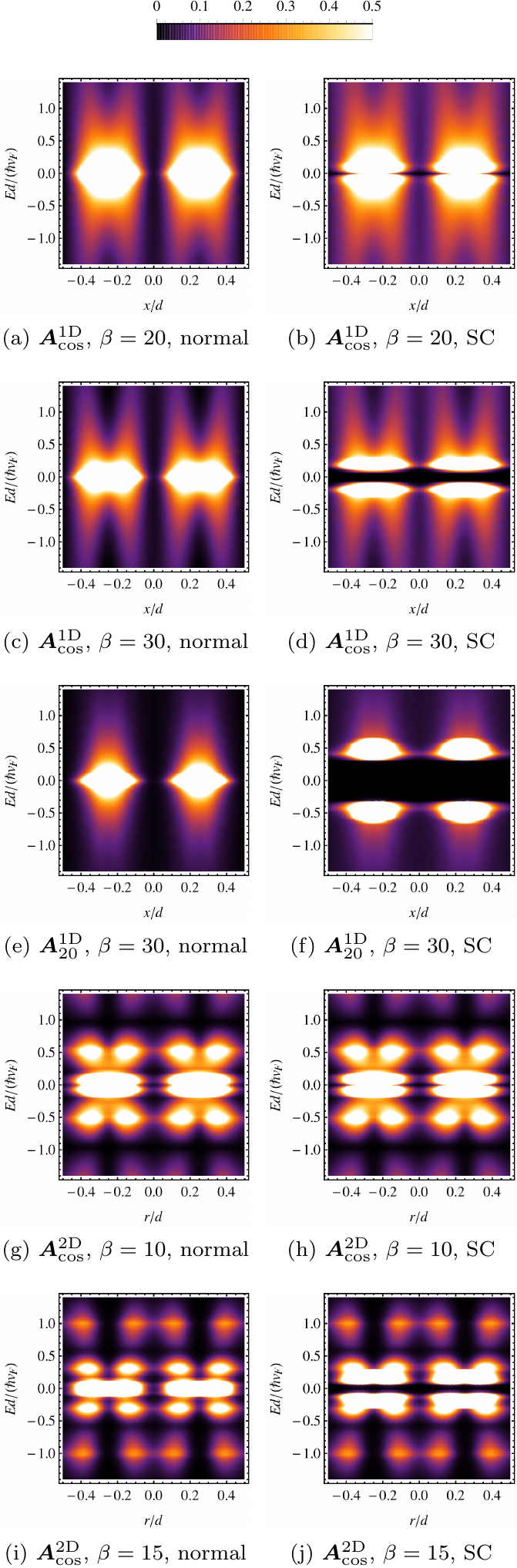}
    \caption{Local density of states (LDOS) at $\mu=0$ and $T=0$ along the line $(x,0)$ [1D potentials] or $r(1,-1)$ [2D potential] both in the (a,c,e,g,i) normal and (b,d,f,h,j) superconducting (SC) states. In the superconducting state $\lambda/(\hbar\vF d)=-0.01$. In each plot the states on the positive (negative) $x$ or $r$ side is coming from the sublattice $A$ ($B$). Each plot is normalized such that the total visible area integrates to unity.}
    \label{fig:LDOSvsEvsr}
\end{figure}

To determine more properties that could be measured \eg by STM \cite{Xie2019,Jiang2019a}, we show in figure~\ref{fig:LDOSvsEvsr} the local densities of states (LDOS) along the line $(x,0)$ [1D potentials] or $r(1,-1)$ [2D potential], which further illustrate the results discussed so far. In the normal state the overall energy dependence shows the clear peak at zero energy for the 1D potentials, as well as the multiple-peak structure for the 2D potential. In the superconducting state the energy dependence also shows the superconducting gap, as already seen in the total DOS in figure~\ref{fig:DOSvsE}. The position dependence gives us more information about the underlying strain field. They clearly show the high density of low-energy states near the points $x=\pm d/4$ (1D potentials) or $r=\pm d/4$ (2D potential), that is, points where $\vect{B}$ has extrema. Furthermore the states on the positive (negative) $x$ or $r$ side are those coming from the sublattice $A$ ($B$), which, by comparison to Fig.~\ref{fig:uvsr+curlAvsr+Deltavsr}(c--d), means that the $A$ ($B$) sublattice states are localized at the minima (maxima) of $\vect{B}$. This kind of localization and sublattice polarization was also experimentally observed by Jiang \etal \cite{Jiang2019a}. Since the low-energy states are the ones contributing to superconductivity, their localization explains the similar localization of the order parameter $\Delta_{A/B}$, as seen in figures~\ref{fig:uvsr+curlAvsr+Deltavsr}(e,f).

In the normal state LDOS we further see the localization pattern splitting at higher energies for the 1D potentials. This is contrasted with the 2D potential, where the higher-energy peaks are separated not only in position but also in energy. Furthermore in the superconducting state LDOS we see the same behavior in the energy gaps as in the total DOS: increasing $\beta$ or $c$ leads to an increasing gap size, with the localization pattern staying the same. Again the 2D potential behaves slightly differently: the gap is largest at $r=\pm d/4$, while for the 1D potentials the gap at $x=\pm d/4$ is smallest.

\subsection{Critical doping level and temperature}

We can in principle calculate the critical doping level $\muc$ and the critical temperature $\Tc$ by solving the self-consistency equation \eqref{eq:self-consistency_equation_kspace} for various $\mu$ and $T$ and by solving for the point where $\Delta$ vanishes. But since the fixed-point iteration scheme converges slowly when $\Delta$ is small, we calculate $\muhalf$ [$\Thalf$] instead, corresponding to the chemical potential [temperature] at which $\max\Delta$ has decreased to $\max\Delta(\mu=0)/2$ [$\max\Delta(T=0)/2$].

\begin{figure}
    \includegraphics[width=\columnwidth]{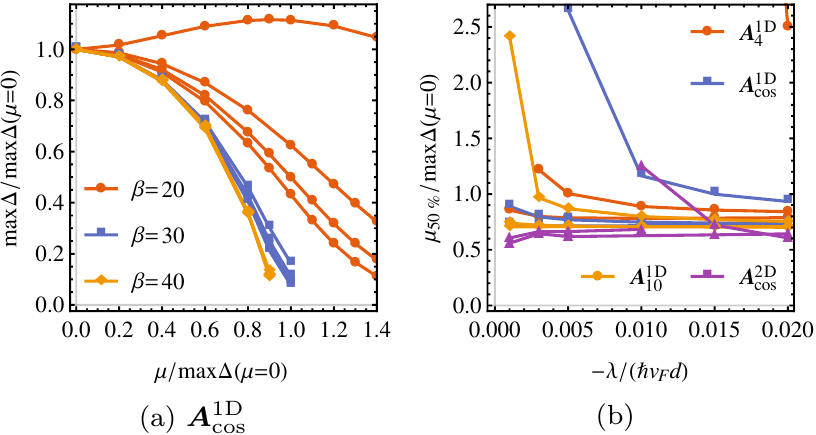}
    \caption{Solving the ``critical'' chemical potential $\muhalf$ at $T=0$, where $\muhalf$ is the chemical potential where $\max\Delta$ has dropped to $\max\Delta(\mu=0)/2$. (a) Normalized order parameter maximum $\max\Delta/\max\Delta(\mu=0)$ as a function of the normalized chemical potential $\mu/\max\Delta(\mu=0)$ for $\vect{A}_{\cos}^\text{1D}$ showing how doping away from the flat band, located at the DOS peak (which is at the zero energy in the flat-band regime and at a nonzero energy in the dispersive regime), kills superconductivity. The behavior is the same for $-\mu$. The four curves for each $\beta$ are those for $-\lambda/(\hbar\vF d) = 0.005,0.01,0.015,0.02$ (from top to bottom). (b) The ratio $\muhalf/\max\Delta(\mu=0)$ as a function of $\lambda$ for different potentials $\vect{A}$. Each 1D potential has three curves corresponding to $\beta=20,30,40$ (from top to bottom). In the flat-band regime the ratio tends approximately to a constant as in \eqref{eq:muhalf_result}.}
    \label{fig:Deltavsmu+mucvslambda}
\end{figure}

We show in figure~\ref{fig:Deltavsmu+mucvslambda}(a) the $\mu$-dependence of $\Delta$ at $T=0$ in the case of $\vect{A}_{\cos}^\text{1D}$, from where $\muhalf$ is determined. We see how doping away from the flat band, which in the flat-band regime is located at zero energy, kills superconductivity. In this case $\muhalf$ approaches $\sim 0.7\max\Delta(\mu=0)$ in the flat-band limit. In the flat-band regime the results fit very well the relation $\max\Delta(\mu)=\sqrt{(\max\Delta(\mu=0))^2-(\mu/b)^2}$ with $b$ as the fitting parameter, as compared to the result \cite{Heikkila2016} $\Delta^\text{FB}(\mu)=\sqrt{\Delta^\text{FB}(\mu=0)^2-\mu^2}$ for exactly flat bands and homogeneous $\Delta^\text{FB}$. On the other hand in the dispersive regime $\Delta$ is not maximized at zero chemical potential, but around $\mu \approx 0.9\max\Delta(\mu=0) \approx \frac{0.9}{1.1}\max\Delta(\mu=\muopt) \approx 0.02\hbar\vF/d$ instead, which corresponds to the DOS peak position shown in figure~\ref{fig:DOSvsE}(a). This is exactly the same behavior as seen in TBG \cite{Peltonen2018,Wu2018}: in the flat-band regime the energy scale of $\Delta$ exceeds the DOS peak separation (the ``bandwidth'') and the smeared DOS is centered at zero energy, while in the dispersive regime $\Delta$ can ``see'' the double-peaked DOS because of the small energy scale of $\Delta$. In TBG this might explain \cite{Peltonen2018,Wu2018} why superconductivity is observed at a nonzero doping level \cite{Cao2018b}, and the same might happen also in PSG if the interaction strength $\lambda$ is small enough. But note that in PSG we can in principle tune $\vect{A}$ (its functional dependence, $\beta$, $c$, and $d$) to move the interface between the flat-band and dispersive regimes so that superconductivity would be observed at zero doping.

To further verify that $\muhalf$ is linear in $\max\Delta(\mu=0)$ in the flat-band regime, 
\begin{equation}
    \muhalf = \eta\max\Delta(\mu=0),
    \label{eq:muhalf_result}
\end{equation}
we show in figure~\ref{fig:Deltavsmu+mucvslambda}(b) the ratio $\eta$ at $T=0$ for a selection of potentials. In the flat-band regime the ratio tends approximately to a constant $\eta\approx 0.7$ as long as $c\lesssim 10$. For $c\gtrsim 10$ we start seeing slight deviations from this, with $\eta\approx 0.6$ and $0.5$ for $c=20$ and $100$, respectively. The critical chemical potential $\muc$ is slightly larger, approximately $\muc \approx \max\Delta(\mu=0)$ for $\vect{A}_{\cos}^\text{1D}$ in the flat-band regime according to figure~\ref{fig:Deltavsmu+mucvslambda}(a). This coincides with the case of perfectly flat bands and a constant $\Delta^\text{FB}$ for which [\onlinecite{Peltonen2018}, Supplemental Material] $\muc^\text{FB}=\Delta^\text{FB}(\mu=0)$.

\begin{figure}
    \includegraphics[width=\columnwidth]{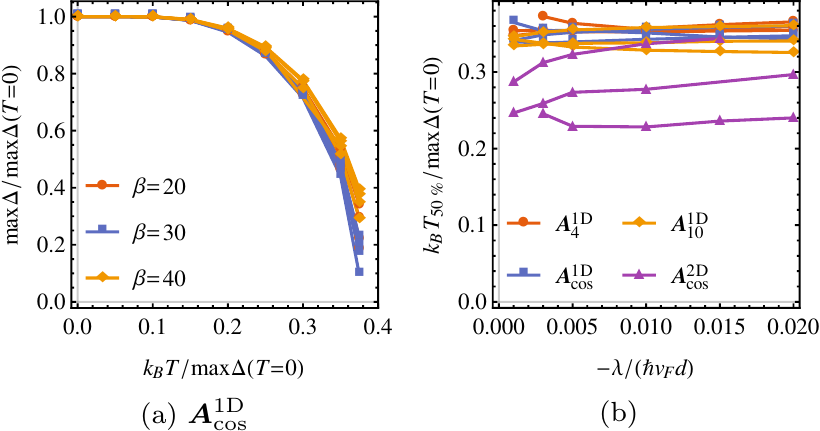}
    \caption{Solving the ``critical'' temperature $\Thalf$ at optimal doping $\mu=\muopt$, where $\Thalf$ is the temperature where $\max\Delta$ has dropped to $\max\Delta(T=0)/2$. (a) Normalized order parameter maximum $\max\Delta/\max\Delta(T=0)$ as a function of the normalized temperature $\kB T/\max\Delta(T=0)$ for $\vect{A}_{\cos}^\text{1D}$. Each $\beta$ has four curves corresponding to $-\lambda/(\hbar\vF d)=0.005,0.01,0.015,0.02$. (b) The ratio $\kB\Thalf/\max\Delta(T=0)$ as a function of $\lambda$ for different potentials $\vect{A}$. Each $\vect{A}$ has three curves corresponding to $\beta=20,30,40$ (1D potentials) or $\beta=10,15,20$ (2D potential), with the outliers being those for the smallest $\beta$. In the flat-band regime the ratio tends approximately to a constant as in \eqref{eq:Thalf_result}.}
    \label{fig:DeltavsT+Tcvslambda}
\end{figure}

In figure~\ref{fig:DeltavsT+Tcvslambda} we show the corresponding plots for determining $\Thalf$ at $\mu=\muopt$. Again the ratio $\xi$ in
\begin{equation}
    \kB\Thalf = \xi\max\Delta(T=0),
    \label{eq:Thalf_result}
\end{equation}
tends approximately to a constant $\xi\approx 0.35$ in the flat-band regime as long as $c\lesssim 10$. For $c\gtrsim 10$ we start seeing deviations from this, with $\xi\approx 0.3$ for $c=20$ and $\xi\approx 0.25$ for $c=100$. The critical temperature $\Tc$ is slightly larger, approximately $\kB\Tc\approx 0.4\max\Delta(T=0)$ for $\vect{A}_{\cos}^\text{1D}$ in the flat-band regime according to figure~\ref{fig:DeltavsT+Tcvslambda}(a). For comparison, in the case of perfectly flat bands and a constant $\Delta^\text{FB}$ we have the result \cite{Heikkila2011} $\kB\Tc^\text{FB}=\frac{1}{2}\Delta^\text{FB}(T=0)$ and in TBG \cite{Peltonen2018} within the same interaction model $\kB\Tc \approx 0.25\max\Delta(T=0)$.

\subsection{Superfluid weight and Berezinskii--Kosterlitz--Thouless transition temperature}

\begin{figure}
    \includegraphics[width=\columnwidth]{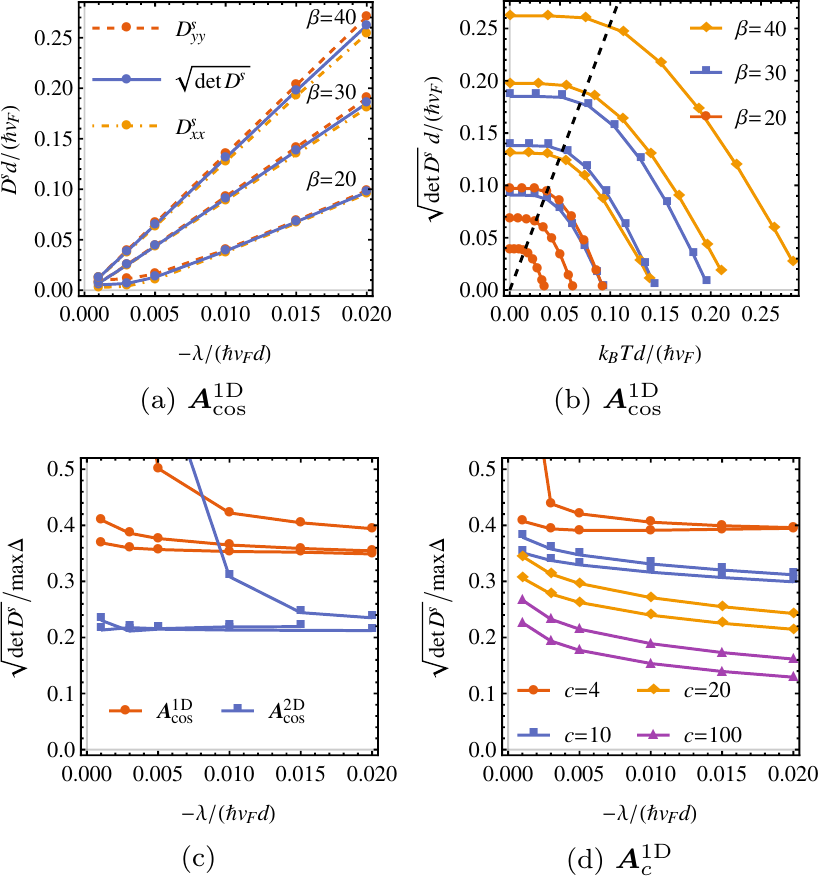}
    \caption{Behavior of the superfluid weight $\Ds$ at optimal doping $\mu=\muopt$ and (a,c,d) $T=0$. (a) $\Ds$ as a function of $\lambda$ for $\vect{A}_{\cos}^\text{1D}$ showing linearity in the flat-band regime. The superfluid weight for the 1D potentials is slightly anisotropic, $\Ds_{xx}\neq\Ds_{yy}$, for large $\beta$ and $\lambda$. For the 2D potential $\vect{A}_{\cos}^\text{2D}$ (not shown) the superfluid weight is isotropic, $\Ds_{xx}=\Ds_{yy}$. The off-diagonal components $\Ds_{xy}=0=\Ds_{xy}$ are zero for all the studied potentials. The (an)isotropy is consistent with the symmetries of the studied potentials. (b) $\sqrt{\det\Ds}$ as a function of temperature $T$ for $\vect{A}_{\cos}^\text{1D}$. Each $\beta$ has three curves corresponding to $-\lambda/(\hbar\vF d) = 0.01,0.015,0.02$, from bottom to top. Also the dashed line $\sqrt{\det\Ds}=8\kB T/\pi$ is shown, from intersections of which $\TBKT$ is determined through \eqref{eq:TBKT}. (c,d) The ratio $\sqrt{\det\Ds}/\max\Delta$ as a function of the interaction strength $\lambda$ for (c) the cosine potentials and (d) $\vect{A}_c^\text{1D}$ with varying $c$. In (c) each $\vect{A}$ has three curves corresponding to (from top to bottom) $\beta=20,30,40$ (1D potentials) or $\beta=10,15,20$ (2D potential), while in (d) each $c$ has two curves corresponding to (from top to bottom) $\beta=30,40$. In the flat-band regime the ratio is approximately a constant depending slightly on the potential, as in \eqref{eq:Ds_result}.}
    \label{fig:Ds}
\end{figure}

To determine the true superconducting transition temperature we calculate the superfluid weight $\Ds$ and the Berezinskii--Kosterlitz--Thouless transition temperature $\TBKT$ from \eqref{eq:Ds} and \eqref{eq:TBKT}. In figure~\ref{fig:Ds}(a) we show the total superfluid weight $\sqrt{\det\Ds}$, together with the different components $\Ds_{\mu\nu}$, as a function of the interaction strength $\lambda$ for $\vect{A}_{\cos}^\text{1D}$. The behavior is very similar to that of $\max\Delta$ in figure~\ref{fig:Deltavslambda}(a): it is linear in the flat-band regime and also increases linearly with increasing $\beta$. To further verify that $\sqrt{\det\Ds}$ is linear in $\max\Delta$,
\begin{equation}
    \sqrt{\det\Ds} = \chi\max\Delta,
    \label{eq:Ds_result}
\end{equation}
we show the ratio $\chi$ in figure~\ref{fig:Ds}(c,d) at $\mu=\muopt$ and $T=0$. In the flat-band regime the ratio tends approximately to a constant $\chi\approx 0.15\dots 0.4$, which has more variation than $\eta$ and $\xi$ for $\muhalf$ and $\Thalf$ in the flat-band regime. For comparison, in TBG we found \cite{Julku2019} within the same interaction model that $\chi\approx 0.35$ in the flat-band regime.

We may again compare \eqref{eq:Ds_result} to the case of exactly flat bands and a constant $\Delta^\text{FB}$. But since the superfluid weight depends heavily on the Hamiltonian itself and not only its eigenvalues, we need to specify which flat-band model to use. We take the ``graphene flat-band limit'', that is, graphene with $\vF\rightarrow 0$. In this case \cite{Kopnin2010,Liang2017} $\Ds_\text{FB}=\frac{2}{\pi}\Delta^\text{FB}$ at $\mu=0\approx\muopt$ and $T=0$, which in fact holds for any $\vF$.

What is intriguing in figure~\ref{fig:Ds}(a) is that for the studied 1D potentials the superfluid weight is almost isotropic although the potentials are highly anisotropic. There is, however, a slight anisotropy, $\Ds_{xx}\neq\Ds_{yy}$ and $\Ds_{xy}=0=\Ds_{yx}$, visible for large $\beta$ and $\lambda$. On the other hand the 2D potential produces an isotropic superfluid weight, $\Ds_{xx}=\Ds_{yy}$ and $\Ds_{xy}=0=\Ds_{yx}$. This (an)isotropy is consistent with the symmetries of the studied potentials. For comparison in TBG it was found \cite{Julku2019} that local interaction always produces an isotropic superfluid weight, while the more complicated resonating valence bond (RVB) interaction was able to produce anisotropy through spontaneous symmetry breaking. The anisotropy could serve as one experimental signature for superconductivity described by the presented model, and it could be measured by radio frequency impedance spectroscopy \cite{Chiodi2011} in a Hall-like four-probe setup \cite{Julku2019}.

Although in this work we do not separate the superfluid weight into the conventional and geometric contributions \cite{Liang2017}, from general knowledge \cite{Liang2017} and calculations in TBG \cite{Julku2019,Hu2019} we expect the geometric contribution to dominate in the flat-band regime.

\begin{figure}
    \includegraphics[width=\columnwidth]{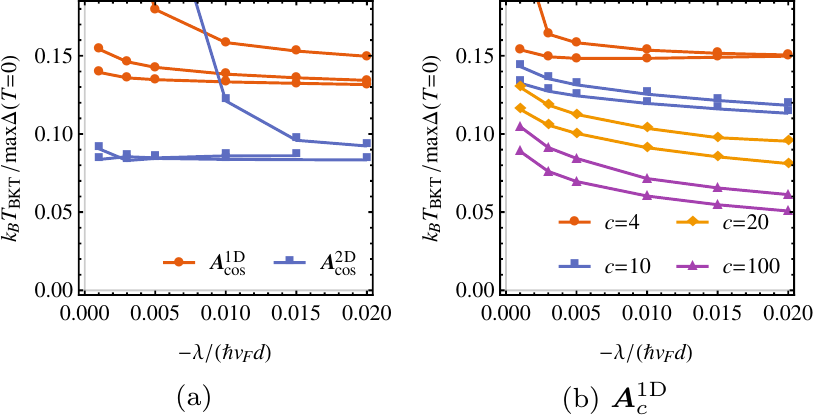}
    \caption{Behavior of the Berezinskii--Kosterlitz--Thouless transition temperature $\TBKT$ at optimal doping $\mu=\muopt$. The ratio $\kB\TBKT/\max\Delta(T=0)$ for (a) the cosine potentials and (b) $\vect{A}_c^\text{1D}$ with varying $c$. In (a) both potentials have three curves corresponding to $\beta=20,30,40$ (1D potential) or $\beta=10,15,20$ (2D potential), from top to bottom in the flat-band regime, while in (b) each $c$ has two curves corresponding to $\beta=30,40$, from top to bottom. In the flat-band regime the ratio tends approximately to a constant depending slightly on the potential, as in \eqref{eq:TBKT_result}. Due to slow convergence the approximation $\Ds(\TBKT) \approx \Ds(0)$ is used for the $\vect{A}_{\cos}^\text{2D}$ potential.}
    \label{fig:TBKT}
\end{figure}

In figure~\ref{fig:Ds}(b) we further show $\sqrt{\det\Ds}$ as a function of temperature $T$ for $\vect{A}_{\cos}^\text{1D}$, from where $\TBKT$ is determined through \eqref{eq:TBKT} by solving for the intersection point with the line $\sqrt{\det\Ds}=8\kB T/\pi$. We immediately see that in the flat-band regime $\Ds(\TBKT)\approx\Ds(0)$ is a rather good approximation so that the self-consistency in \eqref{eq:TBKT} is not essential. This is very different from TBG \cite{Julku2019}, where the temperature dependence is essential due to $\TBKT$ being closer to $\Tc$. We nevertheless need to solve the full self-consistent equation for all the potentials, as the relative magnitude of $\Tc$ and $\TBKT$ is not known beforehand.

The resulting ratio $\kB\TBKT/\max\Delta(T=0)$ is shown in figure~\ref{fig:TBKT} for the different potentials at $\mu=\muopt$, further confirming that $\Ds(\TBKT)\approx\Ds(0)$: apart from the different scale, the $\TBKT$ plots in figure~\ref{fig:TBKT} are very similar to the $\Ds$ plots in figures~\ref{fig:Ds}(c,d). Furthermore in the linear relation
\begin{equation}
    \kB\TBKT = \kappa\max\Delta(T=0),
    \label{eq:TBKT_result}
\end{equation}
the ratio $\kappa$ tends approximately to a constant $\kappa\approx 0.05\dots 0.15$ in the flat-band regime. Again in \eqref{eq:TBKT_result} we see similarity to the ``graphene flat-band limit'' result with a homogeneous $\Delta^\text{FB}$, for which $\kB\TBKT^\text{FB} = \frac{\pi}{8}\Ds_\text{FB}(\TBKT^\text{FB}) \approx \frac{1}{4}\Delta^\text{FB}(T=0)$ at $\mu=0\approx\muopt$ if we furthermore assume $\Ds_\text{FB}(\TBKT^\text{FB}) \approx \Ds_\text{FB}(0)$.

Combining \eqref{eq:Ds_result} and \eqref{eq:TBKT_result} we get in the flat-band regime at $\mu=\muopt$ the ratio $\TBKT/\Thalf=\kappa/\xi\approx 0.2\dots 0.4$ depending on the potential. For $\vect{A}_{\cos}^\text{1D}$ this yields $\TBKT/\Thalf\approx 0.4$, and within the same accuracy $\TBKT/\Tc\approx 0.4$. For comparison in TBG we found within the same interaction model in the flat-band regime $\kB\TBKT\approx 0.16\dots 0.2\max\Delta(T=0)$ \cite{Julku2019} (depending slightly on $\lambda$), $\kB\Tc\approx 0.25\max\Delta(T=0)$ \cite{Peltonen2018}, and thus $\TBKT/\Tc\approx 0.6\dots 0.8$.

By combining \eqref{eq:Delta_result} and \eqref{eq:TBKT_result} we get $\kB\TBKT=-\kappa\zeta B\lambda$ at $\mu=\muopt$. Let us calculate an estimate of $\TBKT$ by using $\lambda=-\SI{1}{eV}a^2\approx-\SI{6}{eV\angstrom^2}$, which roughly corresponds \cite{Peltonen2018,Julku2019} to $\TBKT\approx\SI{1}{K}$ measured in TBG \cite{Cao2018b}. Here $a = \sqrt{3}a_0 \approx \SI{2.46}{\angstrom}$ is the graphene lattice constant. For $\vect{A}_{\cos}^\text{1D}$ we have $B=2\pi\beta/d^2$ and in the flat-band regime $\kappa=0.15$ and $\zeta=0.15$, yielding a similar $\TBKT\approx\SI{1}{K}$ if we apply strain for example such that $\beta=40$ and $d=\SI{60}{nm}$ [then $\lambda/(\hbar\vF d)\approx -0.002$ if using $\vF=\SI{1e6}{m/s}$, which is in the flat-band regime according to figures~\ref{fig:Deltavslambda}(b) and \ref{fig:TBKT}(a)]. In the case of the in-plane displacement field $\vect{u}_{\cos}^\text{1D}$ \eqref{eq:u_cos1D} this corresponds to the displacement amplitude $\beta a_0/(\gruneisen\pi)\approx\SI{1}{nm}$ if $\gruneisen=2$. Since in this case the elasticity theory assumes $\beta/\gruneisen\ll d/a_0$ and $d/a_0\gg 1$ \cite{supplement}, we are very well in the validity regime. On the other hand, if we are able to decrease the strain period to $d=\SI{10}{nm}$ [then $\lambda/(\hbar\vF d)=-0.009$], we get to a high-temperature superconductor value of $\TBKT\approx\SI{40}{K}$, which is still in the validity regime. Note the optimization problem in increasing $\TBKT$: decreasing $d$ directly enhances $\TBKT$ but at the same time it makes the validity limit for $\beta$ tighter, while at the same time we should have as large $\beta$ as possible. But this might only be a limiting factor in our linear elasticity theory, while a more complete microscopic theory could, perhaps, yield a result that increasing $\beta$ or decreasing $d$ always increases $\TBKT$.

The experiments of Jiang \etal \cite{Jiang2019a} with $\frac{\hbar}{e}B \approx \SI{100}{T}$ and $d=\SI{14}{nm}$ can be described by the 1D cosine potential with $\beta\approx 5$. When $\lambda=\SI{-6}{eV\angstrom^2}$, $\lambda/(\hbar\vF d)\approx -0.007$ is not in the flat-band regime. Hence $\TBKT$ cannot be obtained from the simple estimate used above, and is likely much lower than $\SI{1}{K}$. Increasing the strain amplitude by a factor of $4$, so that $\beta=20$, would yield $\zeta=0.05$, $\kappa=0.17$, and thus $\TBKT \approx 0.007\hbar\vF/(d\kB) \approx \SI{4}{K}$. Further decreasing the period to $d=\SI{8}{nm}$, a period which was already observed by Jiang \etal, would yield already $\lambda/(\hbar\vF d)\approx -0.01$ and thus $\TBKT\approx\SI{11}{K}$.


\section{Conclusions}
\label{sec:conclusions}
We have studied both the normal and superconducting $s$-wave state properties of periodically strained graphene (PSG) in the continuum low-energy model. We have shown that periodic strain might be a mechanism that allows increasing the critical temperature $\Tc$ higher than a few kelvin, observed in doped graphene and in twisted bilayer graphene (TBG), or possibly even to tens of kelvins. Especially we have generalized the results of Kauppila \etal \cite{Kauppila2016}, where the authors studied the same problem in the case of a 1D cosine-like pseudo vector potential $\vect{A}$, to potentials with arbitrary shape and dimension. We furthermore calculated the superfluid weight and the Berezinskii--Kosterlitz--Thouless transition temperature $\TBKT$ to determine the true transition temperature observed in experiments. In the normal state we observed flat bands in the spectrum and localization of low-energy states near the extremum points of the effective magnetic field $\vect{B}=\nabla\cross\vect{A}$.

We modelled the superconducting state by the Bogoliubov--de Gennes mean-field theory assuming a local interaction between the Cooper pair leading to $s$-wave pairing. Because of the inhomogeneous strain field we observed a highly inhomogeneous order parameter $\Delta_{A/B}$ that is localized near the extremum points of $\vect{B}$, similarly to the localization of the low-energy states. We also noticed how the superconducting $\Tc$ or $\TBKT$ can be linearly increased by increasing the strain strength $\beta$, decreasing the period $d$, or by increasing the slope (near the extremum points of $\vect{B}$) of the corresponding pseudo vector potential $\vect{A}$. On the other hand increasing the slope makes the order parameter also more localized.

While between the 1D potentials we observed only quantitative differences in the results, for the 2D cosine potential we saw also some qualitative differences when compared to the 1D potentials. The main differences are the localization pattern of $\Delta_{A/B}$, and the more peaked structure of the (local) density of states both in the normal and superconducting states. In the 2D case we studied only the cosine potential, but on the other hand the qualitative similarity in the results between the different 1D potentials gives us certainty that changing the shape of the potential would not change the qualitative results in the 2D case neither. However, it should be noted that it is the shape of $\vect{B}$ that matters and not that of the potential $\vect{A}$ itself, and thus even a 2D potential can produce results that are effectively those of a 1D potential.

We chose all our potentials to be periodic in a square (super)lattice, but note that any other lattice could be chosen as well, with different shapes and different periodicities in the two directions. Properties of this lattice are then directly seen in the dispersion, as well as in the localization of $\vect{B}$ and $\Delta_{A/B}$. We also observed the symmetry $\Delta_B(\vect{r})=\Delta_A(-\vect{r})$ of the order parameter for all the chosen potentials. This is due to the inversion symmetry $\vect{A}(\vect{r})=\vect{A}(-\vect{r})$ present in all of them. The relative magnitude between $\Delta_A$ and $\Delta_B$ can then be tuned by breaking this symmetry, $\eg$ by using a sawtooth-wave potential.

We also observed some very peculiar structures in the (local) density of states, which could serve as an experimental fingerprint of the physics described by this model. We furthermore found that in the flat-band regime the superconducting order parameter maximum $\max\Delta$ at $\mu=\muopt$ and $T=0$, the ``critical'' chemical potential $\muhalf$ at $T=0$, the ``critical'' temperature $\Thalf$ at $\mu=\muopt$, the superfluid weight $\sqrt{\det\Ds}$ at $\mu=\muopt$ and $T=0$, and the BKT transition temperature $\TBKT$ at $\mu=\muopt$ are all approximately linear in the interaction strength $\lambda$. The linear relations, instead of exponential ones in usual bulk superconductors, suggest that high-temperature superconductivity might be possible in PSG. 

As is known from TBG, also other phases like correlated insulators might be present due to the flat bands. These are obviously excluded from the present study, but as we showed in a previous study \cite{Peltonen2018}, the superconductivity-only model gives a plausible explanation for the observed superconducting states in TBG. This view of competing phases is supported by recent experiments where superconductivity could be seen without the correlated insulating phases \cite{Stepanov2019,Saito2019}. Thus we expect our similar model to work also in PSG when concentrating only on superconductivity. If the competing phase (if any) is magnetic, we know from a recent study \cite{Ojajarvi2018} that in a pure flat-band system superconductivity is favored over magnetism whenever (in the weak coupling regime) the effective attractive electron--electron interaction strength $\hat{\lambda}\hbar\omegac = [g^2/(\hbar\omegac)](\Omega_\text{FB}/\Omega_\text{BZ})$ is stronger than the repulsive one $u = U\Omega_\text{FB}/\Omega_\text{BZ}$. Here $g$ is the electron--phonon coupling constant, $U$ is the repulsive Hubbard coupling constant, $\hbar\omegac$ is the characteristic phonon energy (in this case the Einstein energy $\hbar\omega_E$), and $\Omega_\text{FB}/\Omega_\text{BZ}$ is the ratio of the flat-band area to the Brillouin zone area.

An interesting future prospect would be to study the other phases which, by the analogue of TBG, are highly probable. Secondly the combination of \moire \cite{Peltonen2018} and strain [this work] physics would perhaps advance the understanding of superconductivity in TBG, where intrinsic periodic strain is inevitable. From the experimental point of view the challenge is to manufacture periodically strained graphene samples with large amplitudes and small periods and to perform low-temperature conductivity measurements in this (electrically doped) system to reveal the possible superconducting and/or correlated insulator states. The periodic strain and flat bands observed by Jiang \etal \cite{Jiang2019a} are already an intriguing starting point, but according to our calculations a $\TBKT$ of the order of $\SI{4}{K}$ would need a strain amplitude $4$ times larger than in the experiment. On the other hand, further
decreasing the period to $\SI{8}{nm}$ would yield already $\TBKT\approx\SI{11}{K}$.


\begin{acknowledgments}
We thank Risto Ojajärvi for discussions. T.J.P. acknowledges funding from the Emil Aaltonen foundation and T.T.H. from the Academy of Finland via its project number 317118. We acknowledge grants of computer capacity from the Finnish Grid and Cloud Infrastructure (persistent identifier urn:nbn:fi:research-infras-2016072533).
\end{acknowledgments}


\bibliography{library,library_footnotes}

\begin{thebibliography}{67}%
\makeatletter
\providecommand \@ifxundefined [1]{%
 \@ifx{#1\undefined}
}%
\providecommand \@ifnum [1]{%
 \ifnum #1\expandafter \@firstoftwo
 \else \expandafter \@secondoftwo
 \fi
}%
\providecommand \@ifx [1]{%
 \ifx #1\expandafter \@firstoftwo
 \else \expandafter \@secondoftwo
 \fi
}%
\providecommand \natexlab [1]{#1}%
\providecommand \enquote  [1]{``#1''}%
\providecommand \bibnamefont  [1]{#1}%
\providecommand \bibfnamefont [1]{#1}%
\providecommand \citenamefont [1]{#1}%
\providecommand \href@noop [0]{\@secondoftwo}%
\providecommand \href [0]{\begingroup \@sanitize@url \@href}%
\providecommand \@href[1]{\@@startlink{#1}\@@href}%
\providecommand \@@href[1]{\endgroup#1\@@endlink}%
\providecommand \@sanitize@url [0]{\catcode `\\12\catcode `\$12\catcode
  `\&12\catcode `\#12\catcode `\^12\catcode `\_12\catcode `\%12\relax}%
\providecommand \@@startlink[1]{}%
\providecommand \@@endlink[0]{}%
\providecommand \url  [0]{\begingroup\@sanitize@url \@url }%
\providecommand \@url [1]{\endgroup\@href {#1}{\urlprefix }}%
\providecommand \urlprefix  [0]{URL }%
\providecommand \Eprint [0]{\href }%
\providecommand \doibase [0]{http://dx.doi.org/}%
\providecommand \selectlanguage [0]{\@gobble}%
\providecommand \bibinfo  [0]{\@secondoftwo}%
\providecommand \bibfield  [0]{\@secondoftwo}%
\providecommand \translation [1]{[#1]}%
\providecommand \BibitemOpen [0]{}%
\providecommand \bibitemStop [0]{}%
\providecommand \bibitemNoStop [0]{.\EOS\space}%
\providecommand \EOS [0]{\spacefactor3000\relax}%
\providecommand \BibitemShut  [1]{\csname bibitem#1\endcsname}%
\let\auto@bib@innerbib\@empty
\bibitem [{\citenamefont {Ludbrook}\ \emph {et~al.}(2015)\citenamefont
  {Ludbrook}, \citenamefont {Levy}, \citenamefont {Nigge}, \citenamefont
  {Zonno}, \citenamefont {Schneider}, \citenamefont {Dvorak}, \citenamefont
  {Veenstra}, \citenamefont {Zhdanovich}, \citenamefont {Wong}, \citenamefont
  {Dosanjh}, \citenamefont {Stra{\ss}er}, \citenamefont {St{\"{o}}hr},
  \citenamefont {Forti}, \citenamefont {Ast}, \citenamefont {Starke},\ and\
  \citenamefont {Damascelli}}]{Ludbrook2015}%
  \BibitemOpen
  \bibfield  {author} {\bibinfo {author} {\bibfnamefont {B.~M.}\ \bibnamefont
  {Ludbrook}}, \bibinfo {author} {\bibfnamefont {G.}~\bibnamefont {Levy}},
  \bibinfo {author} {\bibfnamefont {P.}~\bibnamefont {Nigge}}, \bibinfo
  {author} {\bibfnamefont {M.}~\bibnamefont {Zonno}}, \bibinfo {author}
  {\bibfnamefont {M.}~\bibnamefont {Schneider}}, \bibinfo {author}
  {\bibfnamefont {D.~J.}\ \bibnamefont {Dvorak}}, \bibinfo {author}
  {\bibfnamefont {C.~N.}\ \bibnamefont {Veenstra}}, \bibinfo {author}
  {\bibfnamefont {S.}~\bibnamefont {Zhdanovich}}, \bibinfo {author}
  {\bibfnamefont {D.}~\bibnamefont {Wong}}, \bibinfo {author} {\bibfnamefont
  {P.}~\bibnamefont {Dosanjh}}, \bibinfo {author} {\bibfnamefont
  {C.}~\bibnamefont {Stra{\ss}er}}, \bibinfo {author} {\bibfnamefont
  {A.}~\bibnamefont {St{\"{o}}hr}}, \bibinfo {author} {\bibfnamefont
  {S.}~\bibnamefont {Forti}}, \bibinfo {author} {\bibfnamefont {C.~R.}\
  \bibnamefont {Ast}}, \bibinfo {author} {\bibfnamefont {U.}~\bibnamefont
  {Starke}}, \ and\ \bibinfo {author} {\bibfnamefont {A.}~\bibnamefont
  {Damascelli}},\ }\href {\doibase 10.1073/pnas.1510435112} {\bibfield
  {journal} {\bibinfo  {journal} {Proc. Natl. Acad. Sci.}\ }\textbf {\bibinfo
  {volume} {112}},\ \bibinfo {pages} {11795} (\bibinfo {year}
  {2015})}\BibitemShut {NoStop}%
\bibitem [{\citenamefont {Chapman}\ \emph {et~al.}(2016)\citenamefont
  {Chapman}, \citenamefont {Su}, \citenamefont {Howard}, \citenamefont
  {Kundys}, \citenamefont {Grigorenko}, \citenamefont {Guinea}, \citenamefont
  {Geim}, \citenamefont {Grigorieva},\ and\ \citenamefont
  {Nair}}]{Chapman2016}%
  \BibitemOpen
  \bibfield  {author} {\bibinfo {author} {\bibfnamefont {J.}~\bibnamefont
  {Chapman}}, \bibinfo {author} {\bibfnamefont {Y.}~\bibnamefont {Su}},
  \bibinfo {author} {\bibfnamefont {C.~A.}\ \bibnamefont {Howard}}, \bibinfo
  {author} {\bibfnamefont {D.}~\bibnamefont {Kundys}}, \bibinfo {author}
  {\bibfnamefont {A.~N.}\ \bibnamefont {Grigorenko}}, \bibinfo {author}
  {\bibfnamefont {F.}~\bibnamefont {Guinea}}, \bibinfo {author} {\bibfnamefont
  {A.~K.}\ \bibnamefont {Geim}}, \bibinfo {author} {\bibfnamefont {I.~V.}\
  \bibnamefont {Grigorieva}}, \ and\ \bibinfo {author} {\bibfnamefont {R.~R.}\
  \bibnamefont {Nair}},\ }\href {\doibase 10.1038/srep23254} {\bibfield
  {journal} {\bibinfo  {journal} {Sci. Rep.}\ }\textbf {\bibinfo {volume}
  {6}},\ \bibinfo {pages} {23254} (\bibinfo {year} {2016})}\BibitemShut
  {NoStop}%
\bibitem [{\citenamefont {Ichinokura}\ \emph {et~al.}(2016)\citenamefont
  {Ichinokura}, \citenamefont {Sugawara}, \citenamefont {Takayama},
  \citenamefont {Takahashi},\ and\ \citenamefont {Hasegawa}}]{Ichinokura2016}%
  \BibitemOpen
  \bibfield  {author} {\bibinfo {author} {\bibfnamefont {S.}~\bibnamefont
  {Ichinokura}}, \bibinfo {author} {\bibfnamefont {K.}~\bibnamefont
  {Sugawara}}, \bibinfo {author} {\bibfnamefont {A.}~\bibnamefont {Takayama}},
  \bibinfo {author} {\bibfnamefont {T.}~\bibnamefont {Takahashi}}, \ and\
  \bibinfo {author} {\bibfnamefont {S.}~\bibnamefont {Hasegawa}},\ }\href
  {\doibase 10.1021/acsnano.5b07848} {\bibfield  {journal} {\bibinfo  {journal}
  {ACS Nano}\ }\textbf {\bibinfo {volume} {10}},\ \bibinfo {pages} {2761}
  (\bibinfo {year} {2016})}\BibitemShut {NoStop}%
\bibitem [{\citenamefont {Tiwari}\ \emph {et~al.}(2017)\citenamefont {Tiwari},
  \citenamefont {Shin}, \citenamefont {Hwang}, \citenamefont {Jung},
  \citenamefont {Park},\ and\ \citenamefont {Lee}}]{Tiwari2017}%
  \BibitemOpen
  \bibfield  {author} {\bibinfo {author} {\bibfnamefont {A.~P.}\ \bibnamefont
  {Tiwari}}, \bibinfo {author} {\bibfnamefont {S.}~\bibnamefont {Shin}},
  \bibinfo {author} {\bibfnamefont {E.}~\bibnamefont {Hwang}}, \bibinfo
  {author} {\bibfnamefont {S.-G.}\ \bibnamefont {Jung}}, \bibinfo {author}
  {\bibfnamefont {T.}~\bibnamefont {Park}}, \ and\ \bibinfo {author}
  {\bibfnamefont {H.}~\bibnamefont {Lee}},\ }\href {\doibase
  10.1088/1361-648X/aa88fb} {\bibfield  {journal} {\bibinfo  {journal} {J.
  Phys. Condens. Matter}\ }\textbf {\bibinfo {volume} {29}},\ \bibinfo {pages}
  {445701} (\bibinfo {year} {2017})}\BibitemShut {NoStop}%
\bibitem [{\citenamefont {Cao}\ \emph {et~al.}(2018)\citenamefont {Cao},
  \citenamefont {Fatemi}, \citenamefont {Fang}, \citenamefont {Watanabe},
  \citenamefont {Taniguchi}, \citenamefont {Kaxiras},\ and\ \citenamefont
  {Jarillo-Herrero}}]{Cao2018b}%
  \BibitemOpen
  \bibfield  {author} {\bibinfo {author} {\bibfnamefont {Y.}~\bibnamefont
  {Cao}}, \bibinfo {author} {\bibfnamefont {V.}~\bibnamefont {Fatemi}},
  \bibinfo {author} {\bibfnamefont {S.}~\bibnamefont {Fang}}, \bibinfo {author}
  {\bibfnamefont {K.}~\bibnamefont {Watanabe}}, \bibinfo {author}
  {\bibfnamefont {T.}~\bibnamefont {Taniguchi}}, \bibinfo {author}
  {\bibfnamefont {E.}~\bibnamefont {Kaxiras}}, \ and\ \bibinfo {author}
  {\bibfnamefont {P.}~\bibnamefont {Jarillo-Herrero}},\ }\href {\doibase
  10.1038/nature26160} {\bibfield  {journal} {\bibinfo  {journal} {Nature}\
  }\textbf {\bibinfo {volume} {556}},\ \bibinfo {pages} {43} (\bibinfo {year}
  {2018})}\BibitemShut {NoStop}%
\bibitem [{\citenamefont {Yankowitz}\ \emph {et~al.}(2016)\citenamefont
  {Yankowitz}, \citenamefont {Watanabe}, \citenamefont {Taniguchi},
  \citenamefont {San-Jose},\ and\ \citenamefont {LeRoy}}]{Yankowitz2016}%
  \BibitemOpen
  \bibfield  {author} {\bibinfo {author} {\bibfnamefont {M.}~\bibnamefont
  {Yankowitz}}, \bibinfo {author} {\bibfnamefont {K.}~\bibnamefont {Watanabe}},
  \bibinfo {author} {\bibfnamefont {T.}~\bibnamefont {Taniguchi}}, \bibinfo
  {author} {\bibfnamefont {P.}~\bibnamefont {San-Jose}}, \ and\ \bibinfo
  {author} {\bibfnamefont {B.~J.}\ \bibnamefont {LeRoy}},\ }\href {\doibase
  10.1038/ncomms13168} {\bibfield  {journal} {\bibinfo  {journal} {Nat.
  Commun.}\ }\textbf {\bibinfo {volume} {7}},\ \bibinfo {pages} {13168}
  (\bibinfo {year} {2016})}\BibitemShut {NoStop}%
\bibitem [{\citenamefont {Lu}\ \emph {et~al.}()\citenamefont {Lu},
  \citenamefont {Stepanov}, \citenamefont {Yang}, \citenamefont {Xie},
  \citenamefont {Aamir}, \citenamefont {Das}, \citenamefont {Urgell},
  \citenamefont {Watanabe}, \citenamefont {Taniguchi}, \citenamefont {Zhang},
  \citenamefont {Bachtold}, \citenamefont {MacDonald},\ and\ \citenamefont
  {Efetov}}]{Lu2019}%
  \BibitemOpen
  \bibfield  {author} {\bibinfo {author} {\bibfnamefont {X.}~\bibnamefont
  {Lu}}, \bibinfo {author} {\bibfnamefont {P.}~\bibnamefont {Stepanov}},
  \bibinfo {author} {\bibfnamefont {W.}~\bibnamefont {Yang}}, \bibinfo {author}
  {\bibfnamefont {M.}~\bibnamefont {Xie}}, \bibinfo {author} {\bibfnamefont
  {M.~A.}\ \bibnamefont {Aamir}}, \bibinfo {author} {\bibfnamefont
  {I.}~\bibnamefont {Das}}, \bibinfo {author} {\bibfnamefont {C.}~\bibnamefont
  {Urgell}}, \bibinfo {author} {\bibfnamefont {K.}~\bibnamefont {Watanabe}},
  \bibinfo {author} {\bibfnamefont {T.}~\bibnamefont {Taniguchi}}, \bibinfo
  {author} {\bibfnamefont {G.}~\bibnamefont {Zhang}}, \bibinfo {author}
  {\bibfnamefont {A.}~\bibnamefont {Bachtold}}, \bibinfo {author}
  {\bibfnamefont {A.~H.}\ \bibnamefont {MacDonald}}, \ and\ \bibinfo {author}
  {\bibfnamefont {D.~K.}\ \bibnamefont {Efetov}},\ }\href
  {http://arxiv.org/abs/1903.06513} {\ }\Eprint
  {http://arxiv.org/abs/1903.06513} {arXiv:1903.06513} \BibitemShut {NoStop}%
\bibitem [{\citenamefont {Moriyama}\ \emph {et~al.}()\citenamefont {Moriyama},
  \citenamefont {Morita}, \citenamefont {Komatsu}, \citenamefont {Endo},
  \citenamefont {Iwasaki}, \citenamefont {Nakaharai}, \citenamefont {Noguchi},
  \citenamefont {Wakayama}, \citenamefont {Watanabe}, \citenamefont {Tsuya},
  \citenamefont {Watanabe},\ and\ \citenamefont {Taniguchi}}]{Moriyama2019}%
  \BibitemOpen
  \bibfield  {author} {\bibinfo {author} {\bibfnamefont {S.}~\bibnamefont
  {Moriyama}}, \bibinfo {author} {\bibfnamefont {Y.}~\bibnamefont {Morita}},
  \bibinfo {author} {\bibfnamefont {K.}~\bibnamefont {Komatsu}}, \bibinfo
  {author} {\bibfnamefont {K.}~\bibnamefont {Endo}}, \bibinfo {author}
  {\bibfnamefont {T.}~\bibnamefont {Iwasaki}}, \bibinfo {author} {\bibfnamefont
  {S.}~\bibnamefont {Nakaharai}}, \bibinfo {author} {\bibfnamefont
  {Y.}~\bibnamefont {Noguchi}}, \bibinfo {author} {\bibfnamefont
  {Y.}~\bibnamefont {Wakayama}}, \bibinfo {author} {\bibfnamefont
  {E.}~\bibnamefont {Watanabe}}, \bibinfo {author} {\bibfnamefont
  {D.}~\bibnamefont {Tsuya}}, \bibinfo {author} {\bibfnamefont
  {K.}~\bibnamefont {Watanabe}}, \ and\ \bibinfo {author} {\bibfnamefont
  {T.}~\bibnamefont {Taniguchi}},\ }\href {http://arxiv.org/abs/1901.09356} {\
  }\Eprint {http://arxiv.org/abs/1901.09356} {arXiv:1901.09356} \BibitemShut
  {NoStop}%
\bibitem [{\citenamefont {Bardeen}\ \emph {et~al.}(1957)\citenamefont
  {Bardeen}, \citenamefont {Cooper},\ and\ \citenamefont
  {Schrieffer}}]{Bardeen1957}%
  \BibitemOpen
  \bibfield  {author} {\bibinfo {author} {\bibfnamefont {J.}~\bibnamefont
  {Bardeen}}, \bibinfo {author} {\bibfnamefont {L.~N.}\ \bibnamefont {Cooper}},
  \ and\ \bibinfo {author} {\bibfnamefont {J.~R.}\ \bibnamefont {Schrieffer}},\
  }\href {\doibase 10.1103/PhysRev.108.1175} {\bibfield  {journal} {\bibinfo
  {journal} {Phys. Rev.}\ }\textbf {\bibinfo {volume} {108}},\ \bibinfo {pages}
  {1175} (\bibinfo {year} {1957})}\BibitemShut {NoStop}%
\bibitem [{\citenamefont {Heikkil{\"{a}}}\ \emph {et~al.}(2011)\citenamefont
  {Heikkil{\"{a}}}, \citenamefont {Kopnin},\ and\ \citenamefont
  {Volovik}}]{Heikkila2011}%
  \BibitemOpen
  \bibfield  {author} {\bibinfo {author} {\bibfnamefont {T.~T.}\ \bibnamefont
  {Heikkil{\"{a}}}}, \bibinfo {author} {\bibfnamefont {N.~B.}\ \bibnamefont
  {Kopnin}}, \ and\ \bibinfo {author} {\bibfnamefont {G.~E.}\ \bibnamefont
  {Volovik}},\ }\href {\doibase 10.1134/S0021364011150045} {\bibfield
  {journal} {\bibinfo  {journal} {JETP Lett.}\ }\textbf {\bibinfo {volume}
  {94}},\ \bibinfo {pages} {233} (\bibinfo {year} {2011})}\BibitemShut
  {NoStop}%
\bibitem [{\citenamefont {Yankowitz}\ \emph {et~al.}(2019)\citenamefont
  {Yankowitz}, \citenamefont {Chen}, \citenamefont {Polshyn}, \citenamefont
  {Zhang}, \citenamefont {Watanabe}, \citenamefont {Taniguchi}, \citenamefont
  {Graf}, \citenamefont {Young},\ and\ \citenamefont {Dean}}]{Yankowitz2019}%
  \BibitemOpen
  \bibfield  {author} {\bibinfo {author} {\bibfnamefont {M.}~\bibnamefont
  {Yankowitz}}, \bibinfo {author} {\bibfnamefont {S.}~\bibnamefont {Chen}},
  \bibinfo {author} {\bibfnamefont {H.}~\bibnamefont {Polshyn}}, \bibinfo
  {author} {\bibfnamefont {Y.}~\bibnamefont {Zhang}}, \bibinfo {author}
  {\bibfnamefont {K.}~\bibnamefont {Watanabe}}, \bibinfo {author}
  {\bibfnamefont {T.}~\bibnamefont {Taniguchi}}, \bibinfo {author}
  {\bibfnamefont {D.}~\bibnamefont {Graf}}, \bibinfo {author} {\bibfnamefont
  {A.~F.}\ \bibnamefont {Young}}, \ and\ \bibinfo {author} {\bibfnamefont
  {C.~R.}\ \bibnamefont {Dean}},\ }\href {\doibase 10.1126/science.aav1910}
  {\bibfield  {journal} {\bibinfo  {journal} {Science}\ }\textbf {\bibinfo
  {volume} {363}},\ \bibinfo {pages} {1059} (\bibinfo {year}
  {2019})}\BibitemShut {NoStop}%
\bibitem [{\citenamefont {Bistritzer}\ and\ \citenamefont
  {MacDonald}(2011)}]{Bistritzer2011}%
  \BibitemOpen
  \bibfield  {author} {\bibinfo {author} {\bibfnamefont {R.}~\bibnamefont
  {Bistritzer}}\ and\ \bibinfo {author} {\bibfnamefont {A.~H.}\ \bibnamefont
  {MacDonald}},\ }\href {\doibase 10.1073/pnas.1108174108} {\bibfield
  {journal} {\bibinfo  {journal} {Proc. Natl. Acad. Sci.}\ }\textbf {\bibinfo
  {volume} {108}},\ \bibinfo {pages} {12233} (\bibinfo {year}
  {2011})}\BibitemShut {NoStop}%
\bibitem [{\citenamefont {{Lopes dos Santos}}\ \emph
  {et~al.}(2012)\citenamefont {{Lopes dos Santos}}, \citenamefont {Peres},\
  and\ \citenamefont {{Castro Neto}}}]{LopesdosSantos2012}%
  \BibitemOpen
  \bibfield  {author} {\bibinfo {author} {\bibfnamefont {J.~M.~B.}\
  \bibnamefont {{Lopes dos Santos}}}, \bibinfo {author} {\bibfnamefont
  {N.~M.~R.}\ \bibnamefont {Peres}}, \ and\ \bibinfo {author} {\bibfnamefont
  {A.~H.}\ \bibnamefont {{Castro Neto}}},\ }\href {\doibase
  10.1103/PhysRevB.86.155449} {\bibfield  {journal} {\bibinfo  {journal} {Phys.
  Rev. B}\ }\textbf {\bibinfo {volume} {86}},\ \bibinfo {pages} {155449}
  (\bibinfo {year} {2012})}\BibitemShut {NoStop}%
\bibitem [{\citenamefont {Peltonen}\ \emph {et~al.}(2018)\citenamefont
  {Peltonen}, \citenamefont {Ojaj{\"{a}}rvi},\ and\ \citenamefont
  {Heikkil{\"{a}}}}]{Peltonen2018}%
  \BibitemOpen
  \bibfield  {author} {\bibinfo {author} {\bibfnamefont {T.~J.}\ \bibnamefont
  {Peltonen}}, \bibinfo {author} {\bibfnamefont {R.}~\bibnamefont
  {Ojaj{\"{a}}rvi}}, \ and\ \bibinfo {author} {\bibfnamefont {T.~T.}\
  \bibnamefont {Heikkil{\"{a}}}},\ }\href {\doibase 10.1103/PhysRevB.98.220504}
  {\bibfield  {journal} {\bibinfo  {journal} {Phys. Rev. B}\ }\textbf {\bibinfo
  {volume} {98}},\ \bibinfo {pages} {220504(R)} (\bibinfo {year}
  {2018})}\BibitemShut {NoStop}%
\bibitem [{\citenamefont {Guinea}\ \emph {et~al.}(2008)\citenamefont {Guinea},
  \citenamefont {Katsnelson},\ and\ \citenamefont {Vozmediano}}]{Guinea2008}%
  \BibitemOpen
  \bibfield  {author} {\bibinfo {author} {\bibfnamefont {F.}~\bibnamefont
  {Guinea}}, \bibinfo {author} {\bibfnamefont {M.~I.}\ \bibnamefont
  {Katsnelson}}, \ and\ \bibinfo {author} {\bibfnamefont {M.~A.~H.}\
  \bibnamefont {Vozmediano}},\ }\href {\doibase 10.1103/PhysRevB.77.075422}
  {\bibfield  {journal} {\bibinfo  {journal} {Phys. Rev. B}\ }\textbf {\bibinfo
  {volume} {77}},\ \bibinfo {pages} {075422} (\bibinfo {year}
  {2008})}\BibitemShut {NoStop}%
\bibitem [{\citenamefont {Tang}\ and\ \citenamefont {Fu}(2014)}]{Tang2014}%
  \BibitemOpen
  \bibfield  {author} {\bibinfo {author} {\bibfnamefont {E.}~\bibnamefont
  {Tang}}\ and\ \bibinfo {author} {\bibfnamefont {L.}~\bibnamefont {Fu}},\
  }\href {\doibase 10.1038/nphys3109} {\bibfield  {journal} {\bibinfo
  {journal} {Nat. Phys.}\ }\textbf {\bibinfo {volume} {10}},\ \bibinfo {pages}
  {964} (\bibinfo {year} {2014})}\BibitemShut {NoStop}%
\bibitem [{\citenamefont {Venderbos}\ and\ \citenamefont
  {Fu}(2016)}]{Venderbos2016}%
  \BibitemOpen
  \bibfield  {author} {\bibinfo {author} {\bibfnamefont {J.~W.~F.}\
  \bibnamefont {Venderbos}}\ and\ \bibinfo {author} {\bibfnamefont
  {L.}~\bibnamefont {Fu}},\ }\href {\doibase 10.1103/PhysRevB.93.195126}
  {\bibfield  {journal} {\bibinfo  {journal} {Phys. Rev. B}\ }\textbf {\bibinfo
  {volume} {93}},\ \bibinfo {pages} {195126} (\bibinfo {year}
  {2016})}\BibitemShut {NoStop}%
\bibitem [{\citenamefont {Kauppila}\ \emph {et~al.}(2016)\citenamefont
  {Kauppila}, \citenamefont {Aikebaier},\ and\ \citenamefont
  {Heikkil{\"{a}}}}]{Kauppila2016}%
  \BibitemOpen
  \bibfield  {author} {\bibinfo {author} {\bibfnamefont {V.~J.}\ \bibnamefont
  {Kauppila}}, \bibinfo {author} {\bibfnamefont {F.}~\bibnamefont {Aikebaier}},
  \ and\ \bibinfo {author} {\bibfnamefont {T.~T.}\ \bibnamefont
  {Heikkil{\"{a}}}},\ }\href {\doibase 10.1103/PhysRevB.93.214505} {\bibfield
  {journal} {\bibinfo  {journal} {Phys. Rev. B}\ }\textbf {\bibinfo {volume}
  {93}},\ \bibinfo {pages} {214505} (\bibinfo {year} {2016})}\BibitemShut
  {NoStop}%
\bibitem [{\citenamefont {Tahir}\ \emph {et~al.}()\citenamefont {Tahir},
  \citenamefont {Pinaud},\ and\ \citenamefont {Chen}}]{Tahir2018}%
  \BibitemOpen
  \bibfield  {author} {\bibinfo {author} {\bibfnamefont {M.}~\bibnamefont
  {Tahir}}, \bibinfo {author} {\bibfnamefont {O.}~\bibnamefont {Pinaud}}, \
  and\ \bibinfo {author} {\bibfnamefont {H.}~\bibnamefont {Chen}},\ }\href
  {http://arxiv.org/abs/1808.10046} {\ }\Eprint
  {http://arxiv.org/abs/1808.10046} {arXiv:1808.10046} \BibitemShut {NoStop}%
\bibitem [{\citenamefont {Jiang}\ \emph {et~al.}()\citenamefont {Jiang},
  \citenamefont {An{\dj}elkovi{\'{c}}}, \citenamefont {Milovanovi{\'{c}}},
  \citenamefont {Covaci}, \citenamefont {Lai}, \citenamefont {Cao},
  \citenamefont {Watanabe}, \citenamefont {Taniguchi}, \citenamefont {Peeters},
  \citenamefont {Geim},\ and\ \citenamefont {Andrei}}]{Jiang2019a}%
  \BibitemOpen
  \bibfield  {author} {\bibinfo {author} {\bibfnamefont {Y.}~\bibnamefont
  {Jiang}}, \bibinfo {author} {\bibfnamefont {M.}~\bibnamefont
  {An{\dj}elkovi{\'{c}}}}, \bibinfo {author} {\bibfnamefont {S.~P.}\
  \bibnamefont {Milovanovi{\'{c}}}}, \bibinfo {author} {\bibfnamefont
  {L.}~\bibnamefont {Covaci}}, \bibinfo {author} {\bibfnamefont
  {X.}~\bibnamefont {Lai}}, \bibinfo {author} {\bibfnamefont {Y.}~\bibnamefont
  {Cao}}, \bibinfo {author} {\bibfnamefont {K.}~\bibnamefont {Watanabe}},
  \bibinfo {author} {\bibfnamefont {T.}~\bibnamefont {Taniguchi}}, \bibinfo
  {author} {\bibfnamefont {F.~M.}\ \bibnamefont {Peeters}}, \bibinfo {author}
  {\bibfnamefont {A.~K.}\ \bibnamefont {Geim}}, \ and\ \bibinfo {author}
  {\bibfnamefont {E.~Y.}\ \bibnamefont {Andrei}},\ }\href
  {http://arxiv.org/abs/1904.10147} {\ }\Eprint
  {http://arxiv.org/abs/1904.10147} {arXiv:1904.10147} \BibitemShut {NoStop}%
\bibitem [{sup()}]{supplement}%
  \BibitemOpen
  \bibinfo {note} {See Supplementary Material (SM) for definition of the
  Fourier series in the different schemes, details on the derivation of the
  strained graphene Hamiltonian, details on the derivation of the
  superconductivity theory, and how we determine the initial guess of the order
  parameter. SM cites also Refs.
  \onlinecite{Zheng2007,Nazarov2013,Beenakker2006,Titov2006}.}\BibitemShut
  {Stop}%
\bibitem [{\citenamefont {Suzuura}\ and\ \citenamefont
  {Ando}(2002)}]{Suzuura2002}%
  \BibitemOpen
  \bibfield  {author} {\bibinfo {author} {\bibfnamefont {H.}~\bibnamefont
  {Suzuura}}\ and\ \bibinfo {author} {\bibfnamefont {T.}~\bibnamefont {Ando}},\
  }\href {\doibase 10.1103/PhysRevB.65.235412} {\bibfield  {journal} {\bibinfo
  {journal} {Phys. Rev. B}\ }\textbf {\bibinfo {volume} {65}},\ \bibinfo
  {pages} {235412} (\bibinfo {year} {2002})}\BibitemShut {NoStop}%
\bibitem [{\citenamefont {Vozmediano}\ \emph {et~al.}(2010)\citenamefont
  {Vozmediano}, \citenamefont {Katsnelson},\ and\ \citenamefont
  {Guinea}}]{Vozmediano2010}%
  \BibitemOpen
  \bibfield  {author} {\bibinfo {author} {\bibfnamefont {M.~A.~H.}\
  \bibnamefont {Vozmediano}}, \bibinfo {author} {\bibfnamefont {M.~I.}\
  \bibnamefont {Katsnelson}}, \ and\ \bibinfo {author} {\bibfnamefont
  {F.}~\bibnamefont {Guinea}},\ }\href {\doibase 10.1016/j.physrep.2010.07.003}
  {\bibfield  {journal} {\bibinfo  {journal} {Phys. Rep.}\ }\textbf {\bibinfo
  {volume} {496}},\ \bibinfo {pages} {109} (\bibinfo {year}
  {2010})}\BibitemShut {NoStop}%
\bibitem [{\citenamefont {Ilan}\ \emph {et~al.}()\citenamefont {Ilan},
  \citenamefont {Grushin},\ and\ \citenamefont {Pikulin}}]{Ilan2019}%
  \BibitemOpen
  \bibfield  {author} {\bibinfo {author} {\bibfnamefont {R.}~\bibnamefont
  {Ilan}}, \bibinfo {author} {\bibfnamefont {A.~G.}\ \bibnamefont {Grushin}}, \
  and\ \bibinfo {author} {\bibfnamefont {D.~I.}\ \bibnamefont {Pikulin}},\
  }\href {http://arxiv.org/abs/1903.11088} {\ }\Eprint
  {http://arxiv.org/abs/1903.11088} {arXiv:1903.11088} \BibitemShut {NoStop}%
\bibitem [{\citenamefont {Guinea}\ \emph {et~al.}(2010)\citenamefont {Guinea},
  \citenamefont {Katsnelson},\ and\ \citenamefont {Geim}}]{Guinea2010}%
  \BibitemOpen
  \bibfield  {author} {\bibinfo {author} {\bibfnamefont {F.}~\bibnamefont
  {Guinea}}, \bibinfo {author} {\bibfnamefont {M.~I.}\ \bibnamefont
  {Katsnelson}}, \ and\ \bibinfo {author} {\bibfnamefont {A.~K.}\ \bibnamefont
  {Geim}},\ }\href {\doibase 10.1038/nphys1420} {\bibfield  {journal} {\bibinfo
   {journal} {Nat. Phys.}\ }\textbf {\bibinfo {volume} {6}},\ \bibinfo {pages}
  {30} (\bibinfo {year} {2010})}\BibitemShut {NoStop}%
\bibitem [{\citenamefont {Levy}\ \emph {et~al.}(2010)\citenamefont {Levy},
  \citenamefont {Burke}, \citenamefont {Meaker}, \citenamefont {Panlasigui},
  \citenamefont {Zettl}, \citenamefont {Guinea}, \citenamefont {Neto},\ and\
  \citenamefont {Crommie}}]{Levy2010}%
  \BibitemOpen
  \bibfield  {author} {\bibinfo {author} {\bibfnamefont {N.}~\bibnamefont
  {Levy}}, \bibinfo {author} {\bibfnamefont {S.~A.}\ \bibnamefont {Burke}},
  \bibinfo {author} {\bibfnamefont {K.~L.}\ \bibnamefont {Meaker}}, \bibinfo
  {author} {\bibfnamefont {M.}~\bibnamefont {Panlasigui}}, \bibinfo {author}
  {\bibfnamefont {A.}~\bibnamefont {Zettl}}, \bibinfo {author} {\bibfnamefont
  {F.}~\bibnamefont {Guinea}}, \bibinfo {author} {\bibfnamefont {A.~H.~C.}\
  \bibnamefont {Neto}}, \ and\ \bibinfo {author} {\bibfnamefont {M.~F.}\
  \bibnamefont {Crommie}},\ }\href {\doibase 10.1126/science.1191700}
  {\bibfield  {journal} {\bibinfo  {journal} {Science}\ }\textbf {\bibinfo
  {volume} {329}},\ \bibinfo {pages} {544} (\bibinfo {year}
  {2010})}\BibitemShut {NoStop}%
\bibitem [{\citenamefont {Johansson}\ \emph {et~al.}(2017)\citenamefont
  {Johansson}, \citenamefont {Myllyperki{\"{o}}}, \citenamefont {Koskinen},
  \citenamefont {Aumanen}, \citenamefont {Koivistoinen}, \citenamefont {Tsai},
  \citenamefont {Chen}, \citenamefont {Chang}, \citenamefont {Hiltunen},
  \citenamefont {Manninen}, \citenamefont {Woon},\ and\ \citenamefont
  {Pettersson}}]{Johansson2017}%
  \BibitemOpen
  \bibfield  {author} {\bibinfo {author} {\bibfnamefont {A.}~\bibnamefont
  {Johansson}}, \bibinfo {author} {\bibfnamefont {P.}~\bibnamefont
  {Myllyperki{\"{o}}}}, \bibinfo {author} {\bibfnamefont {P.}~\bibnamefont
  {Koskinen}}, \bibinfo {author} {\bibfnamefont {J.}~\bibnamefont {Aumanen}},
  \bibinfo {author} {\bibfnamefont {J.}~\bibnamefont {Koivistoinen}}, \bibinfo
  {author} {\bibfnamefont {H.-C.}\ \bibnamefont {Tsai}}, \bibinfo {author}
  {\bibfnamefont {C.-H.}\ \bibnamefont {Chen}}, \bibinfo {author}
  {\bibfnamefont {L.-Y.}\ \bibnamefont {Chang}}, \bibinfo {author}
  {\bibfnamefont {V.-M.}\ \bibnamefont {Hiltunen}}, \bibinfo {author}
  {\bibfnamefont {J.~J.}\ \bibnamefont {Manninen}}, \bibinfo {author}
  {\bibfnamefont {W.~Y.}\ \bibnamefont {Woon}}, \ and\ \bibinfo {author}
  {\bibfnamefont {M.}~\bibnamefont {Pettersson}},\ }\href {\doibase
  10.1021/acs.nanolett.7b03530} {\bibfield  {journal} {\bibinfo  {journal}
  {Nano Lett.}\ }\textbf {\bibinfo {volume} {17}},\ \bibinfo {pages} {6469}
  (\bibinfo {year} {2017})}\BibitemShut {NoStop}%
\bibitem [{\citenamefont {Koskinen}\ \emph {et~al.}(2018)\citenamefont
  {Koskinen}, \citenamefont {Karppinen}, \citenamefont {Myllyperki{\"{o}}},
  \citenamefont {Hiltunen}, \citenamefont {Johansson},\ and\ \citenamefont
  {Pettersson}}]{Koskinen2018}%
  \BibitemOpen
  \bibfield  {author} {\bibinfo {author} {\bibfnamefont {P.}~\bibnamefont
  {Koskinen}}, \bibinfo {author} {\bibfnamefont {K.}~\bibnamefont {Karppinen}},
  \bibinfo {author} {\bibfnamefont {P.}~\bibnamefont {Myllyperki{\"{o}}}},
  \bibinfo {author} {\bibfnamefont {V.-M.}\ \bibnamefont {Hiltunen}}, \bibinfo
  {author} {\bibfnamefont {A.}~\bibnamefont {Johansson}}, \ and\ \bibinfo
  {author} {\bibfnamefont {M.}~\bibnamefont {Pettersson}},\ }\href {\doibase
  10.1021/acs.jpclett.8b02461} {\bibfield  {journal} {\bibinfo  {journal} {J.
  Phys. Chem. Lett.}\ }\textbf {\bibinfo {volume} {9}},\ \bibinfo {pages}
  {6179} (\bibinfo {year} {2018})}\BibitemShut {NoStop}%
\bibitem [{\citenamefont {Jia}\ \emph {et~al.}(2019)\citenamefont {Jia},
  \citenamefont {Chen}, \citenamefont {Qiao}, \citenamefont {Zhang},
  \citenamefont {Zheng}, \citenamefont {Xue}, \citenamefont {Liang},
  \citenamefont {Tian}, \citenamefont {He}, \citenamefont {Di},\ and\
  \citenamefont {Wang}}]{Jia2019}%
  \BibitemOpen
  \bibfield  {author} {\bibinfo {author} {\bibfnamefont {P.}~\bibnamefont
  {Jia}}, \bibinfo {author} {\bibfnamefont {W.}~\bibnamefont {Chen}}, \bibinfo
  {author} {\bibfnamefont {J.}~\bibnamefont {Qiao}}, \bibinfo {author}
  {\bibfnamefont {M.}~\bibnamefont {Zhang}}, \bibinfo {author} {\bibfnamefont
  {X.}~\bibnamefont {Zheng}}, \bibinfo {author} {\bibfnamefont
  {Z.}~\bibnamefont {Xue}}, \bibinfo {author} {\bibfnamefont {R.}~\bibnamefont
  {Liang}}, \bibinfo {author} {\bibfnamefont {C.}~\bibnamefont {Tian}},
  \bibinfo {author} {\bibfnamefont {L.}~\bibnamefont {He}}, \bibinfo {author}
  {\bibfnamefont {Z.}~\bibnamefont {Di}}, \ and\ \bibinfo {author}
  {\bibfnamefont {X.}~\bibnamefont {Wang}},\ }\href {\doibase
  10.1038/s41467-019-11038-7} {\bibfield  {journal} {\bibinfo  {journal} {Nat.
  Commun.}\ }\textbf {\bibinfo {volume} {10}},\ \bibinfo {pages} {3127}
  (\bibinfo {year} {2019})}\BibitemShut {NoStop}%
\bibitem [{\citenamefont {Aitken}\ and\ \citenamefont
  {Huang}(2010)}]{Aitken2010}%
  \BibitemOpen
  \bibfield  {author} {\bibinfo {author} {\bibfnamefont {Z.~H.}\ \bibnamefont
  {Aitken}}\ and\ \bibinfo {author} {\bibfnamefont {R.}~\bibnamefont {Huang}},\
  }\href {\doibase 10.1063/1.3437642} {\bibfield  {journal} {\bibinfo
  {journal} {J. Appl. Phys.}\ }\textbf {\bibinfo {volume} {107}},\ \bibinfo
  {pages} {123531} (\bibinfo {year} {2010})}\BibitemShut {NoStop}%
\bibitem [{\citenamefont {Jiang}\ \emph {et~al.}(2017)\citenamefont {Jiang},
  \citenamefont {Mao}, \citenamefont {Duan}, \citenamefont {Lai}, \citenamefont
  {Watanabe}, \citenamefont {Taniguchi},\ and\ \citenamefont
  {Andrei}}]{Jiang2017}%
  \BibitemOpen
  \bibfield  {author} {\bibinfo {author} {\bibfnamefont {Y.}~\bibnamefont
  {Jiang}}, \bibinfo {author} {\bibfnamefont {J.}~\bibnamefont {Mao}}, \bibinfo
  {author} {\bibfnamefont {J.}~\bibnamefont {Duan}}, \bibinfo {author}
  {\bibfnamefont {X.}~\bibnamefont {Lai}}, \bibinfo {author} {\bibfnamefont
  {K.}~\bibnamefont {Watanabe}}, \bibinfo {author} {\bibfnamefont
  {T.}~\bibnamefont {Taniguchi}}, \ and\ \bibinfo {author} {\bibfnamefont
  {E.~Y.}\ \bibnamefont {Andrei}},\ }\href {\doibase
  10.1021/acs.nanolett.6b05228} {\bibfield  {journal} {\bibinfo  {journal}
  {Nano Lett.}\ }\textbf {\bibinfo {volume} {17}},\ \bibinfo {pages} {2839}
  (\bibinfo {year} {2017})}\BibitemShut {NoStop}%
\bibitem [{\citenamefont {Yang}\ \emph {et~al.}(2016)\citenamefont {Yang},
  \citenamefont {Chen}, \citenamefont {Pan}, \citenamefont {Wang},
  \citenamefont {Ma},\ and\ \citenamefont {Liu}}]{Yang2016}%
  \BibitemOpen
  \bibfield  {author} {\bibinfo {author} {\bibfnamefont {K.}~\bibnamefont
  {Yang}}, \bibinfo {author} {\bibfnamefont {Y.}~\bibnamefont {Chen}}, \bibinfo
  {author} {\bibfnamefont {F.}~\bibnamefont {Pan}}, \bibinfo {author}
  {\bibfnamefont {S.}~\bibnamefont {Wang}}, \bibinfo {author} {\bibfnamefont
  {Y.}~\bibnamefont {Ma}}, \ and\ \bibinfo {author} {\bibfnamefont
  {Q.}~\bibnamefont {Liu}},\ }\href {\doibase 10.3390/ma9010032} {\bibfield
  {journal} {\bibinfo  {journal} {Materials}\ }\textbf {\bibinfo {volume}
  {9}},\ \bibinfo {pages} {32} (\bibinfo {year} {2016})}\BibitemShut {NoStop}%
\bibitem [{\citenamefont {Androulidakis}\ \emph {et~al.}(2015)\citenamefont
  {Androulidakis}, \citenamefont {Koukaras}, \citenamefont {Frank},
  \citenamefont {Tsoukleri}, \citenamefont {Sfyris}, \citenamefont
  {Parthenios}, \citenamefont {Pugno}, \citenamefont {Papagelis}, \citenamefont
  {Novoselov},\ and\ \citenamefont {Galiotis}}]{Androulidakis2014}%
  \BibitemOpen
  \bibfield  {author} {\bibinfo {author} {\bibfnamefont {C.}~\bibnamefont
  {Androulidakis}}, \bibinfo {author} {\bibfnamefont {E.~N.}\ \bibnamefont
  {Koukaras}}, \bibinfo {author} {\bibfnamefont {O.}~\bibnamefont {Frank}},
  \bibinfo {author} {\bibfnamefont {G.}~\bibnamefont {Tsoukleri}}, \bibinfo
  {author} {\bibfnamefont {D.}~\bibnamefont {Sfyris}}, \bibinfo {author}
  {\bibfnamefont {J.}~\bibnamefont {Parthenios}}, \bibinfo {author}
  {\bibfnamefont {N.}~\bibnamefont {Pugno}}, \bibinfo {author} {\bibfnamefont
  {K.}~\bibnamefont {Papagelis}}, \bibinfo {author} {\bibfnamefont {K.~S.}\
  \bibnamefont {Novoselov}}, \ and\ \bibinfo {author} {\bibfnamefont
  {C.}~\bibnamefont {Galiotis}},\ }\href {\doibase 10.1038/srep05271}
  {\bibfield  {journal} {\bibinfo  {journal} {Sci. Rep.}\ }\textbf {\bibinfo
  {volume} {4}},\ \bibinfo {pages} {5271} (\bibinfo {year} {2015})}\BibitemShut
  {NoStop}%
\bibitem [{\citenamefont {Koukaras}\ \emph {et~al.}(2016)\citenamefont
  {Koukaras}, \citenamefont {Androulidakis}, \citenamefont {Anagnostopoulos},
  \citenamefont {Papagelis},\ and\ \citenamefont {Galiotis}}]{Koukaras2016}%
  \BibitemOpen
  \bibfield  {author} {\bibinfo {author} {\bibfnamefont {E.~N.}\ \bibnamefont
  {Koukaras}}, \bibinfo {author} {\bibfnamefont {C.}~\bibnamefont
  {Androulidakis}}, \bibinfo {author} {\bibfnamefont {G.}~\bibnamefont
  {Anagnostopoulos}}, \bibinfo {author} {\bibfnamefont {K.}~\bibnamefont
  {Papagelis}}, \ and\ \bibinfo {author} {\bibfnamefont {C.}~\bibnamefont
  {Galiotis}},\ }\href {\doibase 10.1016/j.eml.2016.03.016} {\bibfield
  {journal} {\bibinfo  {journal} {Extreme Mech. Lett.}\ }\textbf {\bibinfo
  {volume} {8}},\ \bibinfo {pages} {191} (\bibinfo {year} {2016})}\BibitemShut
  {NoStop}%
\bibitem [{\citenamefont {Koskinen}(2014)}]{Koskinen2014}%
  \BibitemOpen
  \bibfield  {author} {\bibinfo {author} {\bibfnamefont {P.}~\bibnamefont
  {Koskinen}},\ }\href {\doibase 10.1063/1.4868125} {\bibfield  {journal}
  {\bibinfo  {journal} {Appl. Phys. Lett.}\ }\textbf {\bibinfo {volume}
  {104}},\ \bibinfo {pages} {101902} (\bibinfo {year} {2014})}\BibitemShut
  {NoStop}%
\bibitem [{\citenamefont {Tarruell}\ \emph {et~al.}(2012)\citenamefont
  {Tarruell}, \citenamefont {Greif}, \citenamefont {Uehlinger}, \citenamefont
  {Jotzu},\ and\ \citenamefont {Esslinger}}]{Tarruell2012}%
  \BibitemOpen
  \bibfield  {author} {\bibinfo {author} {\bibfnamefont {L.}~\bibnamefont
  {Tarruell}}, \bibinfo {author} {\bibfnamefont {D.}~\bibnamefont {Greif}},
  \bibinfo {author} {\bibfnamefont {T.}~\bibnamefont {Uehlinger}}, \bibinfo
  {author} {\bibfnamefont {G.}~\bibnamefont {Jotzu}}, \ and\ \bibinfo {author}
  {\bibfnamefont {T.}~\bibnamefont {Esslinger}},\ }\href {\doibase
  10.1038/nature10871} {\bibfield  {journal} {\bibinfo  {journal} {Nature}\
  }\textbf {\bibinfo {volume} {483}},\ \bibinfo {pages} {302} (\bibinfo {year}
  {2012})}\BibitemShut {NoStop}%
\bibitem [{\citenamefont {van Wijk}\ \emph {et~al.}(2015)\citenamefont {van
  Wijk}, \citenamefont {Schuring}, \citenamefont {Katsnelson},\ and\
  \citenamefont {Fasolino}}]{vanWijk2015}%
  \BibitemOpen
  \bibfield  {author} {\bibinfo {author} {\bibfnamefont {M.~M.}\ \bibnamefont
  {van Wijk}}, \bibinfo {author} {\bibfnamefont {A.}~\bibnamefont {Schuring}},
  \bibinfo {author} {\bibfnamefont {M.~I.}\ \bibnamefont {Katsnelson}}, \ and\
  \bibinfo {author} {\bibfnamefont {A.}~\bibnamefont {Fasolino}},\ }\href
  {\doibase 10.1088/2053-1583/2/3/034010} {\bibfield  {journal} {\bibinfo
  {journal} {2D Mater.}\ }\textbf {\bibinfo {volume} {2}},\ \bibinfo {pages}
  {34010} (\bibinfo {year} {2015})}\BibitemShut {NoStop}%
\bibitem [{\citenamefont {Dai}\ \emph {et~al.}(2016)\citenamefont {Dai},
  \citenamefont {Xiang},\ and\ \citenamefont {Srolovitz}}]{Dai2016b}%
  \BibitemOpen
  \bibfield  {author} {\bibinfo {author} {\bibfnamefont {S.}~\bibnamefont
  {Dai}}, \bibinfo {author} {\bibfnamefont {Y.}~\bibnamefont {Xiang}}, \ and\
  \bibinfo {author} {\bibfnamefont {D.~J.}\ \bibnamefont {Srolovitz}},\ }\href
  {\doibase 10.1021/acs.nanolett.6b02870} {\bibfield  {journal} {\bibinfo
  {journal} {Nano Lett.}\ }\textbf {\bibinfo {volume} {16}},\ \bibinfo {pages}
  {5923} (\bibinfo {year} {2016})}\BibitemShut {NoStop}%
\bibitem [{\citenamefont {Nam}\ and\ \citenamefont {Koshino}(2017)}]{Nam2017}%
  \BibitemOpen
  \bibfield  {author} {\bibinfo {author} {\bibfnamefont {N.~N.~T.}\
  \bibnamefont {Nam}}\ and\ \bibinfo {author} {\bibfnamefont {M.}~\bibnamefont
  {Koshino}},\ }\href {\doibase 10.1103/PhysRevB.96.075311} {\bibfield
  {journal} {\bibinfo  {journal} {Phys. Rev. B}\ }\textbf {\bibinfo {volume}
  {96}},\ \bibinfo {pages} {075311} (\bibinfo {year} {2017})}\BibitemShut
  {NoStop}%
\bibitem [{\citenamefont {Lin}\ \emph {et~al.}(2018)\citenamefont {Lin},
  \citenamefont {Liu},\ and\ \citenamefont {Tom{\'{a}}nek}}]{Lin2018a}%
  \BibitemOpen
  \bibfield  {author} {\bibinfo {author} {\bibfnamefont {X.}~\bibnamefont
  {Lin}}, \bibinfo {author} {\bibfnamefont {D.}~\bibnamefont {Liu}}, \ and\
  \bibinfo {author} {\bibfnamefont {D.}~\bibnamefont {Tom{\'{a}}nek}},\ }\href
  {\doibase 10.1103/PhysRevB.98.195432} {\bibfield  {journal} {\bibinfo
  {journal} {Phys. Rev. B}\ }\textbf {\bibinfo {volume} {98}},\ \bibinfo
  {pages} {195432} (\bibinfo {year} {2018})}\BibitemShut {NoStop}%
\bibitem [{\citenamefont {Fang}\ \emph {et~al.}()\citenamefont {Fang},
  \citenamefont {Carr}, \citenamefont {Zhu}, \citenamefont {Massatt},\ and\
  \citenamefont {Kaxiras}}]{Fang2019}%
  \BibitemOpen
  \bibfield  {author} {\bibinfo {author} {\bibfnamefont {S.}~\bibnamefont
  {Fang}}, \bibinfo {author} {\bibfnamefont {S.}~\bibnamefont {Carr}}, \bibinfo
  {author} {\bibfnamefont {Z.}~\bibnamefont {Zhu}}, \bibinfo {author}
  {\bibfnamefont {D.}~\bibnamefont {Massatt}}, \ and\ \bibinfo {author}
  {\bibfnamefont {E.}~\bibnamefont {Kaxiras}},\ }\href
  {http://arxiv.org/abs/1908.00058} {\ }\Eprint
  {http://arxiv.org/abs/1908.00058} {arXiv:1908.00058} \BibitemShut {NoStop}%
\bibitem [{\citenamefont {Yoo}\ \emph {et~al.}(2019)\citenamefont {Yoo},
  \citenamefont {Engelke}, \citenamefont {Carr}, \citenamefont {Fang},
  \citenamefont {Zhang}, \citenamefont {Cazeaux}, \citenamefont {Sung},
  \citenamefont {Hovden}, \citenamefont {Tsen}, \citenamefont {Taniguchi},
  \citenamefont {Watanabe}, \citenamefont {Yi}, \citenamefont {Kim},
  \citenamefont {Luskin}, \citenamefont {Tadmor}, \citenamefont {Kaxiras},\
  and\ \citenamefont {Kim}}]{Yoo2018}%
  \BibitemOpen
  \bibfield  {author} {\bibinfo {author} {\bibfnamefont {H.}~\bibnamefont
  {Yoo}}, \bibinfo {author} {\bibfnamefont {R.}~\bibnamefont {Engelke}},
  \bibinfo {author} {\bibfnamefont {S.}~\bibnamefont {Carr}}, \bibinfo {author}
  {\bibfnamefont {S.}~\bibnamefont {Fang}}, \bibinfo {author} {\bibfnamefont
  {K.}~\bibnamefont {Zhang}}, \bibinfo {author} {\bibfnamefont
  {P.}~\bibnamefont {Cazeaux}}, \bibinfo {author} {\bibfnamefont {S.~H.}\
  \bibnamefont {Sung}}, \bibinfo {author} {\bibfnamefont {R.}~\bibnamefont
  {Hovden}}, \bibinfo {author} {\bibfnamefont {A.~W.}\ \bibnamefont {Tsen}},
  \bibinfo {author} {\bibfnamefont {T.}~\bibnamefont {Taniguchi}}, \bibinfo
  {author} {\bibfnamefont {K.}~\bibnamefont {Watanabe}}, \bibinfo {author}
  {\bibfnamefont {G.-C.}\ \bibnamefont {Yi}}, \bibinfo {author} {\bibfnamefont
  {M.}~\bibnamefont {Kim}}, \bibinfo {author} {\bibfnamefont {M.}~\bibnamefont
  {Luskin}}, \bibinfo {author} {\bibfnamefont {E.~B.}\ \bibnamefont {Tadmor}},
  \bibinfo {author} {\bibfnamefont {E.}~\bibnamefont {Kaxiras}}, \ and\
  \bibinfo {author} {\bibfnamefont {P.}~\bibnamefont {Kim}},\ }\href {\doibase
  10.1038/s41563-019-0346-z} {\bibfield  {journal} {\bibinfo  {journal} {Nat.
  Mater.}\ }\textbf {\bibinfo {volume} {18}},\ \bibinfo {pages} {448} (\bibinfo
  {year} {2019})}\BibitemShut {NoStop}%
\bibitem [{\citenamefont {Peotta}\ and\ \citenamefont
  {T{\"{o}}rm{\"{a}}}(2015)}]{Peotta2015}%
  \BibitemOpen
  \bibfield  {author} {\bibinfo {author} {\bibfnamefont {S.}~\bibnamefont
  {Peotta}}\ and\ \bibinfo {author} {\bibfnamefont {P.}~\bibnamefont
  {T{\"{o}}rm{\"{a}}}},\ }\href {\doibase 10.1038/ncomms9944} {\bibfield
  {journal} {\bibinfo  {journal} {Nat. Commun.}\ }\textbf {\bibinfo {volume}
  {6}},\ \bibinfo {pages} {8944} (\bibinfo {year} {2015})}\BibitemShut
  {NoStop}%
\bibitem [{\citenamefont {Liang}\ \emph {et~al.}(2017)\citenamefont {Liang},
  \citenamefont {Vanhala}, \citenamefont {Peotta}, \citenamefont {Siro},
  \citenamefont {Harju},\ and\ \citenamefont {T{\"{o}}rm{\"{a}}}}]{Liang2017}%
  \BibitemOpen
  \bibfield  {author} {\bibinfo {author} {\bibfnamefont {L.}~\bibnamefont
  {Liang}}, \bibinfo {author} {\bibfnamefont {T.~I.}\ \bibnamefont {Vanhala}},
  \bibinfo {author} {\bibfnamefont {S.}~\bibnamefont {Peotta}}, \bibinfo
  {author} {\bibfnamefont {T.}~\bibnamefont {Siro}}, \bibinfo {author}
  {\bibfnamefont {A.}~\bibnamefont {Harju}}, \ and\ \bibinfo {author}
  {\bibfnamefont {P.}~\bibnamefont {T{\"{o}}rm{\"{a}}}},\ }\href {\doibase
  10.1103/PhysRevB.95.024515} {\bibfield  {journal} {\bibinfo  {journal} {Phys.
  Rev. B}\ }\textbf {\bibinfo {volume} {95}},\ \bibinfo {pages} {024515}
  (\bibinfo {year} {2017})}\BibitemShut {NoStop}%
\bibitem [{\citenamefont {Zhao}\ and\ \citenamefont
  {Paramekanti}(2006)}]{Zhao2006}%
  \BibitemOpen
  \bibfield  {author} {\bibinfo {author} {\bibfnamefont {E.}~\bibnamefont
  {Zhao}}\ and\ \bibinfo {author} {\bibfnamefont {A.}~\bibnamefont
  {Paramekanti}},\ }\href {\doibase 10.1103/PhysRevLett.97.230404} {\bibfield
  {journal} {\bibinfo  {journal} {Phys. Rev. Lett.}\ }\textbf {\bibinfo
  {volume} {97}},\ \bibinfo {pages} {230404} (\bibinfo {year}
  {2006})}\BibitemShut {NoStop}%
\bibitem [{\citenamefont {Uchoa}\ and\ \citenamefont {{Castro
  Neto}}(2007)}]{Uchoa2007}%
  \BibitemOpen
  \bibfield  {author} {\bibinfo {author} {\bibfnamefont {B.}~\bibnamefont
  {Uchoa}}\ and\ \bibinfo {author} {\bibfnamefont {A.~H.}\ \bibnamefont
  {{Castro Neto}}},\ }\href {\doibase 10.1103/PhysRevLett.98.146801} {\bibfield
   {journal} {\bibinfo  {journal} {Phys. Rev. Lett.}\ }\textbf {\bibinfo
  {volume} {98}},\ \bibinfo {pages} {146801} (\bibinfo {year}
  {2007})}\BibitemShut {NoStop}%
\bibitem [{\citenamefont {Kopnin}\ and\ \citenamefont
  {Sonin}(2008)}]{Kopnin2008}%
  \BibitemOpen
  \bibfield  {author} {\bibinfo {author} {\bibfnamefont {N.~B.}\ \bibnamefont
  {Kopnin}}\ and\ \bibinfo {author} {\bibfnamefont {E.~B.}\ \bibnamefont
  {Sonin}},\ }\href {\doibase 10.1103/PhysRevLett.100.246808} {\bibfield
  {journal} {\bibinfo  {journal} {Phys. Rev. Lett.}\ }\textbf {\bibinfo
  {volume} {100}},\ \bibinfo {pages} {246808} (\bibinfo {year}
  {2008})}\BibitemShut {NoStop}%
\bibitem [{\citenamefont {Kopnin}\ and\ \citenamefont
  {Sonin}(2010)}]{Kopnin2010}%
  \BibitemOpen
  \bibfield  {author} {\bibinfo {author} {\bibfnamefont {N.~B.}\ \bibnamefont
  {Kopnin}}\ and\ \bibinfo {author} {\bibfnamefont {E.~B.}\ \bibnamefont
  {Sonin}},\ }\href {\doibase 10.1103/PhysRevB.82.014516} {\bibfield  {journal}
  {\bibinfo  {journal} {Phys. Rev. B}\ }\textbf {\bibinfo {volume} {82}},\
  \bibinfo {pages} {014516} (\bibinfo {year} {2010})}\BibitemShut {NoStop}%
\bibitem [{\citenamefont {Hosseini}(2015)}]{Hosseini2015}%
  \BibitemOpen
  \bibfield  {author} {\bibinfo {author} {\bibfnamefont {M.~V.}\ \bibnamefont
  {Hosseini}},\ }\href {\doibase 10.1209/0295-5075/110/47010} {\bibfield
  {journal} {\bibinfo  {journal} {EPL}\ }\textbf {\bibinfo {volume} {110}},\
  \bibinfo {pages} {47010} (\bibinfo {year} {2015})}\BibitemShut {NoStop}%
\bibitem [{\citenamefont {Wu}\ \emph {et~al.}(2018)\citenamefont {Wu},
  \citenamefont {MacDonald},\ and\ \citenamefont {Martin}}]{Wu2018}%
  \BibitemOpen
  \bibfield  {author} {\bibinfo {author} {\bibfnamefont {F.}~\bibnamefont
  {Wu}}, \bibinfo {author} {\bibfnamefont {A.~H.}\ \bibnamefont {MacDonald}}, \
  and\ \bibinfo {author} {\bibfnamefont {I.}~\bibnamefont {Martin}},\ }\href
  {\doibase 10.1103/PhysRevLett.121.257001} {\bibfield  {journal} {\bibinfo
  {journal} {Phys. Rev. Lett.}\ }\textbf {\bibinfo {volume} {121}},\ \bibinfo
  {pages} {257001} (\bibinfo {year} {2018})}\BibitemShut {NoStop}%
\bibitem [{\citenamefont {Julku}\ \emph {et~al.}()\citenamefont {Julku},
  \citenamefont {Peltonen}, \citenamefont {Liang}, \citenamefont
  {Heikkil{\"{a}}},\ and\ \citenamefont {T{\"{o}}rm{\"{a}}}}]{Julku2019}%
  \BibitemOpen
  \bibfield  {author} {\bibinfo {author} {\bibfnamefont {A.}~\bibnamefont
  {Julku}}, \bibinfo {author} {\bibfnamefont {T.~J.}\ \bibnamefont {Peltonen}},
  \bibinfo {author} {\bibfnamefont {L.}~\bibnamefont {Liang}}, \bibinfo
  {author} {\bibfnamefont {T.~T.}\ \bibnamefont {Heikkil{\"{a}}}}, \ and\
  \bibinfo {author} {\bibfnamefont {P.}~\bibnamefont {T{\"{o}}rm{\"{a}}}},\
  }\href {http://arxiv.org/abs/1906.06313} {\ }\Eprint
  {http://arxiv.org/abs/1906.06313} {arXiv:1906.06313} \BibitemShut {NoStop}%
\bibitem [{\citenamefont {Nelson}\ and\ \citenamefont
  {Kosterlitz}(1977)}]{Nelson1977}%
  \BibitemOpen
  \bibfield  {author} {\bibinfo {author} {\bibfnamefont {D.~R.}\ \bibnamefont
  {Nelson}}\ and\ \bibinfo {author} {\bibfnamefont {J.~M.}\ \bibnamefont
  {Kosterlitz}},\ }\href {\doibase 10.1103/PhysRevLett.39.1201} {\bibfield
  {journal} {\bibinfo  {journal} {Phys. Rev. Lett.}\ }\textbf {\bibinfo
  {volume} {39}},\ \bibinfo {pages} {1201} (\bibinfo {year}
  {1977})}\BibitemShut {NoStop}%
\bibitem [{\citenamefont {Cao}\ \emph {et~al.}(2014)\citenamefont {Cao},
  \citenamefont {Zou}, \citenamefont {Liu}, \citenamefont {Yi}, \citenamefont
  {Long},\ and\ \citenamefont {Hu}}]{Cao2014}%
  \BibitemOpen
  \bibfield  {author} {\bibinfo {author} {\bibfnamefont {Y.}~\bibnamefont
  {Cao}}, \bibinfo {author} {\bibfnamefont {S.-H.}\ \bibnamefont {Zou}},
  \bibinfo {author} {\bibfnamefont {X.-J.}\ \bibnamefont {Liu}}, \bibinfo
  {author} {\bibfnamefont {S.}~\bibnamefont {Yi}}, \bibinfo {author}
  {\bibfnamefont {G.-L.}\ \bibnamefont {Long}}, \ and\ \bibinfo {author}
  {\bibfnamefont {H.}~\bibnamefont {Hu}},\ }\href {\doibase
  10.1103/PhysRevLett.113.115302} {\bibfield  {journal} {\bibinfo  {journal}
  {Phys. Rev. Lett.}\ }\textbf {\bibinfo {volume} {113}},\ \bibinfo {pages}
  {115302} (\bibinfo {year} {2014})}\BibitemShut {NoStop}%
\bibitem [{\citenamefont {Xu}\ and\ \citenamefont {Zhang}(2015)}]{Xu2015}%
  \BibitemOpen
  \bibfield  {author} {\bibinfo {author} {\bibfnamefont {Y.}~\bibnamefont
  {Xu}}\ and\ \bibinfo {author} {\bibfnamefont {C.}~\bibnamefont {Zhang}},\
  }\href {\doibase 10.1103/PhysRevLett.114.110401} {\bibfield  {journal}
  {\bibinfo  {journal} {Phys. Rev. Lett.}\ }\textbf {\bibinfo {volume} {114}},\
  \bibinfo {pages} {110401} (\bibinfo {year} {2015})}\BibitemShut {NoStop}%
\bibitem [{Note1()}]{Note1}%
  \BibitemOpen
  \bibinfo {note} {The \protect \textsc {Mathematica} code we have used to
  compute the normal-state properties, the mean-field order parameter,
  superfluid weight, and the BKT transition temperature will be made available
  upon publication of this manuscript.}\BibitemShut {Stop}%
\bibitem [{\citenamefont {Tarnopolsky}\ \emph {et~al.}(2019)\citenamefont
  {Tarnopolsky}, \citenamefont {Kruchkov},\ and\ \citenamefont
  {Vishwanath}}]{Tarnopolsky2019}%
  \BibitemOpen
  \bibfield  {author} {\bibinfo {author} {\bibfnamefont {G.}~\bibnamefont
  {Tarnopolsky}}, \bibinfo {author} {\bibfnamefont {A.~J.}\ \bibnamefont
  {Kruchkov}}, \ and\ \bibinfo {author} {\bibfnamefont {A.}~\bibnamefont
  {Vishwanath}},\ }\href {\doibase 10.1103/PhysRevLett.122.106405} {\bibfield
  {journal} {\bibinfo  {journal} {Phys. Rev. Lett.}\ }\textbf {\bibinfo
  {volume} {122}},\ \bibinfo {pages} {106405} (\bibinfo {year}
  {2019})}\BibitemShut {NoStop}%
\bibitem [{\citenamefont {Xie}\ \emph {et~al.}(2019)\citenamefont {Xie},
  \citenamefont {Lian}, \citenamefont {J{\"{a}}ck}, \citenamefont {Liu},
  \citenamefont {Chiu}, \citenamefont {Watanabe}, \citenamefont {Taniguchi},
  \citenamefont {Bernevig},\ and\ \citenamefont {Yazdani}}]{Xie2019}%
  \BibitemOpen
  \bibfield  {author} {\bibinfo {author} {\bibfnamefont {Y.}~\bibnamefont
  {Xie}}, \bibinfo {author} {\bibfnamefont {B.}~\bibnamefont {Lian}}, \bibinfo
  {author} {\bibfnamefont {B.}~\bibnamefont {J{\"{a}}ck}}, \bibinfo {author}
  {\bibfnamefont {X.}~\bibnamefont {Liu}}, \bibinfo {author} {\bibfnamefont
  {C.-L.}\ \bibnamefont {Chiu}}, \bibinfo {author} {\bibfnamefont
  {K.}~\bibnamefont {Watanabe}}, \bibinfo {author} {\bibfnamefont
  {T.}~\bibnamefont {Taniguchi}}, \bibinfo {author} {\bibfnamefont {B.~A.}\
  \bibnamefont {Bernevig}}, \ and\ \bibinfo {author} {\bibfnamefont
  {A.}~\bibnamefont {Yazdani}},\ }\href {\doibase 10.1038/s41586-019-1422-x}
  {\bibfield  {journal} {\bibinfo  {journal} {Nature}\ }\textbf {\bibinfo
  {volume} {572}},\ \bibinfo {pages} {101} (\bibinfo {year}
  {2019})}\BibitemShut {NoStop}%
\bibitem [{\citenamefont {Heikkil{\"{a}}}\ and\ \citenamefont
  {Volovik}(2016)}]{Heikkila2016}%
  \BibitemOpen
  \bibfield  {author} {\bibinfo {author} {\bibfnamefont {T.~T.}\ \bibnamefont
  {Heikkil{\"{a}}}}\ and\ \bibinfo {author} {\bibfnamefont {G.~E.}\
  \bibnamefont {Volovik}},\ }in\ \href {\doibase 10.1007/978-3-319-39355-1_6}
  {\emph {\bibinfo {booktitle} {Basic Physics of Functionalized Graphite}}},\
  \bibinfo {editor} {edited by\ \bibinfo {editor} {\bibfnamefont
  {P.}~\bibnamefont {Esquinazi}}}\ (\bibinfo  {publisher} {Springer},\ \bibinfo
  {year} {2016})\ pp.\ \bibinfo {pages} {123--143},\ \Eprint
  {http://arxiv.org/abs/1504.05824} {arXiv:1504.05824} \BibitemShut {NoStop}%
\bibitem [{\citenamefont {Chiodi}\ \emph {et~al.}(2011)\citenamefont {Chiodi},
  \citenamefont {Ferrier}, \citenamefont {Tikhonov}, \citenamefont {Virtanen},
  \citenamefont {Heikkil{\"{a}}}, \citenamefont {Feigelman}, \citenamefont
  {Gu{\'{e}}ron},\ and\ \citenamefont {Bouchiat}}]{Chiodi2011}%
  \BibitemOpen
  \bibfield  {author} {\bibinfo {author} {\bibfnamefont {F.}~\bibnamefont
  {Chiodi}}, \bibinfo {author} {\bibfnamefont {M.}~\bibnamefont {Ferrier}},
  \bibinfo {author} {\bibfnamefont {K.}~\bibnamefont {Tikhonov}}, \bibinfo
  {author} {\bibfnamefont {P.}~\bibnamefont {Virtanen}}, \bibinfo {author}
  {\bibfnamefont {T.~T.}\ \bibnamefont {Heikkil{\"{a}}}}, \bibinfo {author}
  {\bibfnamefont {M.}~\bibnamefont {Feigelman}}, \bibinfo {author}
  {\bibfnamefont {S.}~\bibnamefont {Gu{\'{e}}ron}}, \ and\ \bibinfo {author}
  {\bibfnamefont {H.}~\bibnamefont {Bouchiat}},\ }\href {\doibase
  10.1038/srep00003} {\bibfield  {journal} {\bibinfo  {journal} {Sci. Rep.}\
  }\textbf {\bibinfo {volume} {1}},\ \bibinfo {pages} {3} (\bibinfo {year}
  {2011})}\BibitemShut {NoStop}%
\bibitem [{\citenamefont {Hu}\ \emph {et~al.}()\citenamefont {Hu},
  \citenamefont {Hyart}, \citenamefont {Pikulin},\ and\ \citenamefont
  {Rossi}}]{Hu2019}%
  \BibitemOpen
  \bibfield  {author} {\bibinfo {author} {\bibfnamefont {X.}~\bibnamefont
  {Hu}}, \bibinfo {author} {\bibfnamefont {T.}~\bibnamefont {Hyart}}, \bibinfo
  {author} {\bibfnamefont {D.~I.}\ \bibnamefont {Pikulin}}, \ and\ \bibinfo
  {author} {\bibfnamefont {E.}~\bibnamefont {Rossi}},\ }\href
  {http://arxiv.org/abs/1906.07152} {\ }\Eprint
  {http://arxiv.org/abs/1906.07152} {arXiv:1906.07152} \BibitemShut {NoStop}%
\bibitem [{\citenamefont {Stepanov}\ \emph {et~al.}()\citenamefont {Stepanov},
  \citenamefont {Das}, \citenamefont {Lu}, \citenamefont {Fahimniya},
  \citenamefont {Watanabe}, \citenamefont {Taniguchi}, \citenamefont {Koppens},
  \citenamefont {Lischner}, \citenamefont {Levitov},\ and\ \citenamefont
  {Efetov}}]{Stepanov2019}%
  \BibitemOpen
  \bibfield  {author} {\bibinfo {author} {\bibfnamefont {P.}~\bibnamefont
  {Stepanov}}, \bibinfo {author} {\bibfnamefont {I.}~\bibnamefont {Das}},
  \bibinfo {author} {\bibfnamefont {X.}~\bibnamefont {Lu}}, \bibinfo {author}
  {\bibfnamefont {A.}~\bibnamefont {Fahimniya}}, \bibinfo {author}
  {\bibfnamefont {K.}~\bibnamefont {Watanabe}}, \bibinfo {author}
  {\bibfnamefont {T.}~\bibnamefont {Taniguchi}}, \bibinfo {author}
  {\bibfnamefont {F.~H.~L.}\ \bibnamefont {Koppens}}, \bibinfo {author}
  {\bibfnamefont {J.}~\bibnamefont {Lischner}}, \bibinfo {author}
  {\bibfnamefont {L.}~\bibnamefont {Levitov}}, \ and\ \bibinfo {author}
  {\bibfnamefont {D.~K.}\ \bibnamefont {Efetov}},\ }\href
  {http://arxiv.org/abs/1911.09198} {\ }\Eprint
  {http://arxiv.org/abs/1911.09198} {arXiv:1911.09198} \BibitemShut {NoStop}%
\bibitem [{\citenamefont {Saito}\ \emph {et~al.}()\citenamefont {Saito},
  \citenamefont {Ge}, \citenamefont {Watanabe}, \citenamefont {Taniguchi},\
  and\ \citenamefont {Young}}]{Saito2019}%
  \BibitemOpen
  \bibfield  {author} {\bibinfo {author} {\bibfnamefont {Y.}~\bibnamefont
  {Saito}}, \bibinfo {author} {\bibfnamefont {J.}~\bibnamefont {Ge}}, \bibinfo
  {author} {\bibfnamefont {K.}~\bibnamefont {Watanabe}}, \bibinfo {author}
  {\bibfnamefont {T.}~\bibnamefont {Taniguchi}}, \ and\ \bibinfo {author}
  {\bibfnamefont {A.~F.}\ \bibnamefont {Young}},\ }\href
  {http://arxiv.org/abs/1911.13302} {\ }\Eprint
  {http://arxiv.org/abs/1911.13302} {arXiv:1911.13302} \BibitemShut {NoStop}%
\bibitem [{\citenamefont {Ojaj{\"{a}}rvi}\ \emph {et~al.}(2018)\citenamefont
  {Ojaj{\"{a}}rvi}, \citenamefont {Hyart}, \citenamefont {Silaev},\ and\
  \citenamefont {Heikkil{\"{a}}}}]{Ojajarvi2018}%
  \BibitemOpen
  \bibfield  {author} {\bibinfo {author} {\bibfnamefont {R.}~\bibnamefont
  {Ojaj{\"{a}}rvi}}, \bibinfo {author} {\bibfnamefont {T.}~\bibnamefont
  {Hyart}}, \bibinfo {author} {\bibfnamefont {M.~A.}\ \bibnamefont {Silaev}}, \
  and\ \bibinfo {author} {\bibfnamefont {T.~T.}\ \bibnamefont
  {Heikkil{\"{a}}}},\ }\href {\doibase 10.1103/PhysRevB.98.054515} {\bibfield
  {journal} {\bibinfo  {journal} {Phys. Rev. B}\ }\textbf {\bibinfo {volume}
  {98}},\ \bibinfo {pages} {054515} (\bibinfo {year} {2018})}\BibitemShut
  {NoStop}%
\bibitem [{\citenamefont {Zheng}(2007)}]{Zheng2007}%
  \BibitemOpen
  \bibfield  {author} {\bibinfo {author} {\bibfnamefont {X.}~\bibnamefont
  {Zheng}},\ }\emph {\bibinfo {title} {{Efficient Fourier Transforms on
  Hexagonal Arrays}}},\ \href@noop {} {\bibinfo {type} {{PhD} thesis}},\
  \bibinfo  {school} {University of Florida} (\bibinfo {year}
  {2007})\BibitemShut {NoStop}%
\bibitem [{\citenamefont {Nazarov}\ and\ \citenamefont
  {Danon}(2013)}]{Nazarov2013}%
  \BibitemOpen
  \bibfield  {author} {\bibinfo {author} {\bibfnamefont {Y.~V.}\ \bibnamefont
  {Nazarov}}\ and\ \bibinfo {author} {\bibfnamefont {J.}~\bibnamefont
  {Danon}},\ }\href@noop {} {\emph {\bibinfo {title} {{Advanced Quantum
  Mechanics: A practical guide}}}},\ \bibinfo {edition} {1st}\ ed.\ (\bibinfo
  {publisher} {Cambridge University Press},\ \bibinfo {address} {New York},\
  \bibinfo {year} {2013})\BibitemShut {NoStop}%
\bibitem [{\citenamefont {Beenakker}(2006)}]{Beenakker2006}%
  \BibitemOpen
  \bibfield  {author} {\bibinfo {author} {\bibfnamefont {C.~W.~J.}\
  \bibnamefont {Beenakker}},\ }\href {\doibase 10.1103/PhysRevLett.97.067007}
  {\bibfield  {journal} {\bibinfo  {journal} {Phys. Rev. Lett.}\ }\textbf
  {\bibinfo {volume} {97}},\ \bibinfo {pages} {067007} (\bibinfo {year}
  {2006})}\BibitemShut {NoStop}%
\bibitem [{\citenamefont {Titov}\ and\ \citenamefont
  {Beenakker}(2006)}]{Titov2006}%
  \BibitemOpen
  \bibfield  {author} {\bibinfo {author} {\bibfnamefont {M.}~\bibnamefont
  {Titov}}\ and\ \bibinfo {author} {\bibfnamefont {C.~W.~J.}\ \bibnamefont
  {Beenakker}},\ }\href {\doibase 10.1103/PhysRevB.74.041401} {\bibfield
  {journal} {\bibinfo  {journal} {Phys. Rev. B}\ }\textbf {\bibinfo {volume}
  {74}},\ \bibinfo {pages} {041401(R)} (\bibinfo {year} {2006})}\BibitemShut
  {NoStop}%
\end{thebibliography}%


\begin{thebibliography}{10}%
\makeatletter
\providecommand \@ifxundefined [1]{%
 \@ifx{#1\undefined}
}%
\providecommand \@ifnum [1]{%
 \ifnum #1\expandafter \@firstoftwo
 \else \expandafter \@secondoftwo
 \fi
}%
\providecommand \@ifx [1]{%
 \ifx #1\expandafter \@firstoftwo
 \else \expandafter \@secondoftwo
 \fi
}%
\providecommand \natexlab [1]{#1}%
\providecommand \enquote  [1]{``#1''}%
\providecommand \bibnamefont  [1]{#1}%
\providecommand \bibfnamefont [1]{#1}%
\providecommand \citenamefont [1]{#1}%
\providecommand \href@noop [0]{\@secondoftwo}%
\providecommand \href [0]{\begingroup \@sanitize@url \@href}%
\providecommand \@href[1]{\@@startlink{#1}\@@href}%
\providecommand \@@href[1]{\endgroup#1\@@endlink}%
\providecommand \@sanitize@url [0]{\catcode `\\12\catcode `\$12\catcode
  `\&12\catcode `\#12\catcode `\^12\catcode `\_12\catcode `\%12\relax}%
\providecommand \@@startlink[1]{}%
\providecommand \@@endlink[0]{}%
\providecommand \url  [0]{\begingroup\@sanitize@url \@url }%
\providecommand \@url [1]{\endgroup\@href {#1}{\urlprefix }}%
\providecommand \urlprefix  [0]{URL }%
\providecommand \Eprint [0]{\href }%
\providecommand \doibase [0]{http://dx.doi.org/}%
\providecommand \selectlanguage [0]{\@gobble}%
\providecommand \bibinfo  [0]{\@secondoftwo}%
\providecommand \bibfield  [0]{\@secondoftwo}%
\providecommand \translation [1]{[#1]}%
\providecommand \BibitemOpen [0]{}%
\providecommand \bibitemStop [0]{}%
\providecommand \bibitemNoStop [0]{.\EOS\space}%
\providecommand \EOS [0]{\spacefactor3000\relax}%
\providecommand \BibitemShut  [1]{\csname bibitem#1\endcsname}%
\let\auto@bib@innerbib\@empty
\bibitem [{\citenamefont {Zheng}(2007)}]{Zheng2007}%
  \BibitemOpen
  \bibfield  {author} {\bibinfo {author} {\bibfnamefont {X.}~\bibnamefont
  {Zheng}},\ }\emph {\bibinfo {title} {{Efficient Fourier Transforms on
  Hexagonal Arrays}}},\ \href@noop {} {\bibinfo {type} {{PhD} thesis}},\
  \bibinfo  {school} {University of Florida} (\bibinfo {year}
  {2007})\BibitemShut {NoStop}%
\bibitem [{\citenamefont {Suzuura}\ and\ \citenamefont
  {Ando}(2002)}]{Suzuura2002}%
  \BibitemOpen
  \bibfield  {author} {\bibinfo {author} {\bibfnamefont {H.}~\bibnamefont
  {Suzuura}}\ and\ \bibinfo {author} {\bibfnamefont {T.}~\bibnamefont {Ando}},\
  }\href {\doibase 10.1103/PhysRevB.65.235412} {\bibfield  {journal} {\bibinfo
  {journal} {Phys. Rev. B}\ }\textbf {\bibinfo {volume} {65}},\ \bibinfo
  {pages} {235412} (\bibinfo {year} {2002})}\BibitemShut {NoStop}%
\bibitem [{Note1()}]{Note1}%
  \BibitemOpen
  \bibinfo {note} {Suzuura \& Ando \cite {Suzuura2002} get a reduction factor
  in front due to a different definition of $\vb *{u}$.}\BibitemShut {Stop}%
\bibitem [{Note2()}]{Note2}%
  \BibitemOpen
  \bibinfo {note} {Note that $\delta _{jx},\delta _{jy}$ are the components of
  $\vb *{\delta }_j$, not Kronecker delta symbols.}\BibitemShut {Stop}%
\bibitem [{\citenamefont {Vozmediano}\ \emph {et~al.}(2010)\citenamefont
  {Vozmediano}, \citenamefont {Katsnelson},\ and\ \citenamefont
  {Guinea}}]{Vozmediano2010}%
  \BibitemOpen
  \bibfield  {author} {\bibinfo {author} {\bibfnamefont {M.~A.~H.}\
  \bibnamefont {Vozmediano}}, \bibinfo {author} {\bibfnamefont {M.~I.}\
  \bibnamefont {Katsnelson}}, \ and\ \bibinfo {author} {\bibfnamefont
  {F.}~\bibnamefont {Guinea}},\ }\href {\doibase 10.1016/j.physrep.2010.07.003}
  {\bibfield  {journal} {\bibinfo  {journal} {Phys. Rep.}\ }\textbf {\bibinfo
  {volume} {496}},\ \bibinfo {pages} {109} (\bibinfo {year}
  {2010})}\BibitemShut {NoStop}%
\bibitem [{\citenamefont {Nazarov}\ and\ \citenamefont
  {Danon}(2013)}]{Nazarov2013}%
  \BibitemOpen
  \bibfield  {author} {\bibinfo {author} {\bibfnamefont {Y.~V.}\ \bibnamefont
  {Nazarov}}\ and\ \bibinfo {author} {\bibfnamefont {J.}~\bibnamefont
  {Danon}},\ }\href@noop {} {\emph {\bibinfo {title} {{Advanced Quantum
  Mechanics: A practical guide}}}},\ \bibinfo {edition} {1st}\ ed.\ (\bibinfo
  {publisher} {Cambridge University Press},\ \bibinfo {address} {New York},\
  \bibinfo {year} {2013})\BibitemShut {NoStop}%
\bibitem [{\citenamefont {Beenakker}(2006)}]{Beenakker2006}%
  \BibitemOpen
  \bibfield  {author} {\bibinfo {author} {\bibfnamefont {C.~W.~J.}\
  \bibnamefont {Beenakker}},\ }\href {\doibase 10.1103/PhysRevLett.97.067007}
  {\bibfield  {journal} {\bibinfo  {journal} {Phys. Rev. Lett.}\ }\textbf
  {\bibinfo {volume} {97}},\ \bibinfo {pages} {067007} (\bibinfo {year}
  {2006})}\BibitemShut {NoStop}%
\bibitem [{\citenamefont {Titov}\ and\ \citenamefont
  {Beenakker}(2006)}]{Titov2006}%
  \BibitemOpen
  \bibfield  {author} {\bibinfo {author} {\bibfnamefont {M.}~\bibnamefont
  {Titov}}\ and\ \bibinfo {author} {\bibfnamefont {C.~W.~J.}\ \bibnamefont
  {Beenakker}},\ }\href {\doibase 10.1103/PhysRevB.74.041401} {\bibfield
  {journal} {\bibinfo  {journal} {Phys. Rev. B}\ }\textbf {\bibinfo {volume}
  {74}},\ \bibinfo {pages} {041401(R)} (\bibinfo {year} {2006})}\BibitemShut
  {NoStop}%
\bibitem [{\citenamefont {Kopnin}\ and\ \citenamefont
  {Sonin}(2010)}]{Kopnin2010}%
  \BibitemOpen
  \bibfield  {author} {\bibinfo {author} {\bibfnamefont {N.~B.}\ \bibnamefont
  {Kopnin}}\ and\ \bibinfo {author} {\bibfnamefont {E.~B.}\ \bibnamefont
  {Sonin}},\ }\href {\doibase 10.1103/PhysRevB.82.014516} {\bibfield  {journal}
  {\bibinfo  {journal} {Phys. Rev. B}\ }\textbf {\bibinfo {volume} {82}},\
  \bibinfo {pages} {014516} (\bibinfo {year} {2010})}\BibitemShut {NoStop}%
\bibitem [{\citenamefont {Liang}\ \emph {et~al.}(2017)\citenamefont {Liang},
  \citenamefont {Vanhala}, \citenamefont {Peotta}, \citenamefont {Siro},
  \citenamefont {Harju},\ and\ \citenamefont {T{\"{o}}rm{\"{a}}}}]{Liang2017}%
  \BibitemOpen
  \bibfield  {author} {\bibinfo {author} {\bibfnamefont {L.}~\bibnamefont
  {Liang}}, \bibinfo {author} {\bibfnamefont {T.~I.}\ \bibnamefont {Vanhala}},
  \bibinfo {author} {\bibfnamefont {S.}~\bibnamefont {Peotta}}, \bibinfo
  {author} {\bibfnamefont {T.}~\bibnamefont {Siro}}, \bibinfo {author}
  {\bibfnamefont {A.}~\bibnamefont {Harju}}, \ and\ \bibinfo {author}
  {\bibfnamefont {P.}~\bibnamefont {T{\"{o}}rm{\"{a}}}},\ }\href {\doibase
  10.1103/PhysRevB.95.024515} {\bibfield  {journal} {\bibinfo  {journal} {Phys.
  Rev. B}\ }\textbf {\bibinfo {volume} {95}},\ \bibinfo {pages} {024515}
  (\bibinfo {year} {2017})}\BibitemShut {NoStop}%
\end{thebibliography}%
\end{document}


\title{\titlemain: supplementary material}
\newcommand{\titlemain}{Flat-band superconductivity in periodically strained graphene: mean-field and Berezinskii--Kosterlitz--Thouless transition}
\author{Teemu J. Peltonen}
\author{Tero T. Heikkilä}
\affiliation{University of Jyväskylä, Department of Physics and Nanoscience Center,
P.O. Box 35 (YFL), FI-40014 University of Jyväskylä, Finland}
\maketitle
\onecolumngrid


\section{Notation}
\begin{itemize}
	\item $\alpha\in\{A,B\}$ denotes sublattice, $\sigma\in\{\uparrow,\downarrow\}$ spin, and $\rho\in\{+,-\}$ valley (\ie~ $+\vect{K}$ or $-\vect{K}=\vect{K}'$)
    \item The bar operator exchanges sublattices, spins, valleys, and vectors: $\bar{A}=B$, $\bar{B}=A$, $\bar{\uparrow}=\downarrow$, $\bar{\downarrow}=\uparrow$, $\bar{\rho}=-\rho$, $\bar{\vect{k}}=-\vect{k}$
    \item $s$ is a sign function for $\sigma$: $s(\uparrow)=+1$, $s(\downarrow)=-1$
    \item $\sigma_x$, $\sigma_y$, $\sigma_z$ are the Pauli matrices in sublattice space
    \item $\vect{\sigma}^\rho \coloneqq (\rho\sigma_x,\sigma_y)$ is a vector of Pauli matrices
    \item $\tau_x$, $\tau_y$, $\tau_z$ are the Pauli matrices in Nambu space
    \item $\vect{v}_1\cross\vect{v}_2$ is the $2$-dimensional ``cross product'' of $\vect{v}_1,\vect{v}_2\in\R^2$ ($=$ third component of $(\vect{v}_1,0)\cross(\vect{v}_2,0) =$ signed area of the parallelogram defined by $\vect{v}_1$ and $\vect{v}_2$). 
    \item $z^*$ is the complex conjugate of $z\in\C$ and $L^*$ is the reciprocal lattice of a lattice $L$
    \item $\abs{z}$ is the absolute value of $z\in\C$ and $\abs{K}$ is the ``measure'' of a set $K$, \ie
    	\begin{equation*}
        	\abs{K} =
    		\begin{cases}
            	\# K, &\text{if $K$ is discrete} \\
            	\text{length of $K$}, &\text{if $K$ is continuous and $1$ dimensional} \\
                \text{area of $K$}, &\text{if $K$ is continuous and $2$ dimensional}
    		\end{cases}
    	\end{equation*}
    \item $G/H \coloneqq \{[g]: g\in G\}$ denotes the quotient group of the group $G$ modulo a normal subgroup $H$, consisting of the equivalence classes $[g] \coloneqq g+H = \{g+h: h\in H\}$ of the representative $g\in G$, assuming the group operation of addition. Especially if $G$ is a lattice and $H$ is its sublattice/superlattice, $G/H$ identifies all the lattice points whose difference is in $H$, and thus $G/H$ is isomorphic to the unit cell of $G$ modulo $H$. Thus by dropping the brackets from $[g]$, under an isomorphism $g$ might mean either the equivalence class or the representative in the unit cell, and from the context it should be clear which interpretations are allowed.
    \item $\vect{a}_1$ and $\vect{a}_2$ are the primitive vectors of the graphene lattice $L \coloneqq \spn_\Z\{\vect{a}_1,\vect{a}_2\}$, $a\coloneqq\norm{\vect{a}_1}=\norm{\vect{a}_2}$ is the lattice constant, the nearest neighbor vectors are $\vect{\delta}_1$, $\vect{\delta}_2$, and $\vect{\delta}_3$, the carbon--carbon bond length is $a_0\coloneqq\norm{\vect{\delta}_j}=a/\sqrt{3}$, and $\vect{\delta}_A\coloneqq 0$, $\vect{\delta}_B\coloneqq\vect{\delta}_1$ is the sublattice-translation vector
    \item $\LBK$ denotes the large \BvK lattice, in translations of which the creation/annihilation operators are taken to be periodic
    \item $V\coloneqq\abs{\R^2/\LBK}$ is the area of the continuum \BvK unit cell
    \item $SL\coloneqq\spn_\Z\{\vect{t}_1,\vect{t}_2\}$ denotes the superlattice created by periodic strain, $SL^*=\spn_\Z\{\vect{G}_1,\vect{G}_2\}$ is its reciprocal lattice, $SL_1\coloneqq\spn_\Z\{\vect{t}_1\}$ is a 1-dimensional sublattice of $SL$, $SL_1^*=\spn_\Z\{\vect{G}_1\}$ is a 1-dimensional sublattice of $SL^*$
    \item $S\in\{\text{RZ},\text{MZ}\}$ denotes either the reduced zone scheme or the mixed zone scheme, explained in section~\ref{sec:fourier_series}, and $SL_\text{RZ}\coloneqq SL$, $SL_\text{MZ}\coloneqq SL_1$ are shorthand notations for writing the Fourier series in different schemes
    \item $\Op$ is the space of electron creation/annihilation operators (for the sake of notation)
    \item $\psi_\sigma: [L\cup(L+\vect{\delta}_1)]/\LBK\rightarrow\Op$ is an electron annihilation operator defined in the graphene crystal (in the union of $A$ sublattice $L$ and $B$ sublattice $L+\vect{\delta}_1$). In other words, if $\vect{r}\in L/\LBK$, then $\psi_\sigma(\vect{r})$ annihilates a $\sigma$-spin electron at the ($A$-sub)lattice point $\vect{r}$ and $\psi_\sigma(\vect{r}+\vect{\delta}_1)$ annihilates a $\sigma$-spin electron at the $B$-sublattice point $\vect{r}+\vect{\delta}_1$.
    \item $\psi_{\sigma,A}, \psi_{\sigma,B}: L/\LBK\rightarrow\Op$, $\psi_{\sigma,A}(\vect{r}) \coloneqq \psi_\sigma(\vect{r})$, $\psi_{\sigma,B}(\vect{r}) \coloneqq \psi_\sigma(\vect{r}+\vect{\delta}_1)$ are the sublattice-shifted annihilation operators and $\psi_{\sigma\rho,\alpha}$ is the corresponding valley-specific operator defined in \eqref{eq:phi_sigmarhoalpha_fourier_series}. The continuum limit is taken by replacing $L\rightarrow\R^2$ and $\psi/\sqrt{\abs{\vect{a}_1\cross\vect{a}_2}} \rightarrow \psi$.
    \item $\psi_{\sigma\rho} \coloneqq (\psi_{\sigma\rho,A}, \psi_{\sigma\rho,B})^\transpose$ is a corresponding sublattice spinor
\end{itemize}


\section{Fourier series of a lattice-periodic function}
Using the notation of quotient groups, (discrete) Fourier series can be written elegantly \cite{Zheng2007}. We will be using the term ``series'' for functions that are periodic, and the term ``discrete'' for functions defined on a lattice.

\subsection{Discrete Fourier series}
Let $f:L/SL\rightarrow\C^n$ be a function defined on a 2-dimensional lattice $L\subset\R^2$ and periodic in translations of the 2-dimensional superlattice $SL\subset L$. It can be shown that the 2-dimensional discrete Fourier series and its coefficients can be written respectively as \cite{Zheng2007}
\begin{equation}
    f(\vect{r}) = \sum_{\vect{G}\in SL^*/L^*} \e^{\imag\vect{G}\vdot\vect{r}} \tilde{f}(\vect{G}), \quad \tilde{f}(\vect{G}) = \frac{1}{\abs{L/SL}} \sum_{\vect{r}\in L/SL} \e^{-\imag\vect{G}\vdot\vect{r}} f(\vect{r}),
\end{equation}
where $L/SL$ can be interpreted as any of the discrete superlattice unit cells and $SL^*/L^*$ as any of the unit cells of the reciprocal superlattice (\eg the superlattice Brillouin zone SBZ).

\subsection{Fourier series}
\label{sec:fourier_series}
Let $f:\R^2/SL\rightarrow\C^n$ be a function defined on the continuum $\R^2$ and periodic in translations of the 2-dimensional lattice $SL = \spn_\Z\{\vect{t}_1,\vect{t}_2\} \subset \R^2$ with the reciprocal lattice $SL^* = \spn_\Z\{\vect{G}_1,\vect{G}_2\}$. Assuming the 1-dimensional Fourier series is known, it can be shown that the 2-dimensional Fourier series and its coefficients can be written respectively as
\begin{equation}
    f(\vect{r}) = \sum_{\vect{G}\in SL^*} \e^{\imag\vect{G}\vdot\vect{r}} \tilde{f}(\vect{G}), \quad \tilde{f}(\vect{G}) = \frac{1}{\abs{\R^2/SL}} \int_{\R^2/SL} \dd{\vect{r}} \e^{-\imag\vect{G}\vdot\vect{r}} f(\vect{r}),
\end{equation}
where the integral is calculated as a 2-dimensional volume integral which, by interpreting $\R^2/SL$ as the parallelogram defined by $\vect{t}_1$ and $\vect{t}_2$ (one of the superlattice continuum unit cells) and by change of variables, can be written as
\begin{equation}
    \tilde{f}(\vect{G}) = \tilde{f}(m_1\vect{G}_1+m_2\vect{G}_2) = \int_{-\frac{1}{2}}^\frac{1}{2} \dd{x_1} \int_{-\frac{1}{2}}^\frac{1}{2} \dd{x_2} \e^{-\imag 2\pi(m_1 x_1+m_2 x_2)} f(x_1\vect{t}_1+x_2\vect{t}_2).
    \label{eq:fourier_coefficients_changeofvariables}
\end{equation}
Writing the Fourier series this way we call the \emph{reduced zone scheme}, for reasons explained in section~\ref{sec:BdG_equation_in_fourier_space}.

As a special case, if $f$ is constant in the $\vect{t}_2$ direction, we may use the result
\begin{equation}
    \tilde{f}(\vect{G}) = \tilde{f}(m_1\vect{G}_1+m_2\vect{G}_2) = \delta_{m_2,0} \tilde{f}(m_1\vect{G}_1) = \delta_{m_2,0} \tilde{f}(\vect{G})
\end{equation}
yielding the series
\begin{equation}
    f(\vect{r}) = \sum_{\vect{G}\in SL_1^*} \e^{\imag\vect{G}\vdot\vect{r}} \tilde{f}(\vect{G}).
\end{equation}
If we calculate the Fourier series this way (possible only if $f$ is constant in the $\vect{t}_2$ direction), as a sum over the one-dimensional lattice $SL_1^*$, we call this the \emph{mixed zone scheme}, for reasons explained in section~\ref{sec:BdG_equation_in_fourier_space}.

Summarizing both schemes together, we may write the Fourier series and its coefficients as
\begin{equation}
    f(\vect{r}) = \sum_{\vect{G}\in\SLS^*} \e^{\imag\vect{G}\vdot\vect{r}} \tilde{f}(\vect{G}), \quad \tilde{f}(\vect{G}) = \frac{1}{\abs{\R^2/SL}} \int_{\R^2/SL} \dr \e^{-\imag\vect{G}\vdot\vect{r}} f(\vect{r}),
    \label{eq:fourier_series+coefficients}
\end{equation}
with the mixed zone scheme $S=\text{MZ}$ being applicable only in the case of $f$ being constant in the $\vect{t}_2$ direction.

\subsection{Fourier components of selected functions}
If we calculate the Fourier components of the pseudo vector potentials chosen in the main paper by \eqref{eq:fourier_series+coefficients} [or more explicitly, \eqref{eq:fourier_coefficients_changeofvariables}], we get
\begin{align}
    \tilde{\vect{A}}_{\cos}^\text{1D}(m_1\vect{G}_1+m_2\vect{G}_2) &= \frac{\beta}{2d}(0, \delta_{m_1,-1}+\delta_{m_1,1})\delta_{m_2,0}, \label{eq:fourier_coefficients_cos1D_reducedzone} \\
    \tilde{\vect{A}}_{\cos}^\text{2D}(m_1\vect{G}_1+m_2\vect{G}_2) &= \frac{\beta}{2d}(\delta_{m_2,-1}+\delta_{m_2,1}, \delta_{m_1,-1}+\delta_{m_1,1}), \label{eq:fourier_coefficients_cos2D_reducedzone} \\
    \tilde{\vect{A}}_{1/c}^\text{1D}(m_1\vect{G}_1+m_2\vect{G}_2) &=
    \begin{cases}
        (0,0), &\text{if $m_1=0$}, \\
        \left( 0, \e^{\imag\pi m_1/2} \frac{\e^{-\imag2\pi m_1(1+c)}}{c(2\pi m_1)^2} (-1+\e^{\imag\pi m_1}) (-1+\e^{\imag 2\pi m_1c}) (-\e^{\imag\pi m_1}+\e^{\imag 2\pi m_1c }) \right) \delta_{m_2,0}, &\text{otherwise}
    \end{cases}
    \label{eq:fourier_coefficients_triangleSquare1D_reducedzone}
\end{align}
in the reduced zone scheme, where $m_1\vect{G}_1+m_2\vect{G}_2\in SL_\text{RZ}^*$. As discussed in the previous section, because of the $\delta_{m_2,0}$ factor in the 1D potentials we may as well use a one-dimensional Fourier series and write
\begin{equation}
    \tilde{\vect{A}}_{\cos}^\text{1D}(m_1\vect{G}_1) = \frac{\beta}{2d}(0, \delta_{m_1,-1}+\delta_{m_1,1})
    \label{eq:fourier_coefficients_cos1D_mixedzone}
\end{equation}
in the mixed zone scheme (and similarly for $\tilde{\vect{A}}_{1/c}^\text{1D}$), where $m_1\vect{G}_1\in SL_\text{MZ}^*$.


\section{Strained graphene low energy effective BCS Hamiltonian}
Let us first fix the lattice vectors. Note that these are needed only when deriving the continuum theory, but after moving to the continuum theory the lattice is not anymore present, except of its orientation. We take the graphene lattice primitive vectors
\begin{equation}
    \vect{a}_1 = \frac{a}{2}(1,\sqrt{3}), \quad \vect{a}_2 = \frac{a}{2}(-1,\sqrt{3}),
\end{equation}
and the nearest neighbor vectors
\begin{equation}
    \vect{\delta}_1 = \frac{1}{3}(\vect{a}_1+\vect{a}_2), \quad \vect{\delta}_2 = \frac{1}{3} (\vect{a}_2-2\vect{a}_1), \quad \vect{\delta}_3 = \frac{1}{3} (\vect{a}_1-2\vect{a}_2).
\end{equation}
With these definitions the zigzag direction is in the $x$ direction and the $\vect{K}$ point is located at
\begin{equation}
    \vect{K} = \frac{4\pi}{3a^2}(\vect{a}_1-\vect{a}_2) = \frac{4\pi}{3a}(1,0).
\end{equation}
 
In the nearest neighbour tight binding model the interacting Hamiltonian of strained graphene is
\begin{equation}
	H_\text{BdG} = H_\text{p} + \delta H_\text{s} + H_\text{int} \eqqcolon H + H_\text{int},
\end{equation}
where the noninteracting pristine graphene part is
\begin{align}
	H_\text{p} &= -t\sum_{\sigma\in\{\uparrow,\downarrow\}} \sum_{j=1}^3 \sum_{\vect{r}\in L/\LBK} \psi_\sigma^\dagger(\vect{r}) \psi_\sigma(\vect{r}+\vect{\delta}_j) +\hc -\mu\sum_{\substack{\alpha\in\{A,B\},\\ \sigma\in\{\uparrow,\downarrow\}}} \sum_{\vect{r}\in L/\LBK} \psi_\sigma^\dagger(\vect{r}+\vect{\delta}_\alpha) \psi_\sigma(\vect{r}+\vect{\delta}_\alpha) \\
	&= -t\sum_{\sigma\in\{\uparrow,\downarrow\}} \sum_{j=1}^3 \sum_{\vect{r}\in L/\LBK} \psi_{\sigma,1A}^\dagger(\vect{r}) \psi_{\sigma,1B}(\vect{r}+\vect{\delta}_j-\vect{\delta}_1) +\hc -\mu\sum_{\substack{\alpha\in\{A,B\},\\ \sigma\in\{\uparrow,\downarrow\}}} \sum_{\vect{r}\in L/\LBK} \psi_{\sigma,1\alpha}^\dagger(\vect{r}) \psi_{\sigma,1\alpha}(\vect{r}),
\end{align}
the small change to this due to strain is
\begin{align}
    \delta H_\text{s} &= -\sum_{\sigma\in\{\uparrow,\downarrow\}} \sum_{j=1}^3 \sum_{\vect{r}\in L/\LBK} \delta t_j(\vect{r}) \psi_\sigma^\dagger(\vect{r}) \psi_\sigma(\vect{r}+\vect{\delta}_j) +\hc \\
    &= -\sum_{\sigma\in\{\uparrow,\downarrow\}} \sum_{j=1}^3 \sum_{\vect{r}\in L/\LBK} \delta t_j(\vect{r}) \psi_{\sigma,1A}^\dagger(\vect{r}) \psi_{\sigma,1B}(\vect{r}+\vect{\delta}_j-\vect{\delta}_1) +\hc,
\end{align}
and the interacting part is
\begin{align}
	H_\text{int} &= \frac{\lambda}{2} \sum_{\substack{\alpha\in\{A,B\},\\\sigma\in\{\uparrow,\downarrow\}}} \sum_{\vect{r}\in L/\LBK} \psi_\sigma^\dagger(\vect{r}+\vect{\delta}_\alpha) \psi_{\bar{\sigma}}^\dagger(\vect{r}+\vect{\delta}_\alpha) \psi_{\bar{\sigma}}(\vect{r}+\vect{\delta}_\alpha) \psi_\sigma(\vect{r}+\vect{\delta}_\alpha) \\
	&= \frac{\lambda}{2} \sum_{\substack{\alpha\in\{A,B\},\\\sigma\in\{\uparrow,\downarrow\}}} \sum_{\vect{r}\in L/\LBK} \psi_{\sigma,\alpha}^\dagger(\vect{r}) \psi_{\bar{\sigma},\alpha}^\dagger(\vect{r}) \psi_{\bar{\sigma},\alpha}(\vect{r}) \psi_{\sigma,\alpha}(\vect{r})
\end{align}
assuming local (also in sublattice) interaction of strength $\lambda$ (negative for attractive interaction considered here) that is independent of sublattice, spin, and position. Here $t$ is the graphene nearest neighbour hopping energy, $\delta t_j(\vect{r})$ is a small change to this due to strain in the bond from $\vect{r}$ to $\vect{r}+\vect{\delta}_j$, and $\mu$ is the chemical potential.

Because of the periodicity of $\psi_{\sigma,\alpha}$ we may expand it as a discrete Fourier series
\begin{equation}
	\psi_{\sigma,\alpha}(\vect{r}) = \sum_{\vect{k}\in\LBK^*/L^*} \e^{\imag\vect{k}\vdot\vect{r}} c_{\sigma,\alpha}(\vect{k}).
\end{equation}
Dividing the sum in parts near and far from the Dirac points yields
\begin{equation}
 	\psi_{\sigma,\alpha}(\vect{r}) = \sum_{\rho\in\{+,-\}} \sum_{\substack{\vect{k}\in\LBK^*/L^*\\ (\vect{k}\text{ near }\rho\vect{K})}} \e^{\imag\vect{k}\vdot\vect{r}} c_{\sigma,\alpha}(\vect{k}) + \sum_{\substack{\vect{k}\in\LBK^*/L^*\\ (\vect{k}\text{ not near }\vect{K},\vect{K}')}} \e^{\imag\vect{k}\vdot\vect{r}} c_{\sigma,\alpha}(\vect{k}),
    \label{eq:phi_sigmaalpha_fourier_series_division}
\end{equation}
where we can drop the last term by going into effective low-energy theory where terms far from the Dirac points are uninteresting. By defining the fermionic valley-specific annihilation operators in Fourier space,
\begin{equation}
	c_{\sigma\rho,\alpha}(\vect{k}) \coloneqq \begin{cases}
    	c_{\sigma,\alpha}(\vect{k}+\rho\vect{K}), \quad &\text{if $\vect{k}$ small}, \\
        0, \quad &\text{otherwise},
	\end{cases}
\end{equation}
and its corresponding discrete Fourier series
\begin{equation}
	\psi_{\sigma\rho,\alpha}(\vect{r}) = \sum_{\substack{\vect{k}\in\LBK^*/L^*\\ (\vect{k}\text{ small})}} \e^{\imag\vect{k}\vdot\vect{r}} c_{\sigma\rho,\alpha}(\vect{k}), \quad
	c_{\sigma\rho,\alpha}(\vect{k}) = \frac{1}{\abs{L/\LBK}} \sum_{\vect{r}\in L/\LBK} \e^{-\imag\vect{k}\vdot\vect{r}} \psi_{\sigma\rho,\alpha}(\vect{r}),
    \label{eq:phi_sigmarhoalpha_fourier_series}
\end{equation}
equation \eqref{eq:phi_sigmaalpha_fourier_series_division} can be written as
\begin{equation}
	\psi_{\sigma,\alpha}(\vect{r}) = \sum_\rho \sum_{\substack{\vect{k}\in\LBK^*/L^*\\ (\vect{k}\text{ small})}} \e^{\imag(\vect{k}+\rho\vect{K})\vdot\vect{r}} c_{\sigma\rho,\alpha}(\vect{k}) = \sum_\rho \e^{\imag\rho\vect{K}\vdot\vect{r}} \psi_{\sigma\rho,\alpha}(\vect{r})
    \label{eq:phi_sigmaalpha_valley_expansion}
\end{equation}
which is the expansion to use when we want to go to the low-energy theory and express the original operators in the valley-operator basis.

\subsection{Strained graphene Hamiltonian}
The derivation of the strained Hamiltonian has been already done in \cite{Suzuura2002} in the case of carbon nanotubes, but for transparency we repeat the calculation here. Writing the annihilation/creation operators as the valley expansion \eqref{eq:phi_sigmaalpha_valley_expansion} and linearizing $\psi_{\sigma\rho,B}(\vect{r}+\vect{\delta}_j) \approx \psi_{\sigma\rho,B}(\vect{r}) + (\vect{\delta}_j-\vect{\delta}_1)\vdot\nabla\psi_{\sigma\rho,B}(\vect{r})$ the pristine graphene Hamiltonian becomes
\begin{align}
	H_\text{p} = \hbar\vF &\sum_{\sigma\rho\rho'} \sum_{\vect{r}\in L/\LBK} \e^{\imag(\rho'-\rho)\vect{K}\vdot\vect{r}} \psi_{\sigma\rho,A}^\dagger(\vect{r})(-\imag)(\rho'\partial_x-\imag\partial_y) \psi_{\sigma\rho',B}(\vect{r}) +\hc \notag\\
    -\mu &\sum_{\sigma\rho\rho'\alpha} \sum_{\vect{r}\in L/\LBK} \e^{\imag(\rho'-\rho)\vect{K}\vdot\vect{r}} \psi_{\sigma\rho,\alpha}^\dagger(\vect{r}) \psi_{\sigma\rho',\alpha}(\vect{r}),
\end{align}
where we used $\sum_{j=1}^3 \e^{\imag\rho\vect{K}\vdot\vect{\delta}_j}=0$ and defined the Fermi velocity by $\hbar\vF \coloneqq \sqrt{3}at/2$. The exponential factor gives simply $\delta_{\rho\rho'}$. This can be seen by going into Fourier space by using \eqref{eq:phi_sigmarhoalpha_fourier_series}, after which the overall exponential gives $\abs{L/\LBK}\delta_{\vect{k}+\rho\vect{K},\vect{k}'+\rho'\vect{K}}$ after calculating the $\vect{r}$ sum. By using the property that $\vect{k}$ and $\vect{k}'$ are small, this is equal to $\abs{L/\LBK}\delta_{\vect{k}\vect{k}'}\delta_{\rho\rho'}$. Then after coming back to real space the Hamiltonian reads
\begin{equation}
	H_\text{p} = \hbar\vF \sum_{\sigma\rho} \sum_{\vect{r}\in L/\LBK} \psi_{\sigma\rho,A}^\dagger(\vect{r}) (-\imag)(\rho\partial_x-\imag\partial_y) \psi_{\sigma\rho,B}(\vect{r}) +\hc - \mu \sum_{\sigma\rho\alpha} \sum_{\vect{r}\in L/\LBK} \psi_{\sigma\rho,\alpha}^\dagger(\vect{r}) \psi_{\sigma\rho,\alpha}(\vect{r}).
\end{equation}

For the strain Hamiltonian we similarly write the sublattice-shifted annihilation operators as the valley expansion \eqref{eq:phi_sigmaalpha_valley_expansion} and make the zeroth order approximation $\psi_{\sigma\rho,B}(\vect{r}+\vect{\delta}_j) \approx \psi_{\sigma\rho,B}(\vect{r})$. This yields
\begin{equation}
    \delta H_\text{s} = \sum_{\sigma\rho} \sum_{\vect{r}\in L/\LBK} \psi_{\sigma\rho,A}^\dagger(\vect{r}) \rho\hbar\vF A^\rho(\vect{r})^* \psi_{\sigma\rho,B}(\vect{r}) +\hc,
\end{equation}
where
\begin{equation}
    A^\rho(\vect{r}) \coloneqq \rho A_x(\vect{r})+\imag A_y(\vect{r}) \coloneqq -\frac{\rho}{\hbar\vF} \sum_j \e^{-\imag\rho\vect{K}\vdot\vect{\delta}_j} \delta t_j(\vect{r}).
    \label{eq:Arhor}
\end{equation}
The strained graphene Hamiltonian then becomes
\begin{equation}
	H = H_\text{p} + \delta H_\text{s} = \sum_{\sigma\rho} \sum_{\vect{r}\in L/\LBK} \psi_{\sigma\rho}^\dagger(\vect{r}) \mathcal{H}^\rho(\vect{r}) \psi_{\sigma\rho}(\vect{r}),
\end{equation}
where we defined the Hamiltonian matrix element
\begin{equation}
	\mathcal{H}^\rho(\vect{r}) \coloneqq \hbar\vF\vect{\sigma}^\rho\vdot(-\imag\nabla+\rho\vect{A}(\vect{r})) - \mu
    \label{eq:H0r}
\end{equation}
and the pseudo vector potential $\vect{A}\coloneqq(A_x,A_y)$.

We now know the connection \eqref{eq:Arhor} between the pseudo vector potential $\vect{A}$ and the small change $\delta t_j$ in the hopping energy, but we still need to find the connection between $\delta t_j$ and strain. Assuming the atom at $\vect{r}$ to be displaced by a vector $\vect{v}(\vect{r}) = (\vect{u}(\vect{r}),h(\vect{r}))$, where $\vect{u}=(u_x,u_y)$ is the in-plane and $h$ is the out-of-plane displacement field, the change in the bond length of the $\vect{\delta}_j$ bond due to strain is
\begin{align}
    \delta u_j(\vect{r}) &\coloneqq \norm{\vect{r}+\vect{\delta}_j+\vect{v}(\vect{r}+\vect{\delta}_j) - [\vect{r}+\vect{v}(\vect{r})]} - \norm{\vect{r}+\vect{\delta}_j-\vect{r}} \\
    &\approx \frac{1}{\norm{\vect{\delta}_j}} \left[ \vect{\delta}_j\vdot(\vect{u}(\vect{r}+\vect{\delta}_j)-\vect{u}(\vect{r})) + \frac{1}{2}(h(\vect{r}+\vect{\delta}_j)-h(\vect{r}))^2 \right],
\end{align}
where in the last step we linearized in $\norm{\vect{u}(\vect{r}+\vect{\delta}_j)-\vect{u}(\vect{r})}, \abs{h(\vect{r}+\vect{\delta}_j)-h(\vect{r})} \ll \norm{\vect{\delta}_j}$. Furthermore in the linear order we may approximate \footnote{Suzuura \& Ando \cite{Suzuura2002} get a reduction factor in front due to a different definition of $\vect{u}$.}
\begin{equation}
    \vect{u}(\vect{r}+\vect{\delta}_j)-\vect{u}(\vect{r}) \approx (\vect{\delta}_j\vdot\nabla)\vect{u}(\vect{r}), \quad h(\vect{r}+\vect{\delta}_j)-h(\vect{r}) \approx \vect{\delta}_j\vdot\nabla h(\vect{r}),
\end{equation}
and if we define the strain tensor as
\begin{equation}
    u_{ij} \coloneqq \frac{1}{2}(\partial_i u_j+\partial_j u_i) + \frac{1}{2}\partial_i h\partial_j h,
\end{equation}
the change in the bond length becomes \footnote{Note that $\delta_{jx},\delta_{jy}$ are the components of $\vect{\delta}_j$, not Kronecker delta symbols.}
\begin{equation}
    \delta u_j(\vect{r}) = \frac{1}{\norm{\vect{\delta}_j}} \left[ \delta_{jx}^2 u_{xx}(\vect{r}) + 2\delta_{jx}\delta_{jy}u_{xy}(\vect{r}) + \delta_{jy}^2 u_{yy}(\vect{r}) \right].
    \label{eq:deltaujr}
\end{equation}

Now that the change in the hopping energy can be linearized to
\begin{equation}
    \delta t_j(\vect{r}) \approx \dv{t}{a_0} \delta u_j(\vect{r}) = -\frac{t\gruneisen}{a_0} \delta u_j(\vect{r}),
    \label{eq:deltatjr}
\end{equation}
where $\beta_\text{G} \coloneqq -\dv*{\ln{t}}{\ln{a_0}}\approx 2$ is the graphene Gr\"uneisen parameter \cite{Vozmediano2010}, equations \eqref{eq:Arhor}, \eqref{eq:deltaujr}, and \eqref{eq:deltatjr} yield for the pseudo vector potential
\begin{equation}
    \vect{A} = -\frac{\gruneisen}{2a_0} (u_{xx}-u_{yy},-2u_{xy}).
\end{equation}

Finally we can extend the annihilation operators to the continuum $\R^2/\LBK$ by the discrete Fourier series \eqref{eq:phi_sigmarhoalpha_fourier_series}, and everything else is trivially extended. By furthermore redefining the continuum annihilation operator density as $\psi_{\sigma\rho,\alpha} / \sqrt{\abs{\vect{a}_1\cross\vect{a}_2}} \rightarrow \psi_{\sigma\rho,\alpha}$ we arrive at the continuum Hamiltonian
\begin{equation}
	H = \sum_{\sigma\rho} \int_{\R^2/\LBK} \dr \psi_{\sigma\rho}^\dagger(\vect{r}) \mathcal{H}^\rho(\vect{r}) \psi_{\sigma\rho}(\vect{r}).
	\label{eq:H0}
\end{equation}
Note that while a normal vector potential would break time-reversal symmetry, this Hamiltonian is time-reversal symmetric, $\mathcal{H}^{\bar{\rho}*} = \mathcal{H}^\rho$, because of the valley-odd pseudo vector potential.

\subsubsection{Limits in the theory}
The only assumptions regarding strain in deriving the Hamiltonian \eqref{eq:H0} were $\norm{\vect{u}(\vect{r}+\vect{\delta}_j)-\vect{u}(\vect{r})}, \abs{h(\vect{r}+\vect{\delta}_j)-h(\vect{r})} \ll \norm{\vect{\delta}_j}$. For the cosine displacement fields $\vect{u}_{\cos}^\text{1D}$ and $\vect{u}_{\cos}^\text{2D}$ of the main paper this assumption reads
\begin{align}
    &\vect{u}_{\cos}^\text{1D}: \qquad \frac{\beta}{\gruneisen} \ll \frac{d}{a_0} \qqtext{and} \frac{d}{a_0} \gg 1 \\
    &\vect{u}_{\cos}^\text{2D}: \qquad \frac{\beta}{\gruneisen} \ll \frac{d}{2a_0} \qqtext{and} \frac{d}{a_0} \gg 1.
\end{align}
Note that this is equivalent to assuming that the strain is $\abs{u_{ij}} \ll 1$.
 
\subsection{Interaction Hamiltonian}
The derivation regarding superconductivity is based on the book by Nazarov \& Danon \cite{Nazarov2013}. First making the mean-field approximation in the Cooper channel for $H_\text{int}$ yields
\begin{equation}
	H_\text{int} \approx \frac{1}{2} \sum_{\sigma\alpha} \sum_{\vect{r}\in L/\LBK} \Delta_{\sigma,\alpha}(\vect{r}) \psi_{\sigma,\alpha}^\dagger(\vect{r}) \psi_{\bar{\sigma},\alpha}^\dagger(\vect{r}) +\hc +E^0
\end{equation}
where the superconducting order parameter is
\begin{equation}
    \Delta_{\sigma,\alpha} \coloneqq \lambda \expval{ \psi_{\bar\sigma,\alpha} \psi_{\sigma,\alpha}}
\end{equation}
with the angle brackets denoting the thermal average and the constant term is
\begin{equation}
    E^0 \coloneqq -\frac{1}{2\lambda} \sum_{\sigma\alpha} \sum_{\vect{r}\in L/\LBK} \abs{ \Delta_{\sigma,\alpha}(\vect{r}) }^2.
\end{equation}
Using the valley expansion \eqref{eq:phi_sigmaalpha_valley_expansion} and \emph{assuming only intervalley interaction} gives
\begin{align}
	H_\text{int} &= \frac{1}{2} \sum_{\sigma\rho\alpha} \sum_{\vect{r}\in L/\LBK} \Delta_{\sigma,\alpha}(\vect{r}) \psi_{\sigma\rho,\alpha}^\dagger(\vect{r}) \psi_{\bar\sigma\bar\rho,\alpha}^\dagger(\vect{r}) +\hc +E^0 \\
	&= \frac{1}{2} \sum_{\sigma\rho} \sum_{\vect{r}\in L/\LBK} \psi_{\sigma\rho}^\dagger(\vect{r}) \Delta_\sigma(\vect{r}) \psi_{\bar\sigma\bar\rho}^{\dagger\transpose}(\vect{r}) +\hc +E^0,
\end{align}
\begin{equation}
    \Delta_{\sigma,\alpha} \coloneqq \lambda \sum_\rho \expval{ \psi_{\bar\sigma\bar\rho,\alpha} \psi_{\sigma\rho,\alpha}}, \quad \Delta_\sigma \coloneqq \diag(\Delta_{\sigma,A}, \Delta_{\sigma,B})
	\label{eq:Delta_psi}
\end{equation}
and
\begin{equation}
	E^0 = -\frac{1}{2\lambda} \sum_{\sigma} \sum_{\vect{r}\in L/\LBK} \Tr(\Delta_\sigma^*(\vect{r})\Delta_\sigma(\vect{r})),
\end{equation}
where the trace is over the sublattice structure.

Further taking the continuum limit $L\rightarrow\R^2$ gives
\begin{equation}
	H_\text{int} = \frac{1}{2} \sum_{\sigma\rho} \int_{\R^2/\LBK}\dr \psi_{\sigma\rho}^\dagger(\vect{r}) \Delta_\sigma(\vect{r}) \psi_{\bar{\sigma}\bar{\rho}}^{\dagger\transpose}(\vect{r}) +\hc +E^0,
\end{equation}
if we at the same time replace $\psi_{\sigma\rho} / \sqrt{\abs{\vect{a}_1\cross\vect{a}_2}} \rightarrow \psi_{\sigma\rho}$ and $\lambda / \abs{\vect{a}_1\cross\vect{a}_2} \rightarrow \lambda$. The constant term is then
\begin{equation}
	E^0 = -\frac{1}{2\lambda} \sum_\sigma \int_{\R^2/\LBK}\dr \Tr(\Delta_\sigma^*(\vect{r})\Delta_\sigma(\vect{r})).
\end{equation}


\section{Diagonalizing the Hamiltonian}

\subsection{Writing the Hamiltonian in Nambu basis}
Utilizing the anticommutation relations and the identity
\begin{equation}
    \int\dd{\vect{r}} \psi_{\sigma\rho}^\dagger(\vect{r}) \mathcal{H}^\rho(\vect{r}) \psi_{\sigma\rho}(\vect{r}) = -\int\dd{\vect{r}} \psi_{\sigma\rho}^\transpose(\vect{r}) \mathcal{H}^{\bar{\rho}}(\vect{r}) \psi_{\sigma\rho}^{\dagger\transpose}(\vect{r}) -2\mu V\delta(0),
\end{equation}
where the $-2\mu V\delta(0)$ term comes from anticommuting the annihilation/creation operators in the particle number operator, we may bring the total Hamiltonian into the form
\begin{align}
	H_\text{BdG} &= \frac{1}{2} \sum_{\sigma\rho} \int\dd{\vect{r}}
    \begin{pmatrix}
    	\psi_{\sigma\rho}^\dagger(\vect{r}) & \psi_{\bar{\sigma}\bar{\rho}}^\transpose(\vect{r})
    \end{pmatrix}
    \begin{pmatrix}
    	\mathcal{H}^\rho(\vect{r}) & \Delta_\sigma(\vect{r}) \\
        \Delta_\sigma^*(\vect{r}) & -\mathcal{H}^\rho(\vect{r})
    \end{pmatrix}
    \begin{pmatrix}
    	\psi_{\sigma\rho}(\vect{r}) \\
        \psi_{\bar{\sigma}\bar{\rho}}^{\dagger\transpose}(\vect{r})
    \end{pmatrix} + E^0 - 4\mu V\delta(0) \\
    &= \frac{1}{2} \sum_{\sigma\rho} \int\dd{\vect{r}} \Psi_{\sigma\rho}^\dagger(\vect{r}) \mathcal{H}_\text{BdG}^\rho(\vect{r}) \Psi_{\sigma\rho}(\vect{r}) + E^0 - 4\mu V\delta(0),
    \label{eq:H_Psi}
\end{align}
where in the last step we defined the spin-independent Bogoliubov--de Gennes Hamiltonian in Nambu space, the spin-independent order parameter, and the Nambu-spinor operator respectively as
\begin{equation}
    \mathcal{H}_\text{BdG}^\rho \coloneqq
    \begin{pmatrix}
        \mathcal{H}^\rho & \Delta \\
        \Delta^* & -\mathcal{H}^\rho
    \end{pmatrix},
    \quad \Delta \coloneqq \Delta_\uparrow = -\Delta_\downarrow = s(\sigma)\Delta_\sigma, \quad \Psi_{\sigma\rho} \coloneqq
    \begin{pmatrix}
        \psi_{\sigma\rho} \\ s(\sigma)\psi_{\bar{\sigma}\bar{\rho}}^{\dagger\transpose}
    \end{pmatrix}.
    \label{eq:Delta}
\end{equation}

\subsection{Writing the Hamiltonian in eigenbasis: Bogoliubov--de Gennes equation}
Simply by using the definition of the Dirac delta we may write $H_\text{BdG}$ in \eqref{eq:H_Psi} as
\begin{equation}
	H_\text{BdG} = \frac{1}{2} \sum_{\sigma\rho} \int\dd{\vect{r}} \int\dd{\vect{r}'} \Psi_{\sigma\rho}^\dagger(\vect{r}) \delta(\vect{r}-\vect{r}') \mathcal{H}_\text{BdG}^\rho(\vect{r}) \Psi_{\sigma\rho}(\vect{r}') + E^0 - 4\mu V\delta(0).
    \label{eq:H_Psi_double_integral}
\end{equation}
Now let $\delta(\vect{r}-\vect{r}')\mathcal{H}_\text{BdG}^\rho(\vect{r})$ be the representation of $\hat{\mathcal{H}}_\text{BdG}^\rho$ in position space, that is,
\begin{equation}
	\delta(\vect{r}-\vect{r}') \mathcal{H}_\text{BdG}^\rho(\vect{r}) = \bra{\vect{r}} \hat{\mathcal{H}}_\text{BdG}^\rho \ket{\vect{r}'}.
    \label{eq:Hhat}
\end{equation}
Since $\mathcal{H}_\text{BdG}$ is Hermitian we may use the spectral theorem (following from the resolution of identity)
\begin{equation}
	\identity = \frac{1}{V} \sum_n \ketbra{w_{\rho n}}{w_{\rho n}} \quad\Rightarrow\quad \hat{\mathcal{H}}_\text{BdG}^\rho = \frac{1}{V} \sum_n E_{\rho n} \ketbra{w_{\rho n}}{w_{\rho n}}
    \label{eq:resolution_of_identity+spectral_theorem}
\end{equation}
where $n$ enumerates all the eigenstates $\ket{w_{\rho n}}$ of $\hat{\mathcal{H}}_\text{BdG}^\rho$, \ie
\begin{equation}
	\hat{\mathcal{H}}_\text{BdG}^\rho \ket{w_{\rho n}} = E_{\rho n} \ket{w_{\rho n}} \quad\Leftrightarrow\quad \mathcal{H}_\text{BdG}^\rho(\vect{r}) w_{\rho n}(\vect{r}) = E_{\rho n} w_{\rho n}(\vect{r}).
    \label{eq:BdG_equation}
\end{equation}
Here we fixed the normalization of the eigenstates to $\braket{w_{\rho n}}{w_{\rho n}}=V$ (see section~\ref{sec:normalization_of_eigenvectors}). Equation \eqref{eq:BdG_equation} is usually called the \emph{(Dirac--)Bogoliubov--de Gennes equation} \cite{Beenakker2006,Titov2006,Kopnin2010}. Writing the Nambu structure explicitly it reads
\begin{equation}
	\begin{pmatrix}
		\mathcal{H}^\rho(\vect{r}) & \Delta(\vect{r}) \\
        \Delta^*(\vect{r}) & -\mathcal{H}^\rho(\vect{r})
	\end{pmatrix}
    \begin{pmatrix}
    	u_{\rho n}(\vect{r}) \\
        v_{\rho n}(\vect{r})
    \end{pmatrix} = E_{\rho n}
    \begin{pmatrix}
    	u_{\rho n}(\vect{r}) \\
        v_{\rho n}(\vect{r})
    \end{pmatrix}.
    \label{eq:BdG_equation_Nambu}
\end{equation}
Using the spectral theorem \eqref{eq:resolution_of_identity+spectral_theorem} in \eqref{eq:Hhat} yields
\begin{equation}
	\delta(\vect{r}-\vect{r}') \mathcal{H}_\text{BdG}^\rho(\vect{r}) = \frac{1}{V} \sum_n E_{\rho n} \bra{\vect{r}}\ket{w_{\rho n}}\bra{w_{\rho n}}\ket{\vect{r}'} = \frac{1}{V} \sum_n E_{\rho n} w_{\rho n}(\vect{r}) w_{\rho n}^\dagger(\vect{r}')
\end{equation}
and furthermore substituting this to \eqref{eq:H_Psi_double_integral} brings $H_\text{BdG}$ into the diagonal form
\begin{align}
	H_\text{BdG} &= \frac{1}{2V} \sum_{\sigma\rho n} E_{\rho n} \int\dd{\vect{r}} \Psi_{\sigma\rho}^\dagger(\vect{r}) w_{\rho n}(\vect{r}) \int\dd{\vect{r}'} w_{\rho n}^\dagger(\vect{r}') \Psi_{\sigma\rho}(\vect{r}') + E^0 - 4\mu V\delta(0) \\
    &= \frac{1}{2} \sum_{\sigma\rho n} E_{\rho n} \gamma_{\sigma\rho n}^\dagger \gamma_{\sigma\rho n} + E^0 - 4\mu V\delta(0), \label{eq:H_gamma}
\end{align}
where we defined the \emph{Bogoliubon operator} or the \emph{Bogoliubov transformation}
\begin{equation}
	\gamma_{\sigma\rho n} \coloneqq \frac{1}{\sqrt{V}} \int\dd{\vect{r}} w_{\rho n}^\dagger(\vect{r}) \Psi_{\sigma\rho}(\vect{r}).
    \label{eq:gamma}
\end{equation}
The Bogoliubons are not generally fermionic operators, but as we see in section~\ref{sec:making_the_bogoliubons_fermionic}, concentrating only on positive/only on negative energy operators makes them fermionic.

\subsection{Making the Bogoliubons fermionic}
\label{sec:making_the_bogoliubons_fermionic}

Since the noninteracting Hamiltonian has the time-reversal symmetry $\mathcal{H}^{\bar{\rho}*} = \mathcal{H}^\rho$, we have the symmetry
\begin{equation}
	\tau_y \mathcal{H}_\text{BdG}^{\bar{\rho}*} \tau_y = -\mathcal{H}_\text{BdG}^\rho
\end{equation}
of the BdG Hamiltonian. To see what this implies for the eigenenergies and eigenfunctions, we need to identify the so-far abstract index $n$. Taking the annihilation operators to be periodic in translations of the \BvK lattice $\LBK$ and $\vect{A}$ and $\Delta$ to be periodic in translations of the superlattice $SL$, let us take the ansatz that $n=(b,\vect{k})$, where $\vect{k}\in\LBK^*/\SLS^*$ belongs to the superlattice Brillouin zone (in different schemes) and $b=(\eta,\nu)$ enumerates the bands for each $\vect{k}$ with $\nu\in\{+,-\}$ giving the sign of energy of this band. We show in section~\ref{sec:BdG_equation_in_fourier_space} this ansatz to be consistent. Further assuming the bands to be ordered energy-wise such that the noninteracting energies have the symmetry $\epsilon_{\rho b\vect{k}} = \epsilon_{\bar{\rho} b\bar{\vect{k}}}$, and assuming this symmetry to be inherited to the superconducting state,
\begin{equation}
     E_{\rho b\vect{k}} = E_{\bar{\rho} b\bar{\vect{k}}}, 
\end{equation}
we find that
\begin{equation}
    w_{\rho\eta\bar{\nu}\vect{k}} = \imag\tau_y w_{\bar{\rho}\eta\nu\bar{\vect{k}}}^*.
    \label{eq:w_positive_negative_energy_symmetry}
\end{equation}
This then directly gives the symmetry
\begin{equation}
	\gamma_{\sigma\rho\eta\bar{\nu}\vect{k}} = -s(\sigma)\gamma_{\bar{\sigma}\bar{\rho}\eta\nu\bar{\vect{k}}}^\dagger
    \label{eq:gammaplus_gammaminus_symmetry}
\end{equation}
between the positive/negative energy Bogoliubons.

Choosing an orthogonal eigenbasis
\begin{equation}
	V\delta_{nn'} = \braket{w_{\rho n}}{w_{\rho n'}} = \int\dd{\vect{r}} w_{\rho n}^\dagger(\vect{r}) w_{\rho n'}(\vect{r}) 
    \label{eq:eigenvectors_orthogonality}
\end{equation}
(see section~\ref{sec:normalization_of_eigenvectors} for the chosen normalization) gives the first fermionic anticommutation relation $\acomm{\gamma_{\sigma\rho n}}{\gamma_{\sigma'\rho'n'}^\dagger} = \delta_{\sigma\sigma'} \delta_{\rho\rho'} \delta_{nn'}$ for all $n,n'$ and the second fermionic anticommutation relation $\acomm{\gamma_{\sigma\rho\eta\nu\vect{k}}}{\gamma_{\sigma'\rho'\eta'\nu'\vect{k}'}} = 0$ \emph{provided} $\nu=\nu'$ \ie that they both are either positive or negative energy operators. Concentrating then only on the positive energy ones, we get the desired result that the Bogoliubons are fermionic,
\begin{align}
	\acomm{\gamma_{\sigma\rho\eta+\vect{k}}}{\gamma_{\sigma'\rho'\eta'+\vect{k}'}^\dagger} &= \delta_{\sigma\sigma'} \delta_{\rho\rho'} \delta_{\eta\eta'} \delta_{\vect{k}\vect{k}'}, \label{eq:gamma_gammadagger_anticommutator}\\
    \acomm{\gamma_{\sigma\rho\eta+\vect{k}}}{\gamma_{\sigma'\rho'\eta'+\vect{k}'}} &= 0 \label{eq:gamma_gamma_anticommutator}.
\end{align}

Utilizing then the symmetry \eqref{eq:gammaplus_gammaminus_symmetry} of the positive/negative energy Bogoliubons, the fermionic anticommutation relations \eqref{eq:gamma_gammadagger_anticommutator} and \eqref{eq:gamma_gamma_anticommutator} of the positive energy Bogoliubons, and defining
\begin{equation}
	E^\text{gs} \coloneqq E^0 - 4\mu V\delta(0) - \frac{1}{2} \sum_{\sigma\rho n_+} E_{\rho n_+}
	\label{eq:Egs}
\end{equation}
allows us to finally write the Hamiltonian \eqref{eq:H_gamma} in the diagonal form
\begin{equation}
	H_\text{BdG} = \sum_{\sigma\rho n_+} E_{\rho n_+} \gamma_{\sigma\rho n_+}^\dagger \gamma_{\sigma\rho n_+} + E^\text{gs}
    \label{eq:H_gammaplus}
\end{equation}
with the operators $\gamma_{\sigma\rho n_+}$ being fermionic. According to the calculation above diagonalizing $H_\text{BdG}$ [\ie bringing it to the form \eqref{eq:H_gammaplus}] is equivalent to solving the Bogoliubov--de Gennes equation \eqref{eq:BdG_equation}. Note that since the Hamiltonian is diagonal in the fermionic positive energy Bogoliubons, $E^\text{gs}$ measures the ground-state energy.

\subsection{Self-consistency equation}

To write the definition \eqref{eq:Delta_psi} and \eqref{eq:Delta} of the order parameter $\Delta$ in the same Bogoliubon basis as we did for $H_\text{BdG}$ in \eqref{eq:H_gammaplus}, we need to invert the definition \eqref{eq:gamma}. Using the orthogonality condition \eqref{eq:eigenvectors_orthogonality} together with the resolution of identity \eqref{eq:resolution_of_identity+spectral_theorem} the inverse transformation can be shown to be
\begin{equation}
	\Psi_{\sigma\rho}(\vect{r}) = \frac{1}{\sqrt{V}} \sum_n w_{\rho n}(\vect{r}) \gamma_{\sigma\rho n}.
    \label{eq:Psi_gamma}
\end{equation}
This can also be written as a sum over only the positive-energy states as
\begin{equation}
    \Psi_{\sigma\rho}(\vect{r}) = \frac{1}{\sqrt{V}} \sum_{n_+} \left( w_{\rho n_+}(\vect{r}) \gamma_{\sigma\rho n_+} + \imag\tau_y w_{\bar\rho n_+}^*(\vect{r}) \gamma_{\bar\sigma\bar\rho n_+}^\dagger \right)
\end{equation}
by using the symmetries \eqref{eq:w_positive_negative_energy_symmetry} and \eqref{eq:gammaplus_gammaminus_symmetry}.

Since the positive-energy Bogoliubons are fermionic and we assume no interactions between them, they follow the Fermi--Dirac statistics
\begin{align}
	\expval{\gamma_{\sigma\rho n_+} \gamma_{\sigma'\rho'n_+'}} &= 0, \label{eq:gamma_gamma_expval} \\
    \expval{\gamma_{\sigma\rho n_+}^\dagger \gamma_{\sigma'\rho'n_+'}} &= \delta_{\sigma\sigma'} \delta_{\rho\rho'} \delta_{n_+n_+'} f(E_{\rho n_+}), \label{eq:gammadagger_gamma_expval}
\end{align}
where $f(E) \coloneqq [\e^{E/(\kB T)}+1]^{-1}$ is the Fermi--Dirac distribution at temperature $T$. Substituting then the relation \eqref{eq:Psi_gamma} in the definition \eqref{eq:Delta_psi} and \eqref{eq:Delta} of $\Delta$, using the fermionic anticommutation relations \eqref{eq:gamma_gammadagger_anticommutator} and \eqref{eq:gamma_gamma_anticommutator} of the Bogoliubons, and the thermal averages \eqref{eq:gamma_gamma_expval} and \eqref{eq:gammadagger_gamma_expval} then yields the \emph{self-consistency equation}
\begin{equation}
    \Delta_\alpha(\vect{r}) = -\frac{\lambda}{V} \sum_{\rho n_+} u_{\rho n_+,\alpha}(\vect{r}) v_{\rho n_+,\alpha}^*(\vect{r}) \tanh(\frac{E_{\rho n_+}}{2\kB T})
    \label{eq:self-consistency_equation_realspace}
\end{equation}
for the superconducting order parameter at sublattice $\alpha$.


\section{Equations in Fourier space}

\subsection{Bogoliubov--de Gennes equation in Fourier space}
\label{sec:BdG_equation_in_fourier_space}
Assuming the eigenfunctions $w_{\rho b\vect{k}'}=(u_{\rho b\vect{k}'}, v_{\rho b\vect{k}'})^\transpose:\R^2/\LBK\rightarrow\C^2$ (with $\vect{k}'\in\LBK^*/\SLS^*$ in the superlattice Brillouin zone in different schemes) to be periodic in translations of the large \BvK lattice $\LBK$, the pseudo vector potential $\vect{A}:\R^2/SL\rightarrow\R^2$ to be periodic in translations of the arbitrary superlattice $SL = \spn_\Z\{\vect{t}_1,\vect{t}_2\}$, and the order parameter $\Delta:\R^2/SL\rightarrow\C_{2\times 2}$ to be periodic in translations of the same lattice $SL$, we may expand them by \eqref{eq:fourier_series+coefficients} as the Fourier series
\begin{align}
	w_{\rho b\vect{k}'}(\vect{r}) &= \sum_{\vect{k}\in\LBK^*} \e^{\imag\vect{k}\vdot\vect{r}} \tilde{w}_{\rho b\vect{k}'}(\vect{k}) \qqtext{with} \tilde{w}_{\rho b\vect{k}'}(\vect{k}) = \frac{1}{\abs{\R^2/\LBK}} \int_{\R^2/\LBK}\dr \e^{-\imag\vect{k}\vdot\vect{r}} w_{\rho b\vect{k}'}(\vect{r}), \label{eq:w_fourier_series}\\
    \vect{A}(\vect{r}) &= \sum_{\vect{G}\in\SLS^*} \e^{\imag\vect{G}\vdot\vect{r}} \tilde{\vect{A}}(\vect{G}) \qqtext{with} \tilde{\vect{A}}(\vect{G}) = \frac{1}{\abs{\R^2/SL}} \int_{\R^2/SL}\dr \e^{-\imag\vect{G}\vdot\vect{r}} \vect{A}(\vect{r}), \label{eq:A_fourier_series}\\
    \Delta(\vect{r}) &= \sum_{\vect{G}\in\SLS^*} \e^{\imag\vect{G}\vdot\vect{r}} \tilde{\Delta}(\vect{G}) \qqtext{with} \tilde{\Delta}(\vect{G}) = \frac{1}{\abs{\R^2/SL}} \int_{\R^2/SL}\dr \e^{-\imag\vect{G}\vdot\vect{r}} \Delta(\vect{r}). \label{eq:Delta_fourier_series}
\end{align}
Substituting these Fourier series to the BdG equation \eqref{eq:BdG_equation}, shifting the $\vect{k}$ sums properly, writing the $\vect{k}$ sum over the whole space $\LBK^*$ as a sum over the superlattice Brillouin zone $\LBK^*/\SLS^*$ (in different schemes) plus shifted copies of this,
\begin{equation}
	\sum_{\vect{k}\in\LBK^*} g(\vect{k}) = \sum_{\vect{k}\in\LBK^*/\SLS^*} \sum_{\vect{G}\in\SLS^*} g(\vect{k}+\vect{G})
    \label{eq:k_sum_division}
\end{equation}
($g$ being a test function), and using the linear independence of the exponentials yields for all $\vect{k},\vect{k}'\in\LBK^*/\SLS^*$, $\vect{G}\in\SLS^*$ the BdG equation in Fourier space,
\begin{equation}
	\sum_{\vect{G}'\in\SLS^*} \tilde{\mathcal{H}}_{\text{BdG},\vect{G}\vect{G}'}^\rho(\vect{k}) \tilde{w}_{\rho b\vect{k}'}(\vect{k}+\vect{G}') = E_{\rho b\vect{k}'} \tilde{w}_{\rho b\vect{k}'}(\vect{k}+\vect{G}).
    \label{eq:BdG_equation_kspace_components}
\end{equation}
Here the Nambu-matrix
\begin{equation}
	\tilde{\mathcal{H}}_{\text{BdG},\vect{G}\vect{G}'}^\rho(\vect{k}) \coloneqq
    \begin{pmatrix}
    	\tilde{\mathcal{H}}_{\vect{G}\vect{G}'}^\rho(\vect{k}) & \tilde{\Delta}(\vect{G}-\vect{G}') \\
        \tilde{\Delta}^*(\vect{G}'-\vect{G}) & -\tilde{\mathcal{H}}_{\vect{G}\vect{G}'}^\rho(\vect{k})
    \end{pmatrix}
\end{equation}
is the Fourier-space version of the BdG Hamiltonian matrix element with
\begin{equation}
	\tilde{\mathcal{H}}_{\vect{G}\vect{G}'}^\rho(\vect{k}) \coloneqq \hbar\vF\vect{\sigma}^\rho\vdot \left( (\vect{k}+\vect{G})\delta_{\vect{G}\vect{G}'} + \rho\tilde{\vect{A}}(\vect{G}-\vect{G}') \right) -\mu\delta_{\vect{G}\vect{G}'}
\end{equation}
being the Fourier-space version of the noninteracting Hamiltonian matrix element.

By collecting the $\vect{G},\vect{G}'$ components $\tilde{\mathcal{H}}_{\text{BdG},\vect{G}\vect{G}'}^\rho(\vect{k})$ into a countably infinite $\vect{G}$-space matrix
\begin{equation}
	\tilde{\underline{\mathcal{H}}}_\text{BdG}^\rho(\vect{k}) \coloneqq
    \begin{pmatrix}
        \tilde{\mathcal{H}}_{\text{BdG},\vect{G}\vect{G}'}^\rho(\vect{k})
    \end{pmatrix}_{\vect{G},\vect{G}'\in\SLS^*}
    \label{eq:Htildeunderbar}
\end{equation}
and the $\vect{G}$ components $\tilde{w}_{\rho b\vect{k}'}(\vect{k}+\vect{G})$ into a countably infinite $\vect{G}$-space vector
\begin{equation}
	\tilde{\underline{w}}_{\rho b\vect{k}'}(\vect{k}) \coloneqq
    \begin{pmatrix}
        \tilde{w}_{\rho b\vect{k}'}(\vect{k}+\vect{G})
    \end{pmatrix}_{\vect{G}\in\SLS^*}
    \label{eq:wtildeunderbar}
\end{equation}
the BdG equation \eqref{eq:BdG_equation_kspace_components} becomes a matrix eigenvalue equation
\begin{equation}
	\tilde{\underline{\mathcal{H}}}_\text{BdG}^\rho(\vect{k}) \tilde{\underline{w}}_{\rho b\vect{k}'}(\vect{k}) = E_{\rho b\vect{k}'} \tilde{\underline{w}}_{\rho b\vect{k}'}(\vect{k}).
\end{equation}
Obviously we must have $\vect{k}'=\vect{k}$, so we may set
\begin{equation}
	\tilde{w}_{\rho b\vect{k}'}(\vect{k}+\vect{G}) = \delta_{\vect{k}\vect{k}'}\tilde{w}_{\rho b\vect{k}}(\vect{k}+\vect{G})
    \label{eq:eigenvectors_delta}
\end{equation}
and the BdG equation in Fourier space reads
\begin{equation}
	\tilde{\underline{\mathcal{H}}}_\text{BdG}^\rho(\vect{k}) \tilde{\underline{w}}_{\rho b\vect{k}}(\vect{k}) = E_{\rho b\vect{k}} \tilde{\underline{w}}_{\rho b\vect{k}}(\vect{k}).
	\label{eq:BdG_equation_kspace}
\end{equation}
This is clearly a separate problem for each $\vect{k}\in\LBK^*/\SLS^*$, and for each $\vect{k}$ there are exactly $2\times 2\times\abs{\SLS^*}$ (the matrix size) solutions labelled by the band index $b$. Thus our original ansatz $n=(b,\vect{k})$ is consistent. Equation~\eqref{eq:BdG_equation_kspace} is the form of the BdG equation implemented and solved in the numerics.

Note that in the reduced zone scheme here $\vect{k}=k_1\vect{G}_1+k_2\vect{G}_2\in\LBK^*/SL^*$ is periodic in both $k_1\in[-\frac{1}{2},\frac{1}{2}[$ and $k_2\in[-\frac{1}{2},\frac{1}{2}[$, so that both $k_1$ and $k_2$ are periodic Bloch momenta. In this case the $\vect{G}$-translations in \eqref{eq:Htildeunderbar} in both $\vect{G}_1$ and $\vect{G}_2$ directions are transformed to new bands. This is also traditionally called the reduced zone (or the repeated zone) scheme. However, in the case of $\vect{A}$ and $\Delta$ being constant in the $\vect{t}_2$ direction (the 1D potential case), we are also allowed to choose the mixed zone scheme, as discussed in section~\ref{sec:fourier_series}. In this case $\vect{k}=k_1\vect{G}_1+k_2\vect{G}_2\in\LBK^*/SL_1^*$ is periodic in $k_1\in[-\frac{1}{2},\frac{1}{2}[$ but not in $k_2\in]-\infty,\infty[$, so that $k_1$ is a periodic Bloch momentum while $k_2$ is a nonperiodic real momentum. In this case the $\vect{G}$-translations in \eqref{eq:Htildeunderbar} are done only in $\vect{G}_1$ direction, and only these are transformed to new bands. Traditionally the $k_1$ direction is then called to be in the reduced zone scheme and the $k_2$ direction in the extended zone scheme. This is why we call this the mixed zone scheme.

Equation~\eqref{eq:eigenvectors_delta} can be used to write the Fourier series \eqref{eq:w_fourier_series} of $w$ in the Bloch form. Further dividing the $\vect{k}'$ sum as in \eqref{eq:k_sum_division} gives the Fourier series in the Bloch form
\begin{align}
	w_{\rho b\vect{k}}(\vect{r}) &= \sum_{\vect{k}'\in\LBK^*} \e^{\imag\vect{k}'\vdot\vect{r}} \tilde{w}_{\rho b\vect{k}}(\vect{k}') = \sum_{\vect{k}'\in\LBK^*/\SLS^*} \sum_{\vect{G}\in\SLS^*} \e^{\imag(\vect{k}'+\vect{G})} \tilde{w}_{\rho b\vect{k}}(\vect{k}'+\vect{G}) \\
    &= \e^{\imag\vect{k}\vdot\vect{r}} \sum_{\vect{G}\in\SLS^*} \e^{\imag\vect{G}\vdot\vect{r}} \tilde{w}_{\rho b\vect{k}}(\vect{k}+\vect{G}), \label{eq:w_fourier_series_bloch}
\end{align}
where the function multiplying $\e^{\imag\vect{k}\vdot\vect{r}}$ is periodic in translations of the superlattice $SL$.

\subsection{Normalization of eigenvectors}
\label{sec:normalization_of_eigenvectors}
Equation~\eqref{eq:BdG_equation_kspace} is the form of the BdG equation we are solving numerically. For the eigenvectors we choose the normalization $\norm{\tilde{\underbar{w}}_{\rho b\vect{k}}(\vect{k})}=1$. To see what this means for the eigenstates $\ket{w_{\rho b\vect{k}}}$, we may use Parseval's theorem
\begin{equation}
	\sum_{\vect{k}'\in\LBK^*} \norm{\tilde{w}_{\rho b\vect{k}}(\vect{k}')}^2 = \frac{1}{\abs{\R^2/\LBK}} \int_{\R^2/\LBK}\dd{\vect{r}} \norm{w_{\rho b\vect{k}}(\vect{r})}^2 = \frac{1}{V} \braket{w_{\rho b\vect{k}}}{w_{\rho b\vect{k}}}.
\end{equation}
On the other hand dividing the $\vect{k}$ sum as in \eqref{eq:k_sum_division}, using \eqref{eq:eigenvectors_delta}, and using the definition \eqref{eq:wtildeunderbar}, the l.h.s. gives
\begin{equation}
	\sum_{\vect{k}'\in\LBK^*} \norm{\tilde{w}_{\rho b\vect{k}}(\vect{k}')}^2 = \sum_{\vect{G}\in\SLS^*} \norm{\tilde{w}_{\rho b\vect{k}}(\vect{k}+\vect{G})}^2 = \norm{\tilde{\underbar{w}}_{\rho b\vect{k}}(\vect{k})}^2 = 1,
\end{equation}
which then yields the normalization
\begin{equation}
	\braket{w_{\rho b\vect{k}}}{w_{\rho b\vect{k}}} = V.
    \label{eq:eigenvectors_normalization}
\end{equation}

\subsection{Self-consistency equation in Fourier space}
\label{sec:self-consistency_equation_in_fourier_space}
Using the Fourier series \eqref{eq:A_fourier_series}, \eqref{eq:Delta_fourier_series}, and \eqref{eq:w_fourier_series_bloch} we may write the self-consistency equation \eqref{eq:self-consistency_equation_realspace} in Fourier space as
\begin{equation}
	\tilde{\Delta}_\alpha(\vect{G}) = -\frac{\lambda}{V} \sum_{\rho b_+} \sum_{\vect{k}\in\LBK^*/\SLS^*} \sum_{\vect{G}'\in\SLS^*} \tilde{u}_{\rho b_+\vect{k},\alpha}(\vect{k}+\vect{G}') \tilde{v}_{\rho b_+\vect{k},\alpha}^*(\vect{k}+\vect{G}'-\vect{G}) \tanh(\frac{E_{\rho b_+\vect{k}}}{2\kB T})
\end{equation}
for all $\vect{G}\in\SLS^*$. Assuming the \BvK cell to be large, \ie $\LBK$ to be sparse or $\LBK^*$ to be dense, we may approximate the $\vect{k}$ sum as an integral
\begin{equation}
	\sum_{\vect{k}\in\LBK^*/SL^*} \approx \frac{V}{(2\pi)^2} \int_{\R^2/SL^*} \dd{\vect{k}},
    \label{eq:k_sum_integral_approximation}
\end{equation}
yielding the self-consistency equation
\begin{equation}
	\tilde{\Delta}_\alpha(\vect{G}) = -\frac{\lambda}{(2\pi)^2} \sum_{\rho b_+} \sum_{\vect{G}'\in\SLS^*}  \int_{\R^2/\SLS^*} \dd{\vect{k}} \tilde{u}_{\rho b_+\vect{k},\alpha}(\vect{k}+\vect{G}') \tilde{v}_{\rho b_+\vect{k},\alpha}^*(\vect{k}+\vect{G}'-\vect{G}) \tanh(\frac{E_{\rho b_+\vect{k}}}{2\kB T}).
	\label{eq:self-consistency_equation_kspace}
\end{equation}
This form, where the integration region is the rather abstract $\R^2/\SLS^*$, is convenient when doing analytical calculations. But in numerical calculations it is easier to integrate over simpler areas instead, which is done next by change of variables.

\subsubsection{Reduced zone scheme}
In the reduced zone scheme we have $\SLS=SL_\text{RZ}=SL$, meaning that the integration region $\R^2/\SLS^*=\R^2/SL^*$ can be interpreted as the parallelogram defined by $\vect{G}_1$ and $\vect{G}_2$. Making a change of variables with the function
\begin{equation}
	\phi:\left[-\tfrac{1}{2},\tfrac{1}{2}\right[^2\rightarrow\R^2/SL^*, \quad \phi(k_1,k_2) = k_1\vect{G}_1+k_2\vect{G}_2
\end{equation}
the $\vect{k}$ integral in \eqref{eq:self-consistency_equation_kspace} can be written as
\begin{equation}
	\int_{\R^2/SL^*}\dk g(\vect{k}) = \int_{\phi([-\frac{1}{2},\frac{1}{2}[^2)}\dk g(\vect{k}) = \int_{[-\frac{1}{2},\frac{1}{2}[^2}\dk (g\circ\phi)(\vect{k})\abs{J_\phi(\vect{k})} = \abs{\vect{G}_1\cross\vect{G}_2} \int_{-\frac{1}{2}}^\frac{1}{2}\dd{k}_1 \int_{-\frac{1}{2}}^\frac{1}{2}\dd{k}_2 g(k_1\vect{G}_1+k_2\vect{G}_2)
\end{equation}
($g$ being a test function) where the absolute value of the Jacobian determinant,
\begin{equation}
	\abs{J_\phi(\vect{k})} = \abs{\vect{G}_1\cross\vect{G}_2} = \abs{\R^2/SL^*},
\end{equation}
is the area of the parallelogram defined by $\vect{G}_1$ and $\vect{G}_2$.

The self-consistency equation \eqref{eq:self-consistency_equation_kspace} then becomes
\begin{align}
	\tilde{\Delta}_\alpha(\vect{G}) &= -\frac{\lambda}{(2\pi)^2} \abs{\vect{G}_1\cross\vect{G}_2} \sum_{\rho b_+} \sum_{\vect{G}'\in SL^*} \int_{-\frac{1}{2}}^\frac{1}{2}\dd{k_1} \int_{-\frac{1}{2}}^{\frac{1}{2}}\dd{k_2} \notag\\
    &\tilde{u}_{\rho b_+,\alpha}(k_1\vect{G}_1+k_2\vect{G}_2+\vect{G}') \tilde{v}_{\rho b_+,\alpha}^*(k_1\vect{G}_1+k_2\vect{G}_2+\vect{G}'-\vect{G}) \tanh(\frac{E_{\rho b_+}(k_1\vect{G}_1+k_2\vect{G}_2)}{2\kB T}),
    \label{eq:self-consistency_equation_kspace_reducedzone}
\end{align}
where we dropped $k_1\vect{G}_1+k_2\vect{G}_2$ from the subscripts of $\tilde{u}$ and $\tilde{v}$ and denoted $E_{\rho b_+}(\vect{k}) \coloneqq E_{\rho b_+\vect{k}}$. In the numerics we have to make a cutoff to the countably infinite $\vect{G}'$ and $b_+$ sums, both corresponding to a cutoff at high energies. This cutoff can be seen to come from the electron--phonon coupling.

\subsubsection{Mixed zone scheme}
In the mixed zone scheme we have $\SLS=SL_\text{MZ}=SL_1$, meaning that the integration region $\R^2/\SLS^*=\R^2/SL_1^*$ can be interpreted as a semi-infinite parallelogram, where the finite side is $\vect{G}_1$ and the infinite side is in the $\vect{G}_2$ direction. Making a change of variables with the function
\begin{equation}
	\phi:\left[-\tfrac{1}{2},\tfrac{1}{2}\right[\cross\R\rightarrow\R^2/SL_1^*, \quad \phi(k_1,k_2) = k_1\vect{G}_1+k_2\vect{G}_2
\end{equation}
the $\vect{k}$ integral in \eqref{eq:self-consistency_equation_kspace} can be written as
\begin{align}
	\int_{\R^2/SL_1^*}\dk g(\vect{k}) &= \int_{\phi([-\frac{1}{2},\frac{1}{2}[\cross\R)}\dk g(\vect{k}) = \int_{[-\frac{1}{2},\frac{1}{2}[\cross\R}\dk (g\circ\phi)(\vect{k})\abs{J_\phi(\vect{k})} \\
	&= \abs{\vect{G}_1\cross\vect{G}_2} \int_{-\frac{1}{2}}^\frac{1}{2}\dd{k}_1 \int_{-\infty}^\infty\dd{k}_2 g(k_1\vect{G}_1+k_2\vect{G}_2)
\end{align}
($g$ being a test function) where the absolute value of the Jacobian determinant,
\begin{equation}
	\abs{J_\phi(\vect{k})} = \abs{\vect{G}_1\cross\vect{G}_2} = \abs{\R^2/SL^*},
\end{equation}
is the area of the parallelogram defined by $\vect{G}_1$ and $\vect{G}_2$.

The self-consistency equation \eqref{eq:self-consistency_equation_kspace} then becomes
\begin{align}
	\tilde{\Delta}_\alpha(\vect{G}) &= -\frac{\lambda}{(2\pi)^2} \abs{\vect{G}_1\cross\vect{G}_2} \sum_{\rho b_+} \sum_{\vect{G}'\in SL^{1*}} \int_{-\frac{1}{2}}^\frac{1}{2}\dd{k_1} \int_{-\infty}^\infty\dd{k_2} \notag\\
    &\tilde{u}_{\rho b_+,\alpha}(k_1\vect{G}_1+k_2\vect{G}_2+\vect{G}') \tilde{v}_{\rho b_+,\alpha}^*(k_1\vect{G}_1+k_2\vect{G}_2+\vect{G}'-\vect{G}) \tanh(\frac{E_{\rho b_+}(k_1\vect{G}_1+k_2\vect{G}_2)}{2\kB T}),
    \label{eq:self-consistency_equation_kspace_integral_mixedzone}
\end{align}
where we dropped $k_1\vect{G}_1+k_2\vect{G}_2$ from the subscripts of $\tilde{u}$ and $\tilde{v}$ and denoted $E_{\rho b_+}(\vect{k}) \coloneqq E_{\rho b_+\vect{k}}$. While in the reduced zone scheme in \eqref{eq:self-consistency_equation_kspace_reducedzone} both the $k_1$ and $k_2$ momentum directions are cut-off in the $\vect{G}'$ sum, in this case only the $k_1$ momentum direction is cut-off in the $\vect{G}'$ sum while the $k_2$ direction is handled by a momentum cutoff in the limits of the corresponding improper integral. Also the band sum $b_+$ has a cutoff but it is generally different from the one in the reduced zone scheme, as the meaning of bands is different.

\subsection{Ground state energy expectation values}
Using the Hamiltonian \eqref{eq:H_gammaplus} and equation~\eqref{eq:gammadagger_gamma_expval} the energy density expectation value in the ground state for the order parameter $(\Delta_A,\Delta_B)$ can be shown to be
\begin{align}
	\frac{1}{V} &\expval{H_\text{BdG}(\Delta_A,\Delta_B)} = \frac{1}{V} \sum_{\sigma\rho b_+} \sum_{\vect{k}\in\LBK^*/\SLS^*} E_{\rho b_+\vect{k}}(\Delta_A,\Delta_B) \expval{\gamma_{\sigma\rho b_+\vect{k}}^\dagger \gamma_{\sigma\rho b_+\vect{k}}} + \frac{1}{V} E^\text{gs}(\Delta_A,\Delta_B) \\
    &\approx -\frac{1}{(2\pi)^2} \sum_{\rho b_+} \int_{\R^2/\SLS^*} \dd{\vect{k}} E_{\rho b_+\vect{k}}(\Delta_A,\Delta_B) \tanh(\frac{E_{\rho b_+\vect{k}}(\Delta_A,\Delta_B)}{2\kB T}) + \frac{1}{V} E^0(\Delta_A,\Delta_B) - 4\mu\delta(0),
\end{align}
where in the last step we also approximated the sum as an integral as in \eqref{eq:k_sum_integral_approximation}. We would like to show that a ground state with a zero phase difference between the $\Delta$ components has the lowest energy, and thus we define for each $\delta\in\R$ the \emph{Josephson energy density}
\begin{align}
	&\frac{E_J(\theta)}{V} \coloneqq \frac{1}{V}\expval{H_\text{BdG}(\e^{\imag\theta/2}\delta,\e^{-\imag\theta/2}\delta)} - \frac{1}{V}\expval{H_\text{BdG}(\delta,\delta)} = \\
    &-\frac{1}{(2\pi)^2} \sum_{\rho b_+} \int_{\R^2/\SLS^*} \dd{\vect{k}} \left[ E_{\rho b_+\vect{k}}(\e^{\imag\theta/2}\delta,\e^{-\imag\theta/2}\delta) \tanh(\frac{E_{\rho b_+\vect{k}}(\e^{\imag\theta/2}\delta,\e^{-\imag\theta/2}\delta}{2\kB T}) - E_{\rho b_+\vect{k}}(\delta,\delta) \tanh(\frac{E_{\rho b_+\vect{k}}(\delta,\delta)}{2\kB T}) \right],
    \label{eq:josephson_energy}
\end{align}
where the $E^0/V$ and $4\mu\delta(0)$ terms cancel out because they do not contain the phases of $\Delta_{A/B}$. The integral is then calculated in the different schemes as described in section~\ref{sec:self-consistency_equation_in_fourier_space}.

\subsection{Superfluid weight}
Writing the result of \cite{Liang2017} for the superfluid weight $\Ds$ in the superlattice-folded picture near the Dirac points, we get for the $\mu,\nu\in\{x,y\}$ component
\begin{align}
    \Ds_{\mu\nu} = \frac{1}{V} \sum_{\rho bb'} \sum_{\vect{k}\in\LBK^*/\SLS^*} \mathcal{F}_{\rho bb'\vect{k}} \bigg( &\tilde{\underline{w}}_{\rho b\vect{k}}^\dagger(\vect{k}) \partial_\mu\tilde{\underline{\mathcal{H}}}_\text{BdG}^\rho(\vect{k}) \tau_z \tilde{\underline{w}}_{\rho b'\vect{k}}(\vect{k}) \tilde{\underline{w}}_{\rho b'\vect{k}}^\dagger(\vect{k}) \tau_z \partial_\nu\tilde{\underline{\mathcal{H}}}_\text{BdG}^\rho(\vect{k}) \tilde{\underline{w}}_{\rho b\vect{k}}(\vect{k}) \notag\\
    -&\tilde{\underline{w}}_{\rho b\vect{k}}^\dagger(\vect{k}) \partial_\mu\tilde{\underline{\mathcal{H}}}_\text{BdG}^\rho(\vect{k}) \tilde{\underline{w}}_{\rho b'\vect{k}}(\vect{k}) \tilde{\underline{w}}_{\rho b'\vect{k}}^\dagger(\vect{k}) \partial_\nu\tilde{\underline{\mathcal{H}}}_\text{BdG}^\rho(\vect{k}) \tilde{\underline{w}}_{\rho b\vect{k}}(\vect{k}) \bigg),
\end{align}
where the prefactor is
\begin{equation}
    \mathcal{F}_{\rho bb'\vect{k}} =
    \begin{cases}
        f'(E_{\rho b\vect{k}}), &\quad\text{if $E_{\rho b\vect{k}}=E_{\rho b'\vect{k}}$}, \\
        \frac{f(E_{\rho b\vect{k}})-f(E_{\rho b'\vect{k}})}{E_{\rho b\vect{k}}-E_{\rho b'\vect{k}}}, &\quad\text{otherwise},
    \end{cases}
\end{equation}
the $b,b'$ band sums are calculated over \emph{both the positive and negative energy bands}, the partial derivatives are shortly denoted as $\partial_\mu\coloneqq\partial_{k_\mu}$, and the energies and eigenvectors are calculated from the BdG equation \eqref{eq:BdG_equation_kspace}. Since $\Delta$ is $\vect{k}$ independent, we have $\partial_\mu\tilde{\underline{\mathcal{H}}}_\text{BdG}^\rho(\vect{k}) = \tau_z\partial_\mu\tilde{\underline{\mathcal{H}}}^\rho(\vect{k}) = \hbar\vF\tau_z\sigma_\mu^\rho$, yielding
\begin{align}
    \Ds_{\mu\nu} = \frac{(\hbar\vF)^2}{(2\pi)^2} \sum_{\rho bb'} \int_{\R^2/\SLS^*}\dd{\vect{k}} \mathcal{F}_{\rho bb'\vect{k}} \bigg( &\tilde{\underline{w}}_{\rho b\vect{k}}^\dagger(\vect{k}) \sigma_\mu^\rho \tilde{\underline{w}}_{\rho b'\vect{k}}(\vect{k}) \tilde{\underline{w}}_{\rho b'\vect{k}}^\dagger(\vect{k}) \sigma_\nu^\rho \tilde{\underline{w}}_{\rho b\vect{k}}(\vect{k}) \notag\\
    -&\tilde{\underline{w}}_{\rho b\vect{k}}^\dagger(\vect{k}) \tau_z\sigma_\mu^\rho \tilde{\underline{w}}_{\rho b'\vect{k}}(\vect{k}) \tilde{\underline{w}}_{\rho b'\vect{k}}^\dagger(\vect{k}) \tau_z\sigma_\nu^\rho \tilde{\underline{w}}_{\rho b\vect{k}}(\vect{k}) \bigg),
\end{align}
where we also approximated the $\vect{k}$ sum as an integral.


\section{Details of the numerical calculation}

\subsection{Solving the self-consistency equation}
We start by calculating analytically the Fourier coefficients of $\vect{A}$ by \eqref{eq:A_fourier_series}, which are given in \eqref{eq:fourier_coefficients_cos1D_reducedzone}, \eqref{eq:fourier_coefficients_cos2D_reducedzone}, \eqref{eq:fourier_coefficients_triangleSquare1D_reducedzone}, and \eqref{eq:fourier_coefficients_cos1D_mixedzone}. The combination of the BdG equation \eqref{eq:BdG_equation_kspace} and the self-consistency equation \eqref{eq:self-consistency_equation_kspace} is then solved by the fixed-point iteration method, \ie starting from an initial guess of the pair $(\tilde{\Delta}_A,\tilde{\Delta}_B)$, solving the eigenvectors $\tilde{w}_{\rho b_+\vect{k}} = (\tilde{u}_{\rho b_+\vect{k}}, \tilde{v}_{\rho b_+\vect{k}})^\transpose$ and eigenvalues $E_{\rho b_+\vect{k}}$ from the BdG equation \eqref{eq:BdG_equation_kspace}, using these eigenvectors and eigenvalues to calculate new values for $\tilde{\Delta}_A$ and $\tilde{\Delta}_B$ from \eqref{eq:self-consistency_equation_kspace}, and then solving the BdG equation again with these new $\tilde{\Delta}$'s. This iteration is then continued until convergence to some relative or absolute error in all of the components of $\tilde{\Delta}_{A/B}$.

\subsection{Initial guess of the order parameter}
The initial guess of the order parameter is always chosen such that both sublattice-components are the same real constant in space, $\Delta_A(\vect{r}) = \Delta_B(\vect{r}) = 0.1\beta\abs{\lambda}/d^2$. The exact value of the constant seems to have no effect on the result of the fixed-point iteration, merely affecting the speed of convergence, which is understandable from the Banach fixed-point theorem. In Fourier space the initial guess reads $\tilde{\Delta}_A(\vect{G}) = \tilde{\Delta}_B(\vect{G}) = 0.1\beta\abs{\lambda}/d^2 \delta_{\vect{G},0}$. One should note that above we are fixing the overall phase of $(\Delta_A,\Delta_B)$ to be real, since it can be shown that starting from a given overall phase, the fixed-point iteration conserves that phase at each iteration.

\begin{figure}
    \includegraphics{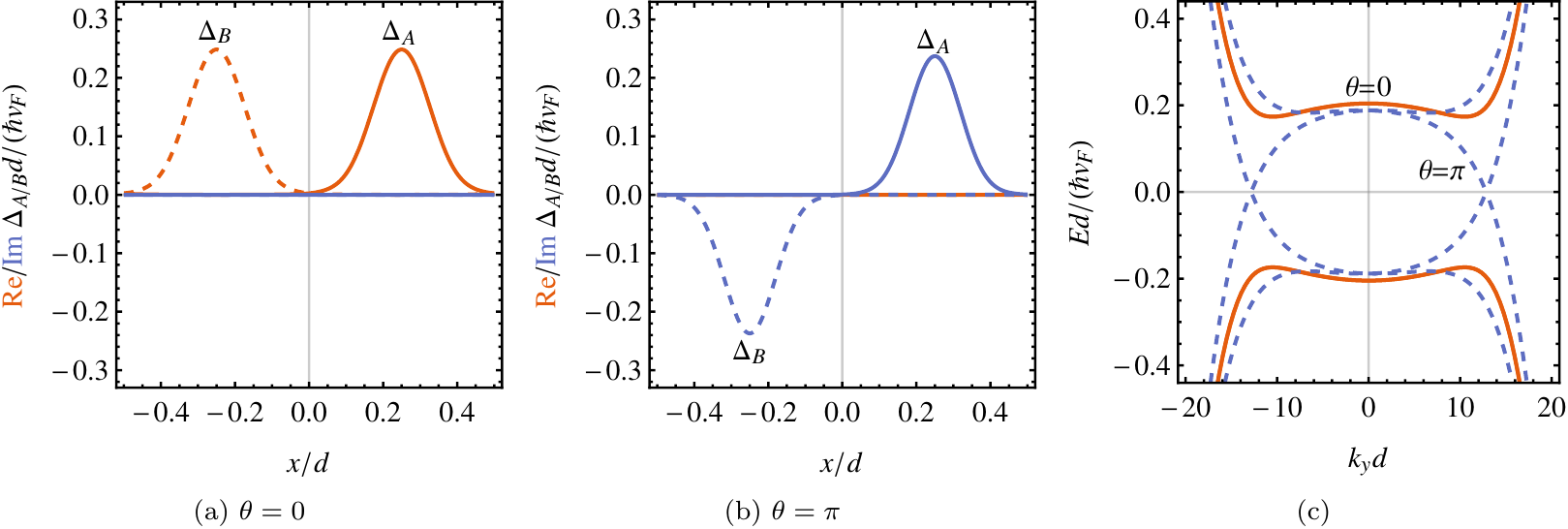}
    \caption{(a,b) Profile of the self-consistent $\Delta_{A/B}$ ($A$ joined, $B$ dashed lines) for the two initial guesses with the relative phases (a) $\theta=0$ and (b) $\theta=\pi$, along the line $(x,0)$. In (a) the imaginary part is zero while in (b) the real part is zero. (c) Profile of the corresponding dispersion relation for $\theta=0$ (joined lines) and $\theta=\pi$ (dashed lines) along the line $(0,k_y)$ in the mixed zone scheme. Here $\vect{A}=\vect{A}_{\cos}^\text{1D}$, $\beta=30$, $\mu=0$, $T=0$, and $\lambda/(\hbar\vF d)=-0.01$.}
    \label{fig:Deltavsx+Evsky_theta}
\end{figure}

We are also fixing the relative phase of $\Delta_A$ and $\Delta_B$ to zero, and the justification for this is discussed next. It can be numerically shown that the initial guess $(\Delta_A,\Delta_B) = (\e^{\imag\theta/2}\delta,\e^{-\imag\theta/2}\delta)$ with $\theta\neq\pi$, $\delta\in\R$ always converges to the $\theta=0$ solution, shown in figure~\ref{fig:Deltavsx+Evsky_theta}(a) for the 1D cosine potential $\vect{A}_{\cos}^\text{1D}$, by the fixed-point iteration. On the other hand the $\theta=\pi$ initial guess converges to a different solution, shown in figure~\ref{fig:Deltavsx+Evsky_theta}(b). The dispersion relations of these two different solutions are shown in figure~\ref{fig:Deltavsx+Evsky_theta}(c), showing how the degeneracy of the $\theta=0$ state is lifted and how the gap is closed in the $\theta=\pi$ state. The dispersion relations alone can be used to calculate the ground state energies of these states by using \eqref{eq:josephson_energy}, showing that the $\theta=0$ solution always yields a lower energy, also at finite temperatures. This allows us to discard the $\theta=\pi$ solution and to concentrate only on the $\theta=0$ solution, which, as discussed above, can always be reached by using the initial guess with $\theta=0$.

\begin{figure}
    \includegraphics{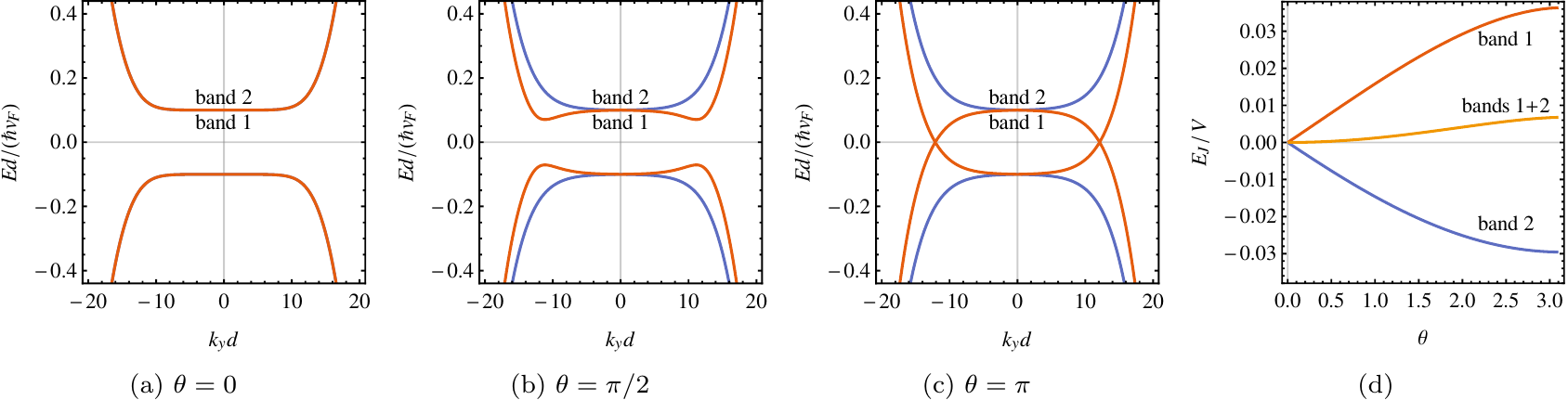}
    \caption{Effect of the relative phase $\theta$ on the non-self-consistent, constant order parameter $(\Delta_A,\Delta_B) = (\e^{\imag\theta/2}\delta, \e^{-\imag\theta/2}\delta)$ for $\vect{A}_{\cos}^\text{1D}$ with $\beta=30$, $\mu=0$, $T=0$, and $\delta=0.1\hbar\vF/d$. (a--c) Dispersion of two of the lowest energy bands with increasing $\theta$ in the mixed zone scheme. (d) Ground state energy density difference [the Josephson energy density \eqref{eq:josephson_energy}] between the $\theta\neq 0$ and $\theta=0$ solutions, by calculating only the contribution from the lowest band 1, the second lowest band 2, and both bands 1 and 2.}
    \label{fig:Evsky_theta+EJvstheta}
\end{figure}

The above behavior as a function of the relative phase $\theta$ can be understood by using a constant, non-self-consistent $(\Delta_A,\Delta_B) = (\e^{\imag\theta/2}\delta, \e^{-\imag\theta/2}\delta)$ with $\delta\in\R$. The dispersion of two of the lowest positive-energy bands is shown in figures~\ref{fig:Evsky_theta+EJvstheta}(a--c) as a function of $\theta$ showing how the finite $\theta$ removes the degeneracy. Two competing effects are observed: energies in band 2 are slightly increased (integral-wise), while the decrease in energy in band 1 is more dominant. This can be seen also in figure~\ref{fig:Evsky_theta+EJvstheta}(d), presenting the ground state energy density difference (the Josephson energy density) between the $\theta\neq 0$ and $\theta=0$ solutions by \eqref{eq:josephson_energy}, whose opposite value at $T=0$ is essentially given by the difference in the $\vect{k}$ integral of the corresponding dispersions. Looking at the contributions from the individual bands 1 and 2 it is clearly seen how the contribution from band 1 is more dominant, giving the net result that the $\theta\neq 0$ solution always gives a higher ground state energy than the $\theta=0$ solution. At finite temperatures the behavior is otherwise the same but with smaller energy differences. The same qualitative behavior is seen for all the tested potentials.

\bibliography{library}